\documentclass{aa}

\usepackage{graphicx}
\usepackage{txfonts}
\usepackage{savesym}
\usepackage{txfonts}
\usepackage{amsmath}
\usepackage{aas_macros}

\usepackage{natbib,twoopt}
\usepackage[draft]{hyperref}
\bibpunct{(}{)}{;}{a}{}{,} %% natbib format for A&A and ApJ

\makeatletter
\newcommandtwoopt{\citeads}[3][][]{\href{http://adsabs.harvard.edu/abs/#3}%
{\def\hyper@linkstart##1##2{}%
\let\hyper@linkend\@empty\citealp[#1][#2]{#3}}}
\newcommandtwoopt{\citepads}[3][][]{\href{http://adsabs.harvard.edu/abs/#3}%
{\def\hyper@linkstart##1##2{}%
\let\hyper@linkend\@empty\citep[#1][#2]{#3}}}
\newcommandtwoopt{\citetads}[3][][]{\href{http://adsabs.harvard.edu/abs/#3}%
{\def\hyper@linkstart##1##2{}%
\let\hyper@linkend\@empty\citet[#1][#2]{#3}}}
\newcommandtwoopt{\citeyearads}[3][][]%
{\href{http://adsabs.harvard.edu/abs/#3}
{\def\hyper@linkstart##1##2{}%
\let\hyper@linkend\@empty\citeyear[#1][#2]{#3}}}
\makeatother

\begin{document}

\title{Hydrogen in diffuse molecular clouds in the Milky Way}
\subtitle{Atomic column densities and molecular fraction along prominent lines of sight}
\author{B. Winkel\inst{1}
        \and
        H. Wiesemeyer\inst{1}
        \and
        K. M. Menten\inst{1}
        \and
        M. Sato\inst{1}
        \and
        A. Brunthaler\inst{1}
        \and
        F. Wyrowski\inst{1}
        \and
        D. Neufeld\inst{2}
        \and
        M. Gerin\inst{3,4}
        \and
        N. Indriolo\inst{5}
       }

\institute{Max-Planck-Institut f\"{u}r Radioastronomie (MPIfR),
              Auf dem H\"{u}gel\,69, 53121 Bonn, Germany\\
              \email{bwinkel@mpifr.de}
            \and
            Department of Physics and Astronomy, Johns Hopkins University, Baltimore, MD 21218, USA
            \and
            LERMA, Observatoire de Paris, PSL Research University, CNRS, UMR8112, Paris, France, F-75014
            \and
            Sorbonne Universit\'es, UPMC Univ. Paris 06, UMR8112, LERMA, Paris, France, F-75005
            \and
            Department of Astronomy, University of Michigan, Ann Arbor, MI 48109, USA
          }

\date{Received ; accepted }

\abstract
{Recent submillimeter and far-infrared wavelength observations of absorption in the rotational ground-state lines of various simple molecules against distant Galactic continuum sources have opened the possibility of studying the chemistry of diffuse molecular clouds throughout the Milky Way. In order to calculate abundances, the column densities of molecular and atomic hydrogen, $\ion{H}{i}$, must be known.}
{We aim at determining the atomic hydrogen column densities for diffuse clouds located on the sight lines toward a sample of prominent high-mass star-forming regions that were intensely studied with the HIFI instrument onboard Herschel.}
{Based on Jansky Very Large Array data, we employ the 21~cm \ion{H}{i} absorption-line technique to construct profiles of the $\ion{H}{i}$ opacity versus radial velocity toward our target sources. These profiles are combined with lower resolution archival data of extended $\ion{H}{i}$ emission to calculate the $\ion{H}{i}$ column densities of the individual clouds along the sight lines. We employ Bayesian inference to estimate the uncertainties of the derived quantities.}
{Our study delivers reliable estimates of the atomic hydrogen column density for a large number of diffuse molecular clouds at various Galactocentric distances\thanks{The data sets are available in electronic form (FITS tables) at the CDS via anonymous ftp to cdsarc.u-strasbg.fr (130.79.128.5) or via\newline\url{http://cdsweb.u-strasbg.fr/cgi-bin/qcat?J/A+A/}}. Together with column densities of molecular hydrogen derived from its surrogates observed with HIFI, the measurements can be used to characterize the clouds and investigate the dependence of their chemistry on the molecular fraction, for example.}

\titlerunning{Hydrogen column densities in diffuse clouds of
spiral arms}
\authorrunning{Winkel et al.}

\maketitle

\section{Introduction}\label{sec:intro}
Between 2009 and 2013, absorption-line measurements with the Heterodyne Instrument for the Far Infrared \citep[HIFI;][]{degraauw10} onboard Herschel\footnote{\textit{Herschel} is an ESA space observatory with science instruments provided by European-led Principal Investigator consortia and with important participation from NASA.} have revolutionized our view of chemistry in the diffuse interstellar medium (ISM). Studies with HIFI have targeted the ground-state rotational transitions of a variety of (mostly) di- and triatomic molecules. For earth-bound astronomy, these are very difficult or impossible to observe as a result of the absorption from H$_2$O vapor and dry constituents of the atmosphere. With excellent spectral resolution ($\sim$$0.1~\mathrm{km\,s}^{-1}$), in particular the Guaranteed Time Observing Programs PRISMAS \citep{gerin12} and HEXOS \citep{bergin10} have delivered a whole series of remarkable results that address H$_2$O itself \citep{lis10, vandishoeck13}, for instance, but also the other reactants leading to its formation, that is, OH$^+$, H$_2$O$^+$, and H$_3$O$^+$ \citep{gerin10a, indriolo15, gerin16}. Now that Herschel has ceased operations, the HIFI instrument finds its continuation and extension in the GREAT receiver \citep{heyminck12} flown on the Stratospheric Observatory for Infrared Astronomy (SOFIA). For example, to the H$_2$/oxygen chemistry path, observations with GREAT have added data with high spectral resolution on the 2.5 THz ground-state transition of OH and the $63~\mu$m [OI] fine-structure line \citep{wiesemeyer12, wiesemeyer16}. The basic pathways of nitrogen and carbon chemistry have been similarly explored through observations of CH, CH$^+$, and C$_3$ \citep{gerin10b, falgarone10, mookerjea12}, and of NH, NH$_2$, and NH$_3$, respectively \citep{persson10,persson12}. It should be added that the rotational ground-state transitions of several hydrides species lie in favorable atmospheric windows and can be explored from the ground, for example with the Atacama Pathfinder Experiment telescope (APEX). These include NH$_2$ \citep{vandishoeck93}, OH$^+$ \citep{wyrowski10}, SH$^+$, HCl, and $^{13}$CH$^+$ \citep[all][]{menten11}.

In optical and UV observations of absorption lines the background sources must be bright stars, which limits these studies mostly to diffuse clouds within 1--2~kpc \citep{snow06}. In contrast, at submillimeter and far-infrared wavelengths, rotational lines observed with HIFI, GREAT, or APEX appear in absorption against the bright continuum radiation from warm dust associated with the hot cores of high-mass star-forming regions. The rotational lines of the aforementioned and many other species are easily excited inside hot molecular cores so that they may appear in emission. In lower density diffuse clouds and the dilute envelopes of hot-core-containing regions, however, these molecules basically remain in their ground state and are seen in absorption at the radial velocities of the spiral arms crossing the sight line, determined by Galactic rotation. The brightest high-mass star-forming regions can be seen in dust and ionized gas throughout the galaxy, therefore sight lines longer than that to the Galactic center can be studied, for instance, to \object{W\,49} at 11~kpc.

To build a picture of the physics and chemistry of the ISM in the diverse clouds along the sight lines, chemical networks are constructed that predict the abundances of various molecules. These theoretical calculations must then be constrained by abundances that are derived from observations. In the low-density ISM of diffuse clouds, the distribution of the rotational levels is generally determined by the temperature of the cosmic microwave background (2.728~K). Thus, it is straightforward to calculate the column densities of the observed species. We note, that the excitation temperature of hydrides can be slightly higher than 2.728~K as a consequence of the far-infrared background. This effect has been described by \citet{flagey13} for o-H$_2$O.

For the derivation of abundances we also need the column densities of atomic and molecular hydrogen, $N_\ion{H}{i}$ and $N_{\mathrm{H}_2}$, respectively, whose relative ratio varies with the cloud density \citep{snow06}. The molecular fraction, that is, the fraction of hydrogen nuclei bound in H$_2$, is given by $f_{\mathrm{H}_2}$ $\equiv 2 N_{\mathrm{H}_2}/(N_{\ion{H}{i}}+ 2 N_{\mathrm{H}_2})$. It is a key parameter for understanding cloud chemistry thanks to the stable $\mathrm{H}_3^+$ ion, which initializes the chemistry of oxygen-bearing species and that of carbon-bearing species in dense clouds.

The electric quadrupole transitions of the homo-nuclear molecule $H_2$ are excited in warm clouds and shocks, but not in diffuse and dense gas. The first detection in diffuse gas was made on the sight line to $\zeta$~Persei in the Lyman resonance absorption bands \citep{carruthers70}. Follow-up work was done with the far-ultraviolet satellites Copernicus \citep{savage77} and FUSE \citep{shull00,rachford02,rachford09} on sight lines toward extragalactic background sources. On longer path lengths, where far-ultraviolet observations are not possible anymore because of high reddening, direct measurements of $N_{\mathrm{H}_2}$ are very difficult. CH and HF have been proposed as surrogates \citep{gerin10b, qin10, sonnentrucker10}. As discussed by \citet{neufeld97} and \citet{monje11}, HF is a good proxy for the H$_2$ column density because fluorine is the only atom that can react exothermically with H$_2$ to form a diatomic hydride and HF is destroyed very slowly. Hence it is expected to be the dominant reservoir of gas-phase fluorine when H$_2$ and HF are the dominant reservoirs of gas-phase H and F nuclei, as shown by \citet{neufeld05}. For more details we refer to \citet[][their Appendix C]{godard12}, \citet[][HF abundance from ro-vibrational spectroscopy]{indriolo13} and \citet[][Appendix E]{wiesemeyer16}, and to the recent review article by \citet{gerin16}. Recently, the abundance of HF has been constrained by \citet{sonnentrucker15} to values in the range from 0.9 to $3.3\cdot10^{-8}$ at the low densities of cloud surfaces and, respectively, in the centers of dense translucent clouds. These models, which use recent measurements of the F+H$_2$ reaction rate \citep{tizniti14}, describe the abundance found by \citet{indriolo13}, who compared the column densities deduced from ro-vibrational near-infrared absorption profiles of HF at $\lambda 2.5~\mu\mathrm{m}$ to directly determined H$_2$ column densities from the far-ultraviolet and near-infrared lines.

In contrast, direct determinations of the atomic hydrogen (\ion{H}{i}) column density, $N_\ion{H}{i}$, from the famous 21~cm line are relatively straightforward, and the main subject of this paper. In the optically thin limit, $N_\ion{H}{i}$ can be inferred from emission spectra alone,
\begin{equation}
N_\ion{H}{i}\left[\mathrm{cm}^{-2}\right] = 1.823\cdot10^{18}\int~\mathrm{d}\varv\,T_\mathrm{B}(\varv)\,\left[\mathrm{K\,km\,s}^{-1}\right],\label{eq:coldens}
\end{equation}
where $T_\mathrm{B}(\varv)$ is the brightness temperature profile of the \ion{H}{i} gas and $\varv$ is the radial velocity. However, especially in the Milky Way (MW) disk, the opacity of the gas is usually non-negligible \citep[e.g.,][]{radhakrishnan60,gibson05a,braun09,martin15}. In such cases, \ion{H}{i} absorption spectroscopy is a great tool to measure the optical depth, which in combination with $T_\mathrm{B}$ can be used to estimate the true column density of the \ion{H}{i} gas and its spin temperature, a method pioneered by \citet{lazareff75}.

Interferometric observations of \ion{H}{i} absorption allow us to study the MW ISM with fantastic angular resolution and sensitivity. In particular, the Canadian Galactic Plane Survey \citep[CGPS,][]{taylor03}, the Southern Galactic Plane Survey \citep[SGPS,][]{mcclure05}, and the VLA Galactic Plane Survey \citep[VGPS,][]{stil06} provide an extremely useful database. For example, the structure of the MW is well traced, as discussed in \citet[][outer spiral arms]{strasser07,dickey09}, and \citet[][inner Galaxy]{dickey03}. With the observation of so-called \ion{H}{i} self-absorption (HISA) features, we can reveal small-scale fluctuations such as filaments, which are not visible in emission data \citep{gibson00,gibson05b,mcclure06}. With the Millennium Arecibo 21~cm absorption-line survey \citep{heiles03a}, the solar neighborhood was studied in detail \citep{heiles03b}. Not only Galactic continuum sources are used, but also extra-galactic sources \citep{strasser04,stanimirovic14,lee15}. This has the great advantage that blending of different features along the lines of sight is much better under control (see also Sect.~\ref{subsec:blending_discussion}). It is even possible to detect \ion{H}{i} absorption with Very Long Baseline Interferometry (VLBI). With this, extremely small-scale features have been discovered, the so-called tiny-scale atomic structure \citep[TSAS;][and references therein]{heiles97}, with characteristic sizes of a few tens of AU.

The main motivation for the study we present here was to provide an \ion{H}{i} dataset, corrected for optical-depth effects, as a complement for previous studies of ISM chemistry: in combination with HF measurements, the total hydrogen column density as well as the ratio between atomic and molecular gas can be inferred, which is a valuable ingredient in the analysis of the complex processes in the ISM, such as hydrogen abstraction reactions in the pathways of diffuse cloud chemistry \citep[][further references therein]{gerin16}. Therefore we performed VLA observations of selected sight lines through the Milky Way disk, mostly targeting sources from the PRISMAS and HEXOS samples, as well as toward \object{G\,31.41$+$0.31}. Preliminary results derived from our observations of \ion{H}{i} along these sight lines have been used in several studies based on PRISMAS, HEXOS, and SOFIA data \citep[e.g.,][]{schilke14, gerin15, indriolo15, neufeld15a, neufeld15b, wiesemeyer16}. They are essential in allowing column densities of species prevalent in primarily atomic gas, including ArH$^+$, C$^+$, OH$^+$, and H$_2$O$^+$, to be converted into relative abundances that can be compared with model predictions. Such comparisons can provide unique information of general importance to our understanding of diffuse interstellar gas, including estimates of the cosmic-ray ionization rate \citep[e.g.,][]{indriolo15}, the gas pressure \citep{gerin15}, and the distribution of cloud sizes \citep[e.g.,][]{neufeld16}. Such investigations will need to be combined with theoretical studies and simulations \citep[e.g.,][]{dobbs12b, valdivia16} to build a consistent picture of the Galactic ISM. In the following we provide a detailed discussion of the derivation of opacity-corrected \ion{H}{i} column densities, to be used as reference for these and similar future studies.

\ion{H}{i} absorption toward some of our targets has been investigated previously. \citet{koo97} used the VLA in its D configuration and the Arecibo 305 m telescope to infer opacities for the \object{W\,51} complex. \citet{brogan01} showed VLA B and D array data of \object{W\,49}. \object{Sgr\,B2} was observed by \citet{liszt83} (VLA, only nine antennas) and recently, in much greater detail, by \citet{lang10}. \citet{roberts97} used the VLA in B configuration to map \object{DR\,21}. Except for \object{DR\,21} and \object{Sgr\,B2}, the cited studies have a spectral resolution that is too low to be of use here. The latter two have even much more sensitive data. Nevertheless, for this work we make use of our own data, for the following reasons:
\begin{itemize}
\item The different sight lines should have been observed and processed in the same way, to allow direct comparison.
\item Not all of the previous studies have calculated column density profiles, which is what we are interested in. This is usually because no suitable \ion{H}{i} emission line data are available.
\end{itemize}
Furthermore, we like to explore the uncertainty of the derived opacity and of the corrected column density spectra. None of the previous studies did this.

It should also be noted here that a VLA C-array survey is currently under way, the \ion{H}{i}/OH/Recombination line survey (THOR, P.I.: H. Beuther). First data products have just been released \citep{beuther16}. Furthermore, a pilot study that analyzes the \object{W\,43} star-forming region was recently published \citep{bihr15}. THOR will produce \ion{H}{i} absorption data for many Galactic lines of sight, but as a large-scale on-the-fly survey, THOR is much less sensitive than pointed observations.

In Sect.~\ref{sec:theory} we present the formalism for deriving the atomic hydrogen column density by combining emission and absorption profiles of $\ion{H}{i}$ line observations. The data sets we used are described in Sect.~\ref{sec:data}, along with some details about data processing. Section~\ref{sec:absorptionspectra} presents the derivation of $N_{\ion{H}{i}}$ and discusses systematic uncertainties. Section~\ref{sec:molfraction} deals with the analysis of the molecular hydrogen fraction, and with its interpretation in terms of Galactic spiral structure. We conclude this work with a summary in Sect.~\ref{sec:summary}.

\section{Deriving opacity-corrected \ion{H}{i} column densities through absorption spectroscopy}\label{sec:theory}

As discussed in the introduction, deriving \ion{H}{i} opacities and corrected column densities has been the topic of many studies for several decades. The method itself and relevant equations are also covered in textbooks. Nevertheless, we here briefly repeat the basics.

We start by considering the simplest case of a single (isothermal) cloud in front of a continuum source. Recording two spectra, one in the direction of the background source, the other one offset but nearby, we find

\begin{align}
T^\mathrm{on}(\varv) &=\left(T_\mathrm{cont}^\mathrm{sou}+T_\mathrm{cont}^\mathrm{bg}\right) e^{-\tau} +T_\mathrm{spin}\left(1- e^{-\tau} \right)\label{eq:t_on}\\
T^\mathrm{off}(\varv) &=T_\mathrm{cont}^\mathrm{bg} e^{-\tau} +T_\mathrm{spin}\left(1- e^{-\tau} \right)\label{eq:t_off}
\end{align}
with the continuum flux density of the background source, $T_\mathrm{cont}^\mathrm{sou}$, and the larger-scale continuum sky background in the area, $T_\mathrm{cont}^\mathrm{bg}$. It is explicitly assumed that the contribution of the foreground cloud is identical for both sight lines, while the $T^\mathrm{off}$ spectrum contains none of the background source's continuum. The brightness temperature $T_\mathrm{B}=T_\mathrm{spin}\left(1- e^{-\tau} \right)$ of the foreground cloud is a function of the spin temperature, $T_\mathrm{spin}$, and of the optical depth, $\tau$, of the cloud.

We left out in Eq.~(\ref{eq:t_on}) the potential effects of (1) a non-uniform coverage of the continuum emission by the absorbing foreground medium and (2) beam filling, that is, the background source being unresolved by the telescope beam. When the background source is smaller than the observing beam, the measured (i.e., beam weighted) $T_\mathrm{cont}^\mathrm{sou}$ contribution is smaller than the true background source's brightness temperature. Likewise, when the foreground cloud is not a sheet with uniform physical and chemical conditions but varies across the beam, we would need to correct for this as well. In general, it will even be the case that relevant structures in fore- and background are completely mismatched because they are not physically related. As a consequence, the correction for the filling factors is non-trivial and often complex \citep{wilson13}. We return to this in Sect.~\ref{sec:absorptionspectra}.

One important step to lessen the impact of structure of the absorber is to place the \textsc{Off} position as close as possible to the \textsc{On} sight line. Data with high angular resolution are obviously preferable for this and also provide a better match to the pencil-beam absorption spectrum. Further improvement can be obtained by means of interpolation, where we try to estimate the $T^\mathrm{off}$ spectrum from the surrounding as it would appear at the spatial position of the \textsc{On} sight line if no absorption was present. For sensitive single-dish observations, where only targeted spectra are measured, \citet{heiles03a} proposed a scheme consisting of 16 Off positions, which makes it possible to correct for side-lobe contamination. Their interpolation is based on Taylor expansion. When we have access to a full spectral data cube (a map of spectra), as in the case of radio interferometry data, we can apply a larger variety of interpolation methods (see, e.g., Sect.~\ref{subsec:emissiondata}).
For convenience, we introduce the baseline (continuum) level for each of the two spectra,
\begin{align}
T_\mathrm{cont}^\mathrm{on}&=T_\mathrm{cont}^\mathrm{sou}+T_\mathrm{cont}^\mathrm{bg}\\
T_\mathrm{cont}^\mathrm{off}&=T_\mathrm{cont}^\mathrm{bg}
\end{align}
In principle, one problem can be that the sky continuum background not necessarily originates from behind the cloud; the cloud may even be located within the diffuse ionized gas that emits the continuum background. However, for the sight lines in this work, $T_\mathrm{cont}^\mathrm{sou}\gg T_\mathrm{cont}^\mathrm{bg}$, such that we can safely ignore this issue.

We note that for the optically thick case ($\tau\gg1$), we find $T_\mathrm{B}=T_\mathrm{spin}$, while for $\tau\ll1,$ we have  $T_\mathrm{B}=\tau T_\mathrm{spin}$. Generally, $T_\mathrm{B}\leq T_\mathrm{spin}$ for all $\tau$. As a consequence, if source- and beam-filling effects can be neglected, $T^\mathrm{on}(\varv)\geq T^\mathrm{off}(\varv)$.

From Eqs.~(\ref{eq:t_on}) and (\ref{eq:t_off}) the optical depth,
\begin{equation}
e^{-\tau(\varv)}= \frac{T^\mathrm{on}-T^\mathrm{off}}{T_\mathrm{cont}^\mathrm{on}-T_\mathrm{cont}^\mathrm{off}} = \frac{T^\mathrm{on}-T^\mathrm{off}}{T_\mathrm{cont}^\mathrm{sou}}\label{eq:tau}\,,
\end{equation}
can be calculated. From the opacity, $\tau$, we can easily infer the spin temperature,
\begin{equation}
T_\mathrm{spin}=\frac{T_\mathrm{B}}{1- e^{-\tau} }\label{eq:tspin}\,,
\end{equation}
and the \ion{H}{i} column density
\begin{equation}
N_\ion{H}{i}\left[\mathrm{cm}^{-2}\right]=1.823\cdot10^{18} \int\mathrm{d} \varv\, \tau(\varv)T_\mathrm{spin}(\varv) \left[\mathrm{K\,km\,s}^{-1}\right]\,.
\end{equation}

While the basic idea and equations are very simple, many problems can arise. In addition to the filling factor issues, the numerical stability of the results is affected by the nonlinearity of Eq.~(\ref{eq:tau}). For very high opacities, the measured on-source spectral intensity is very close to the lowest possible value. Then, even small calibration errors or noise can lead to ``unphysical'' situations, that is, technically, the numerator in Eq.~(\ref{eq:tau}) can become negative. Furthermore, when estimating errors, we cannot assume Normal distributions for several of the quantities. The opacity and the derived column densities in particular have highly asymmetric uncertainty distributions.

\begin{table*}[!t]
\caption{Program sources and JVLA continuum images summary. Column~1: source name, Cols.~2~and~3: phase center equatorial coordinates, Cols.~4~and~5: Galactic coordinates, Col.~6: heliocentric distance with reference, Col.~7: Galactocentric distance, Col.~8: approximate LSR velocity of continuum source, Col.~9: restoring beam size major$\times$minor axis, Col.~10: beam position angle E of N, and continuum-emission RMS noise in Col.~11: brightness temperature, and Col.~12: 1.42-GHz flux density.}\label{tab:vla_sources}
%\centering
\begin{center}
\begin{tabular}{lccrrrrrlrrr}
\hline\hline
\rule{0ex}{3ex}
Source & \multicolumn{2}{c}{Phase center} &  \multicolumn{1}{c}{$l$} &
\multicolumn{1}{c}{$b$} & \multicolumn{1}{c}{$D$} &  \multicolumn{1}{c}{$R$} &
  \multicolumn{1}{c}{$\varv_\mathrm{lsr}$} &
$\theta_\mathrm{maj,min}$ & P.A. & $\sigma (T_\mathrm{B})$ & $\sigma (S_\nu)$ \\
 &  $\alpha_{\rm J2000}$ & $\delta_{\rm J2000}$ &
 \multicolumn{1}{c}{$[\mathrm{deg}]$} & \multicolumn{1}{c}{$[\mathrm{deg}]$} &
\multicolumn{1}{c}{$[\mathrm{kpc}]$} & \multicolumn{1}{c}{$[\mathrm{kpc}]$} &
\multicolumn{1}{c}{$\left[\frac{\mathrm{km}}{\mathrm{s}}\right]$} &
$[\arcsec\times\arcsec]$ & $[\mathrm{deg}]$ & $[\mathrm{K}]$ &
\multicolumn{1}{c}{$\left[\frac{\mathrm{mJy}}{\mathrm{Beam}}\right]$}\\[1ex]
\hline
\rule{0ex}{3ex}\object{Sgr\,B2\,(M)} & $17^\mathrm{h}47^\mathrm{m}20\fs4$ & $-28\degr23\arcmin08\arcsec$ &  0.6666 & $-$0.0362 &  8.3$^\mathrm{(a)}$ & 0.1 & 62$^\mathrm{(i)}$   & $45\times29$ & $ 32$ &  $44$ &  $95$\\
\object{G\,10.62$-$0.39}             & $18^\mathrm{h}10^\mathrm{m}28\fs6$ & $-19\degr55\arcmin51\arcsec$ & 10.6229 & $-$0.3835 &  5.0$^\mathrm{(b)}$ & 3.6 & $-$3$^\mathrm{(a)}$ & $36\times30$ & $ 51$ &  $14$ &  $25$\\
\object{W\,33\,A}                       & $18^\mathrm{h}14^\mathrm{m}39\fs3$ & $-17\degr52\arcmin02\arcsec$ & 12.9077 & $-$0.2596 &  2.4$^\mathrm{(c)}$ & 6.1 & 39$^\mathrm{(a)}$   & $42\times25$ & $ 52$ &  $19$ &  $34$\\
\object{G\,31.41$+$0.31}             & $18^\mathrm{h}47^\mathrm{m}34\fs5$ & $-01\degr12\arcmin44\arcsec$ & 31.4125 & $+$0.3070 &  4.9$^\mathrm{(d)}$ & 4.9 & 96$^\mathrm{(j)}$   & $36\times21$ & $ 75$ &  $12$ &  $15$\\
\object{G\,34.26$+$0.15}             & $18^\mathrm{h}53^\mathrm{m}18\fs6$ & $+01\degr14\arcmin57\arcsec$ & 34.2569 & $+$0.1532 &  1.6$^\mathrm{(e)}$ & 7.1 & 58$^\mathrm{(k)}$   & $36\times21$ & $ 78$ &  $18$ &  $22$\\
\object{W\,49}                       & $19^\mathrm{h}10^\mathrm{m}13\fs1$ & $+09\degr06\arcmin11\arcsec$ & 43.1652 & $+$0.0121 & 11.1$^\mathrm{(f)}$ & 7.6 & 10$^\mathrm{(a)}$   & $34\times20$ & $-86$ &  $76$ &  $84$\\
\object{W\,51}                       & $19^\mathrm{h}23^\mathrm{m}43\fs8$ & $+14\degr30\arcmin30\arcsec$ & 49.4883 & $-$0.3876 &  5.4$^\mathrm{(g)}$ & 6.4 & 57$^\mathrm{(a)}$   & $34\times20$ & $ 84$ & $102$ & $114$\\
\object{DR\,21\,(OH)}                & $20^\mathrm{h}39^\mathrm{m}00\fs6$ & $+42\degr22\arcmin46\arcsec$ & 81.7201 & $+$0.5718 &  1.5$^\mathrm{(h)}$ & 8.3 & $-$3$^\mathrm{(a)}$ & $44\times19$ & $ 67$ &  $12$ &  $17$\\
\hline
\end{tabular}
\end{center}
Notes: References for distances are (a) \citet{reid14}, for this region, we adopted the best-fit Galactic center distance (8.34~kpc) discussed in the study; (b) \citet{sanna14}; (c) \citet{immer13}; (d) we adopt the average value determined by \citet{zhang14} for the source pair \object{G\,31.28$+$0.06}/\object{G\,31.58$+$0.07} that brackets this region in longitude and has a similar LSR velocity, the distance is compatible with the estimate by \citet{raid16}; (e) \citet{kurayama11}, the distance was determined for a source that resides $\sim$$10\arcmin$ of our position in the same infrared dark cloud filament; (f) \citet{zhang13}; (g) \citet{sato10}; (h) \citet{rygl12}; (i) \citet{belloche13} (j) \citet{bronfman96}; (k) \citet{wyrowski12}.
\end{table*}

Furthermore, in most circumstances we do not simply deal with one single isothermal cloud per sight line, but rather with an ensemble of clouds, sheets, or filaments, all having different physical properties ($T_\mathrm{spin}$, $\tau$) and being located at different distances with different radial velocities. Distance and radial velocity can even be uncorrelated. If two clouds at different distances (partly) share radial velocities, then the cloud in front will absorb a certain fraction of the radiation, leaving the cloud in the back. In the extreme case, the continuum source itself might be embedded in absorbing material. In fact, the latter is the case for all of the sight lines in our sample, but as we are mostly interested in the foreground clouds, this is not a major problem since the objects have different radial velocities.

There have been several attempts to correct for these effects. For this, a two-component gas is assumed, the so-called cold and warm neutral medium (CNM, WNM). The former is thought to produce the bulk of the absorption features, while the latter is considered to be diffuse and optically thin. Nevertheless, its higher spin temperature leads to considerable contributions to the total $N_\ion{H}{i}$ \citep[compare, e.g.,][]{heiles03a,stanimirovic14}. The measured brightness temperatures are then separated, accounting for contributions by the CNM and the WNM. For the WNM, a distinction is made for clouds in front and behind each absorber. First attempts by \citet{mebold97} and \citet{dickey00} plotted $T_\mathrm{B}$ vs. $1- e^{-\tau} $ to estimate spin temperatures for different velocity intervals in a spectrum. Later, \citet{heiles03a} significantly improved on this idea by performing a least-squares fit of all CNM and WNM components and their physical parameters, by means of a Gaussian decomposition.

In our case, however, the continuum emitters are within the Milky Way disk. Hence, for most of our sight lines, the Galactic rotation leads to superposition of features from before and behind the continuum source in the emission spectrum. The method discussed above allows WNM components this freedom, but it assumes that high-opacity features are all accounted for in the absorption spectrum. This was the case for previous studies, which used extra-galactic background sources, but it is certainly not the case for our sight lines.

The two effects, blending of CNM/WNM features from the farther fraction of the sight line and neglecting the WNM components situated in front of the continuum emitter, act independently: consider (1) CNM and WNM with one component each, covering the same radial velocities. By associating the complete brightness temperature spectrum with the CNM alone, the inferred spin temperature will be too high, as is also true of $N_\ion{H}{i}$; (2) two CNM components, in front and behind the continuum source. Again, $T_\mathrm{spin}$ will be overestimated (because the brightness temperature falsely associated with the absorber is overestimated). $N_\ion{H}{i}$ can either be over- or underestimated, however, depending entirely on the ratio of the opacities of the two components. If the component in front of the continuum source has higher $\tau$, then $N_\ion{H}{i}$ will be overpredicted, if its $\tau$ is smaller, $N_\ion{H}{i}$ will be underestimated.

Lacking any further information, we can only treat each sight line as if the absorbing material in each spectral channel were associated to a single isothermal cloud, and introduce the the concept of efficient spin temperature
and total optical depth \citep[see][and references therein]{chengalur13}. This approach usually leads to estimates close to the true value by a factor of $\lesssim2$, as was shown by \citet{chengalur13} using Monte Carlo simulations.

\section{\ion{H}{i} absorption and emission line data}\label{sec:data}
\subsection{JVLA observations and data reduction}\label{subsec:jvla_data}

We observed the \ion{H}{i} absorption line at 21~cm (at a rest frequency of 1420.405752 MHz) toward eight high-mass star-forming regions in the Galaxy (\object{Sgr\,B2}, \object{G\,10.62$-$0.39} (\object{W\,31C}), \object{W\,33}, \object{G\,31.41$+$0.31}, \object{G\,34.3$+$0.1}, \object{W\,49N}, \object{W\,51}, and \object{DR\,21}) on January 20, 2012, using the National Radio Astronomy Observatory (NRAO) Karl G. Jansky Very Large Array (JVLA) in DnC configuration (program 11B-236). The phase center coordinates, which are identical to the positions Herschel was pointed at in the PRISMAS and HEXOS projects, for each of the maps are given in Table~\ref{tab:vla_sources}.

During the total observing time of 3~hours, we first observed flux calibrators (e.g., \object{3C\,286}) for 30~minutes, and then placed two observing blocks of about 1~hour duration each. Each 1-hour block consisted of three repetitions of 22-minute groups, each consisting of $\sim$$5$-minute scans toward four target sources and a 3-minute scan toward the phase calibrator \object{1751$-$2524}. We employed a frequency band of 2~MHz bandwidth in both right- and left-circular polarizations. The 2-MHz band was split into 4096 spectral channels, yielding a channel separation of 0.49~kHz and a velocity resolution of $0.10~\mathrm{km\,s}^{-1}$ for the \ion{H}{i} rest frequency.

The data calibration was performed using the NRAO Astronomical Image Processing System (AIPS) in the standard manner. Bandpass amplitude and phase corrections were determined from the calibrator \object{3C\,286} and applied to all sources. We removed instrumental delay and phase offsets using the primary flux calibrator \object{3C\,286} and the phase calibrator \object{1751$-$2524} and corrected the target source data for these offsets. Spectral-line image cubes were then made with the AIPS task IMAGR, and the images were cleaned subsequently.

\begin{figure*}[!tp]
\centering%
\includegraphics[width=0.49\textwidth,viewport=0 13 478 475,clip=]{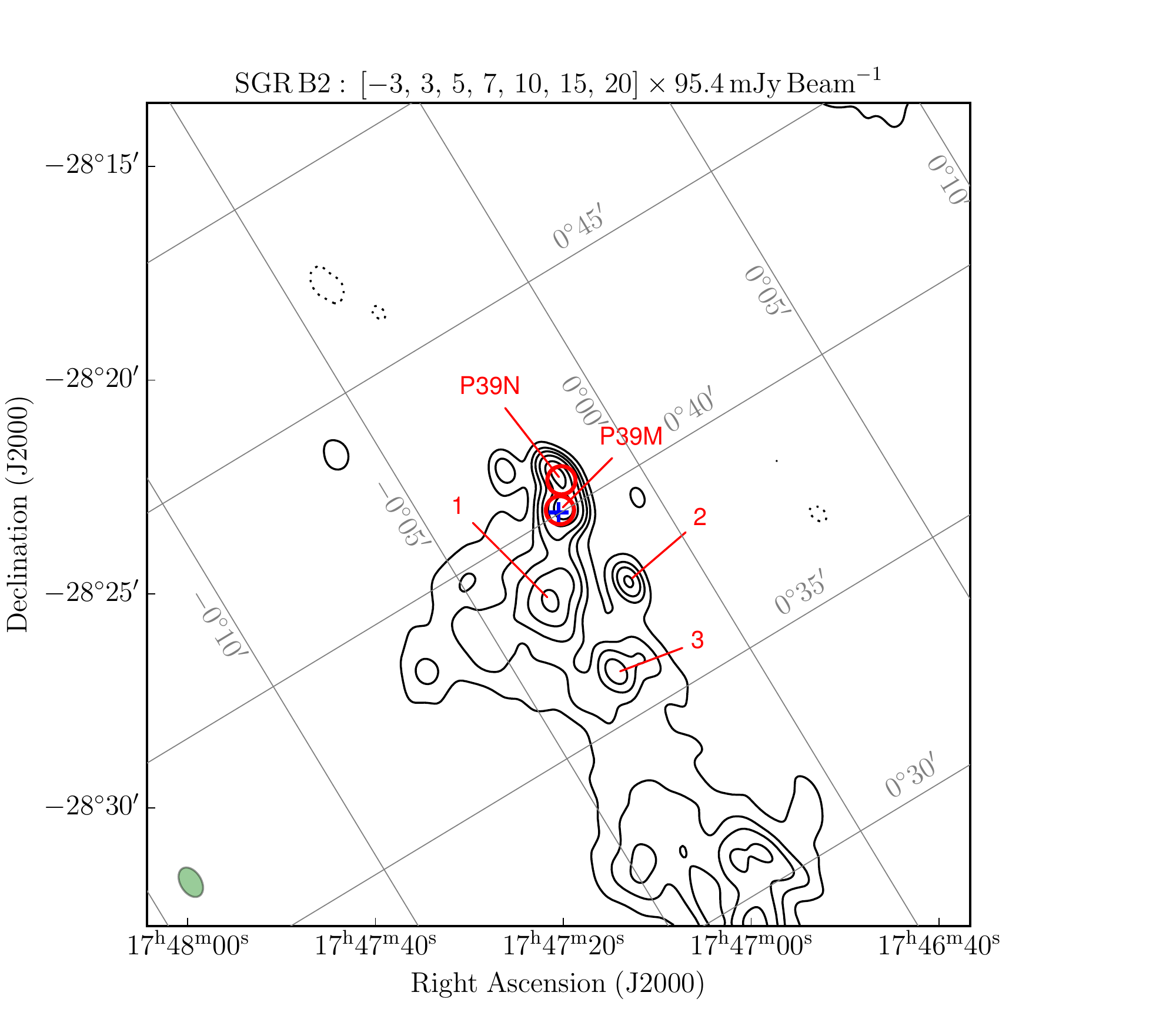}\hfill
\includegraphics[width=0.49\textwidth,viewport=0 13 478 475,clip=]{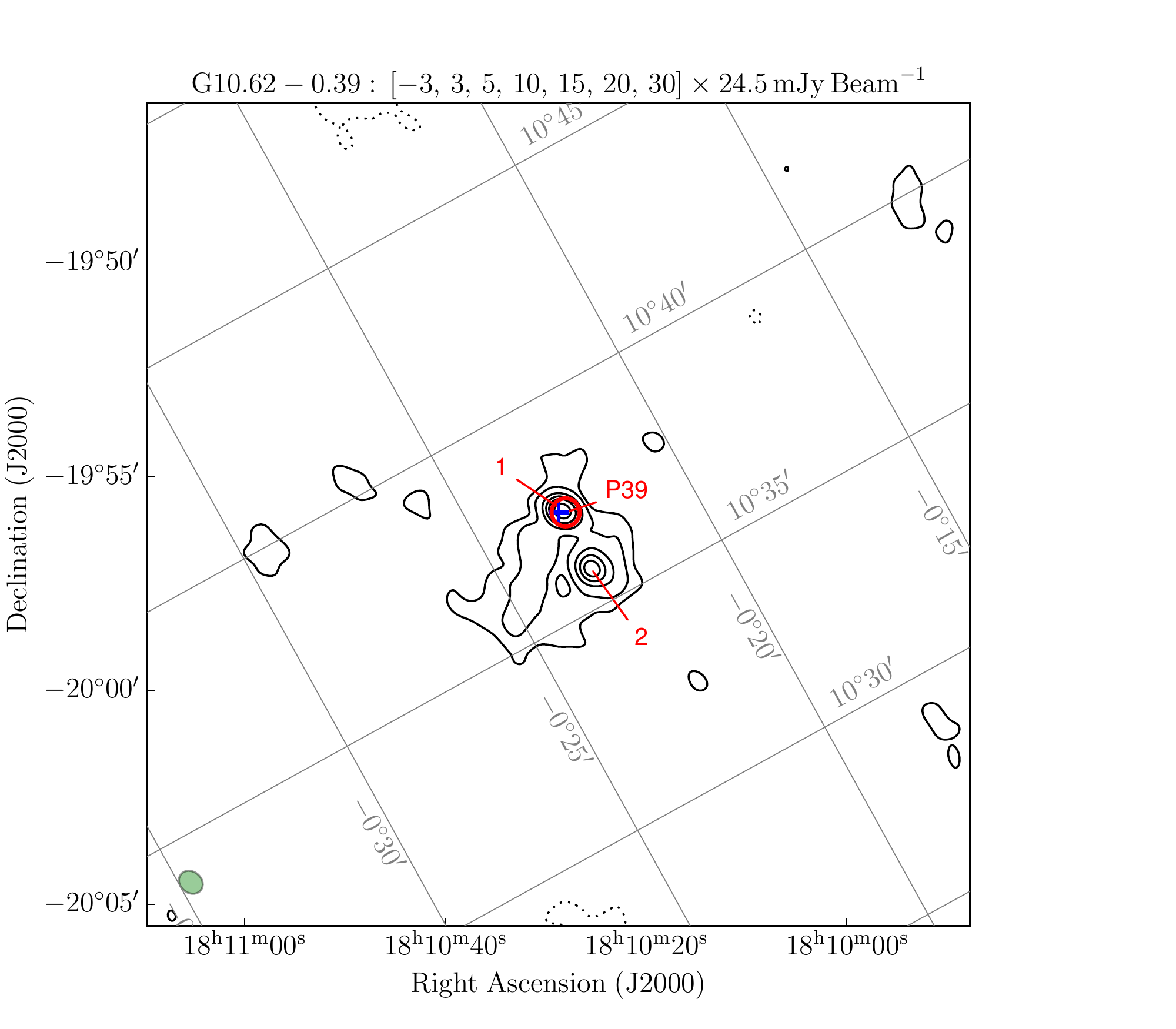}\\[1ex]
\includegraphics[width=0.49\textwidth,viewport=0 13 478 475,clip=]{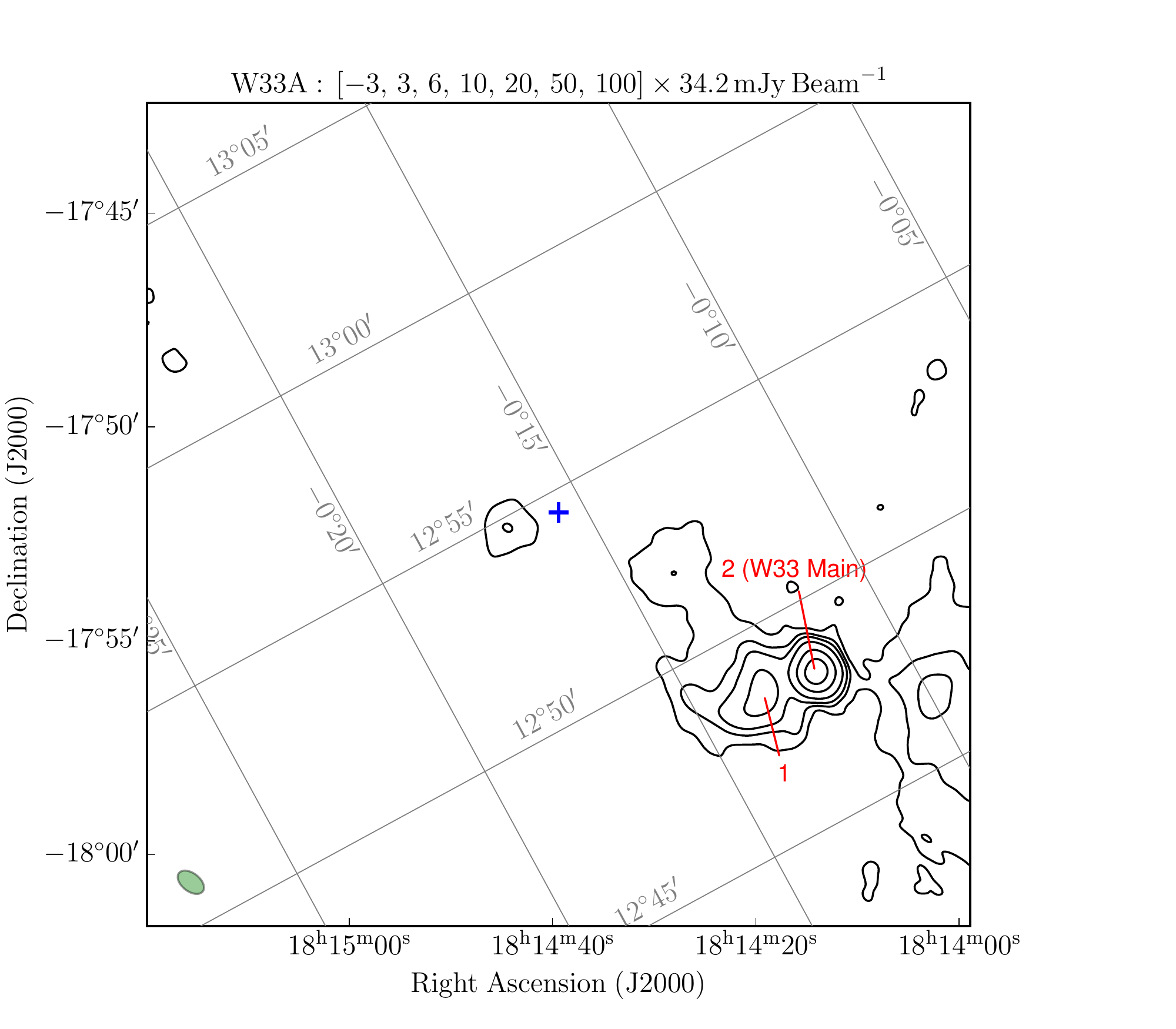}\hfill
\includegraphics[width=0.49\textwidth,viewport=0 13 478 475,clip=]{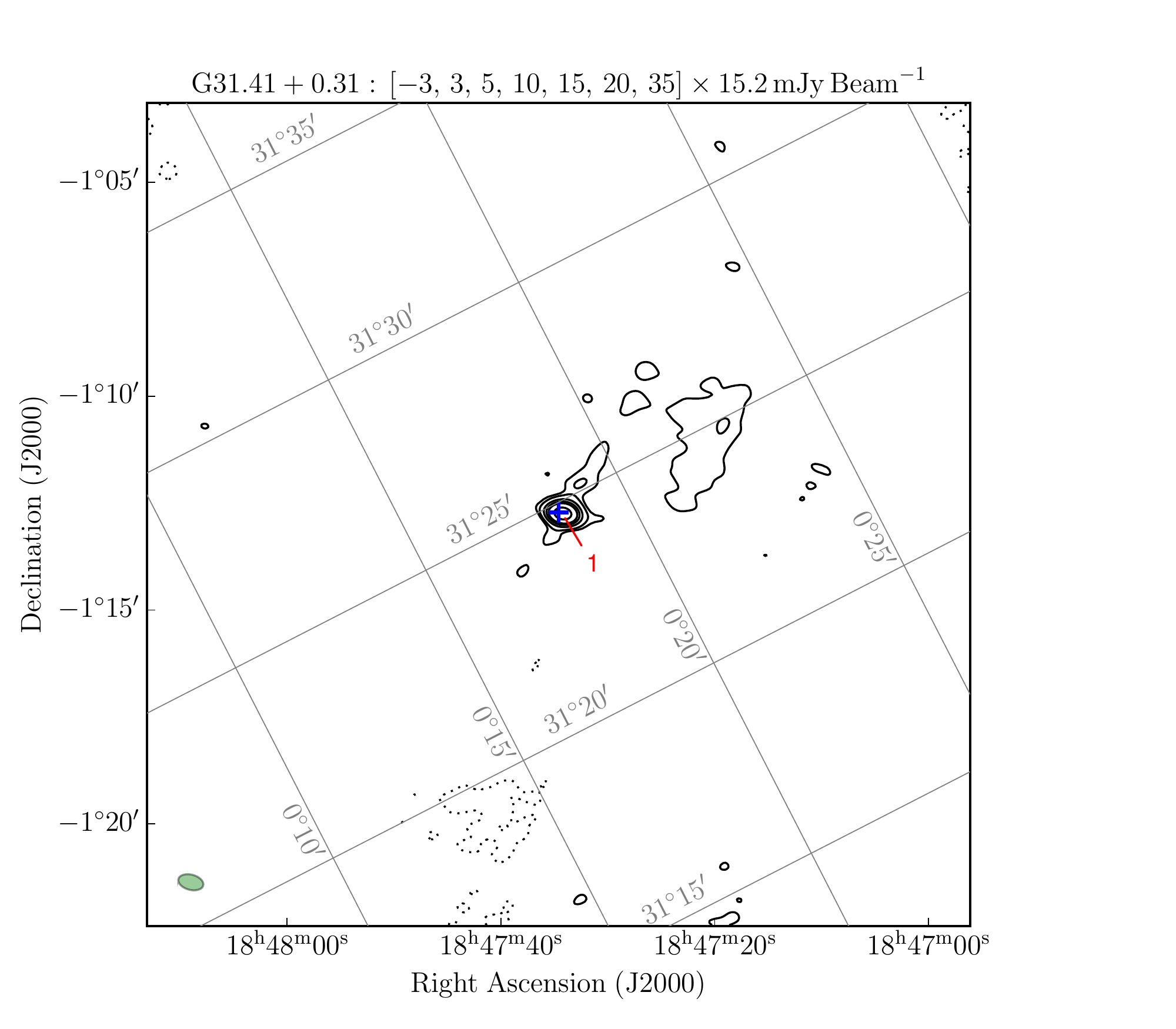}
\caption{Continuum maps at 1.4 GHz of the regions \object{Sgr\,B2}, \object{G\,10.62$-$0.39}, \object{W\,33\,A}, and \object{G\,31.41$+$0.31} as observed with JVLA. The sight lines extracted for our analysis are annotated. The spatial coordinates of the sight lines are compiled in Table~\ref{tab:extracted_positions} for reference. Contour lines are in multiples of the RMS noise level as stated in the title of each panel. Negative contour lines are dotted, positive contours have solid line style. The beam size is indicated by the green ellipse in the lower left corner of each map. Inside graticules display Galactic coordinates. All maps were primary-beam corrected. Note that for the sake of better visualization only the central part of each map is displayed (about half the size of the primary beam). The blue crosses mark the phase centers of the JVLA observations. PRISMAS/HEXOS program positions are labeled with `P39', and the PRISMAS/HEXOS 550-GHz beam size of $\sim$$39\arcsec$ is indicated by red circles. For \object{Sgr\,B2} two sight lines were observed within the HEXOS program, which we annotate `P39M' and `P39N'. Note that the JVLA phase centers do not necessarily match the highest continuum intensity positions.}%
\label{fig:cont_maps1}%
\end{figure*}

\begin{figure*}[!tp]
\centering%
\includegraphics[width=0.49\textwidth,viewport=0 13 478 475,clip=]{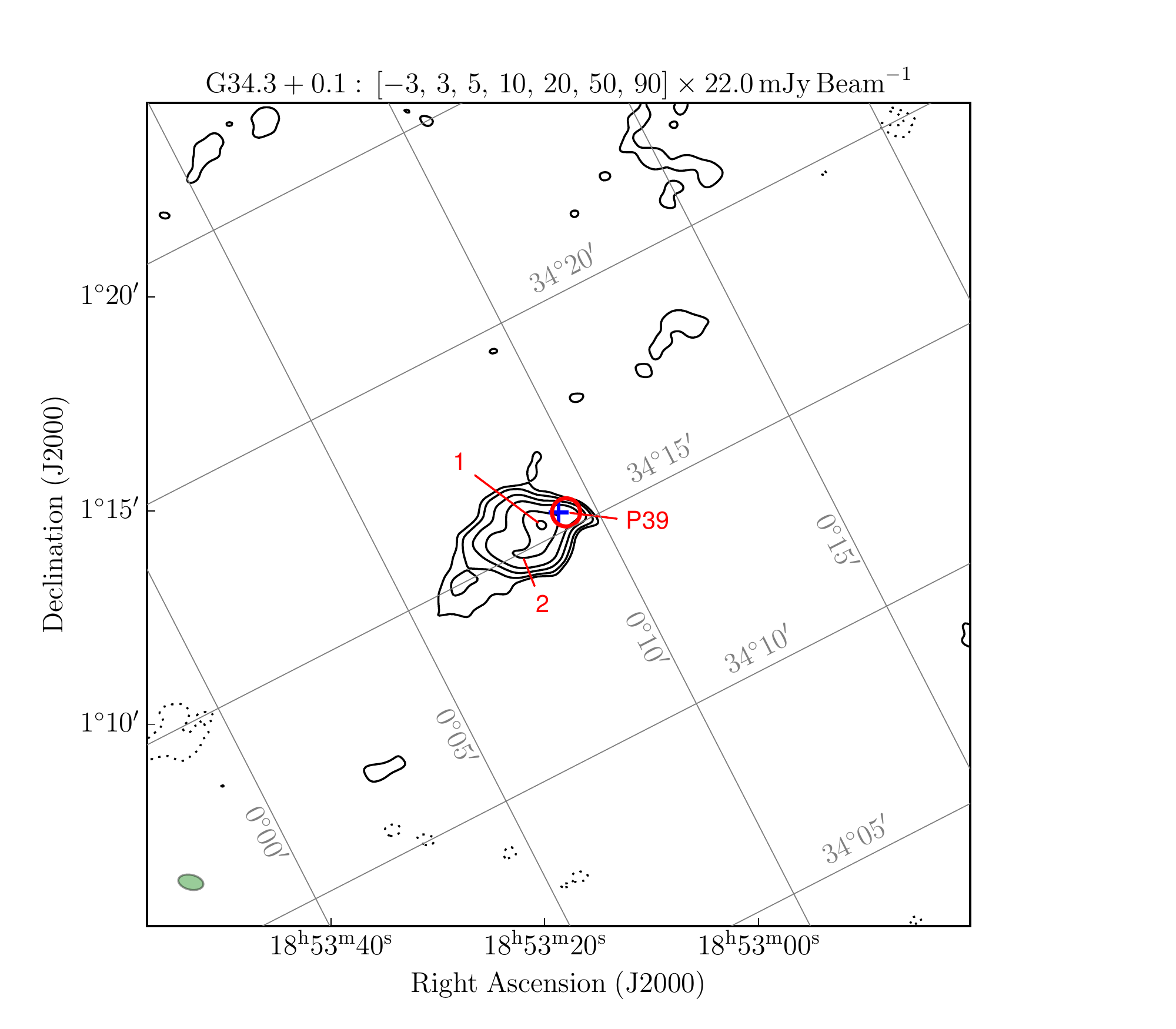}\hfill
\includegraphics[width=0.49\textwidth,viewport=0 13 478 475,clip=]{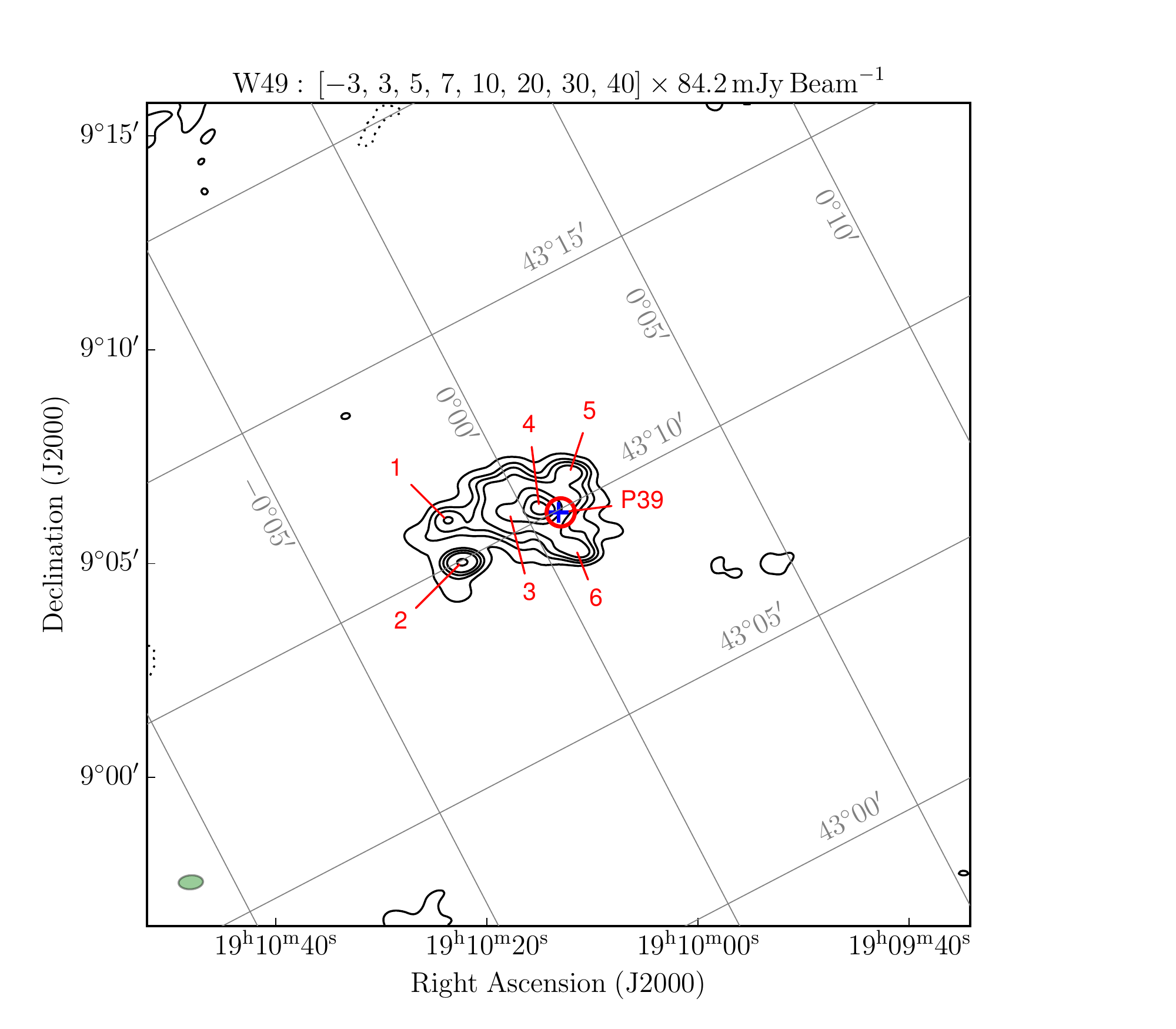}\\[1ex]
\includegraphics[width=0.49\textwidth,viewport=0 13 478 475,clip=]{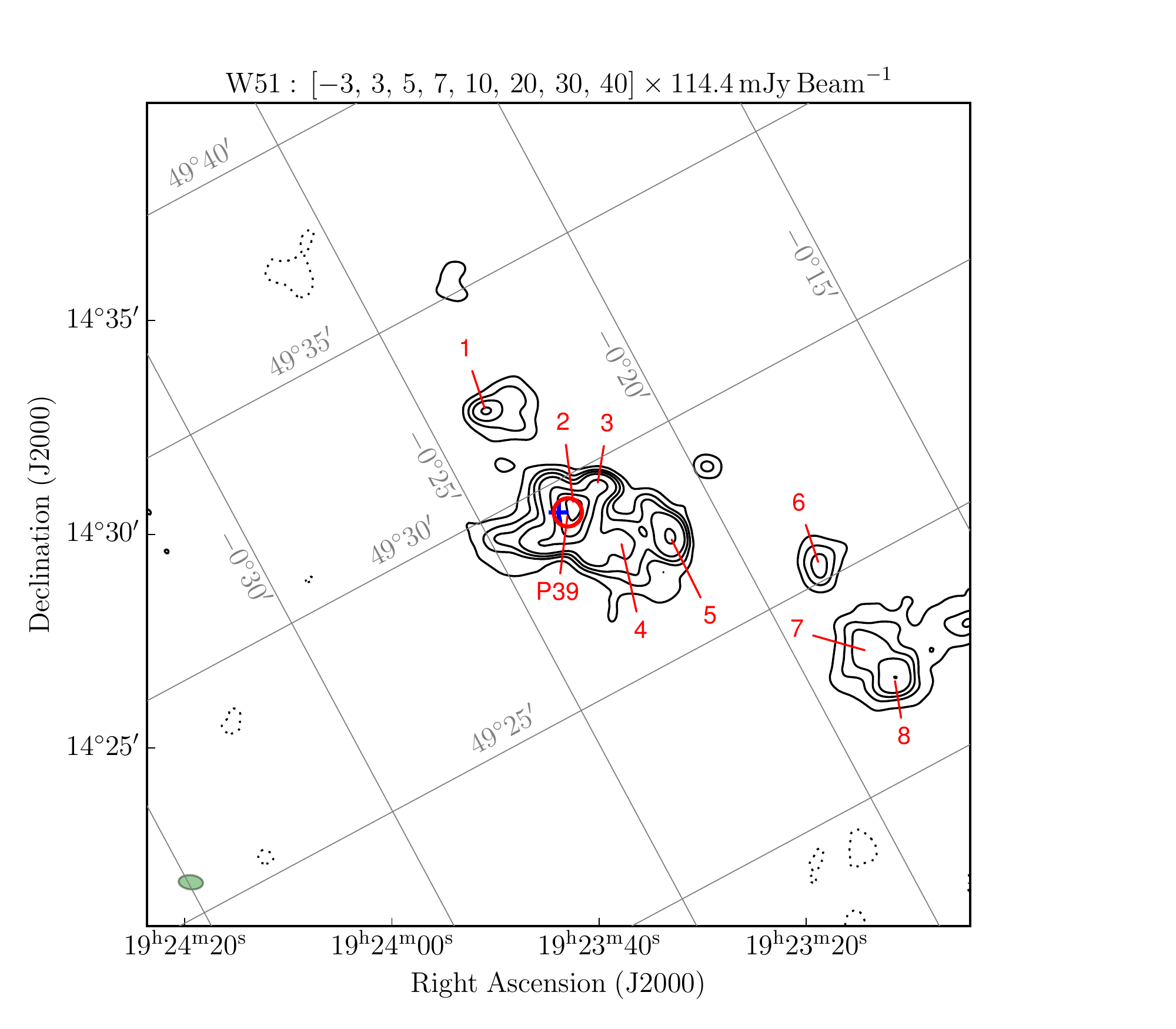}\hfill
\includegraphics[width=0.49\textwidth,viewport=0 13 478 475,clip=]{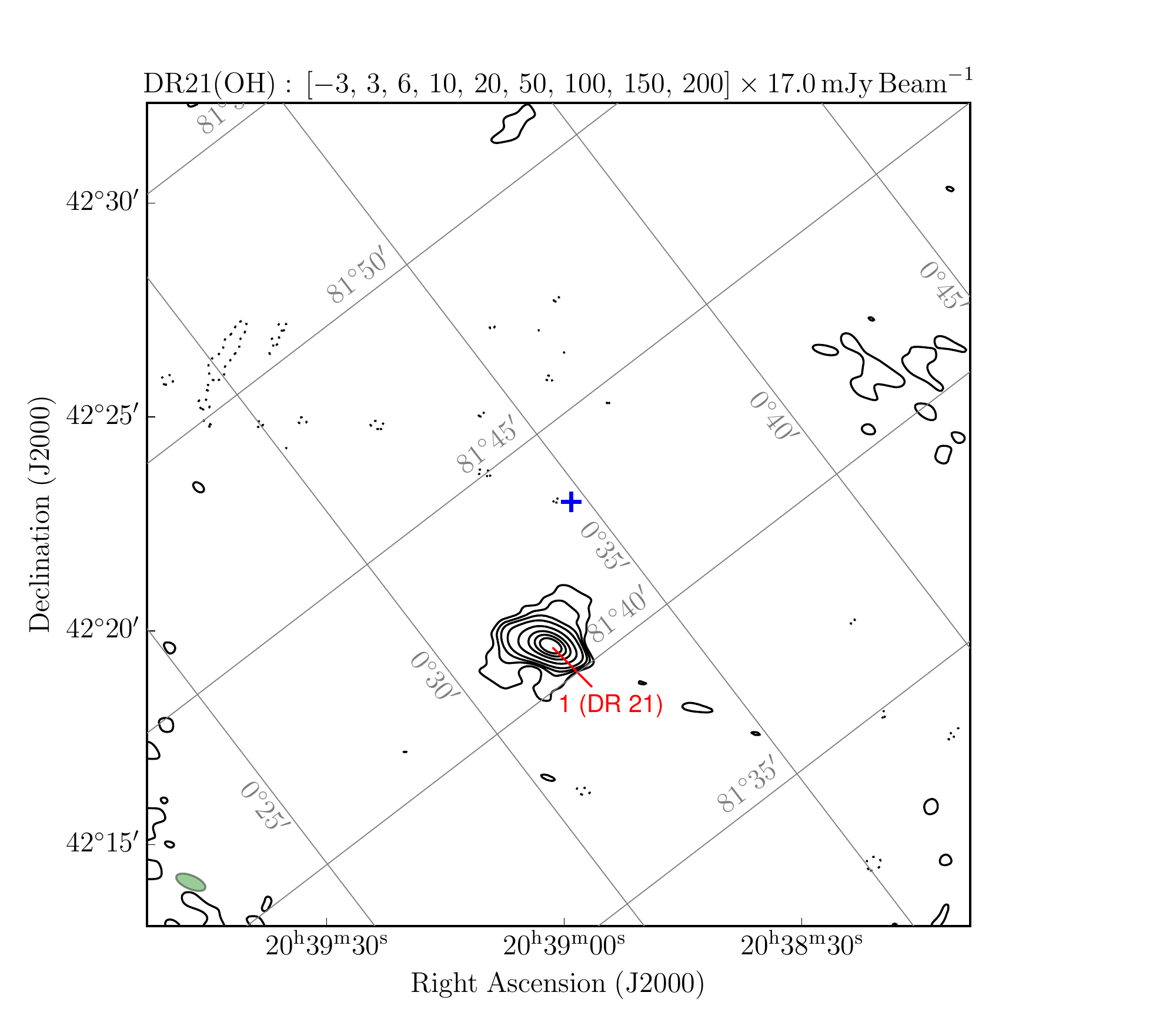}
\caption{As Fig.~\ref{fig:cont_maps1}, showing the regions \object{G\,34.3$+$0.1}, \object{W\,49N}, \object{W\,51}, and \object{DR\,21}. For \object{W\,51} the P39 position coincides with the ultra-compact \ion{H}{ii} regions \object{W\,51}\,e1 to e4.}%
\label{fig:cont_maps2}%
\end{figure*}

In Figs.~\ref{fig:cont_maps1} and \ref{fig:cont_maps2} we show 1.4-GHz continuum maps as observed with JVLA. For all sources, data for one or more (sufficiently bright) sight lines are extracted for subsequent analysis. The spatial coordinates for each of the extracted sight lines are compiled in Table~\ref{tab:extracted_positions}. The continuum root-mean-square (RMS) was calculated following an iterative approach, masking values in excess of $3\sigma_\mathrm{rms}$ in each step. The RMS estimate converged after few iterations. The derived continuum RMS values are given in Table~\ref{tab:vla_sources}.

\begin{table*}[!t]
\caption{Designations, J2000 equatorial and Galactic coordinates coordinates of the positions, as marked in Figs.~\ref{fig:cont_maps1} and \ref{fig:cont_maps2}, toward which the \ion{H}{i} column density vs. $\varv_\mathrm{lsr}$ ``spectra'' presented in the main text and the appendices were extracted.}\label{tab:extracted_positions}
%\centering
\begin{center}
\begin{tabular}{lccccc}
\hline\hline
\rule{0ex}{3ex}
Source & Sight line & \multicolumn{2}{c}{Equatorial coordinates} &  $l$ & $b$\\
 & &  $\alpha_{\rm J2000}$ & $\delta_{\rm J2000}$ & \multicolumn{1}{c}{$[\mathrm{deg}]$} & \multicolumn{1}{c}{$[\mathrm{deg}]$}\\
\hline
\rule{0ex}{3ex}\object{Sgr\,B2} & 1 & $17^\mathrm{h}47^\mathrm{m}21\fs4$ & $-28\degr25\arcmin09\arcsec$ & $0.6397$   & $-$$0.0569$\\
\object{Sgr\,B2} & 2 & $17^\mathrm{h}47^\mathrm{m}13\fs0$ & $-28\degr24\arcmin42\arcsec$ & $0.6301$   & $-$$0.0267$\\
\object{Sgr\,B2} & 3 & $17^\mathrm{h}47^\mathrm{m}14\fs4$ & $-28\degr26\arcmin51\arcsec$ & $0.6020$   & $-$$0.0495$\\
\object{Sgr\,B2} & P39M & $17^\mathrm{h}47^\mathrm{m}20\fs4$ & $-28\degr23\arcmin03\arcsec$ & $0.6676$   & $-$$0.0355$\\
\object{Sgr\,B2} & P39N & $17^\mathrm{h}47^\mathrm{m}20\fs2$ & $-28\degr22\arcmin21\arcsec$ & $0.6773$   & $-$$0.0290$\\
\object{G\,10.62$-$0.39} & 1 & $18^\mathrm{h}10^\mathrm{m}28\fs3$ & $-19\degr55\arcmin47\arcsec$ & $10.6233$  & $-$$0.3820$\\
\object{G\,10.62$-$0.39} & 2 & $18^\mathrm{h}10^\mathrm{m}25\fs5$ & $-19\degr57\arcmin08\arcsec$ & $10.5984$  & $-$$0.3833$\\
\object{G\,10.62$-$0.39} & P39 & $18^\mathrm{h}10^\mathrm{m}28\fs0$ & $-19\degr55\arcmin50\arcsec$ & $10.6221$  & $-$$0.3814$\\
\object{W\,33} & 1 & $18^\mathrm{h}14^\mathrm{m}19\fs2$ & $-17\degr56\arcmin15\arcsec$ & $12.8079$  & $-$$0.2234$\\
\object{W\,33} & 2 & $18^\mathrm{h}14^\mathrm{m}14\fs2$ & $-17\degr55\arcmin45\arcsec$ & $12.8057$  & $-$$0.2019$\\
\object{G\,31.41$+$0.31} & 1 & $18^\mathrm{h}47^\mathrm{m}34\fs2$ & $-01\degr12\arcmin46\arcsec$ & $31.4116$  & $+$$0.3074$\\
\object{G\,34.26$+$0.15} & 1 & $18^\mathrm{h}53^\mathrm{m}20\fs3$ & $+01\degr14\arcmin40\arcsec$ & $34.2561$  & $+$$0.1447$\\
\object{G\,34.26$+$0.15} & 2 & $18^\mathrm{h}53^\mathrm{m}22\fs1$ & $+01\degr13\arcmin58\arcsec$ & $34.2491$  & $+$$0.1327$\\
\object{G\,34.26$+$0.15} & P39 & $18^\mathrm{h}53^\mathrm{m}18\fs0$ & $+01\degr14\arcmin58\arcsec$ & $34.2562$  & $+$$0.1555$\\
\object{W\,49} & 1 & $19^\mathrm{h}10^\mathrm{m}23\fs7$ & $+09\degr06\arcmin00\arcsec$ & $43.1829$  & $-$$0.0281$\\
\object{W\,49} & 2 & $19^\mathrm{h}10^\mathrm{m}22\fs3$ & $+09\degr05\arcmin03\arcsec$ & $43.1661$  & $-$$0.0303$\\
\object{W\,49} & 3 & $19^\mathrm{h}10^\mathrm{m}17\fs9$ & $+09\degr06\arcmin12\arcsec$ & $43.1747$  & $-$$0.0052$\\
\object{W\,49} & 4 & $19^\mathrm{h}10^\mathrm{m}15\fs0$ & $+09\degr06\arcmin18\arcsec$ & $43.1708$  & $+$$0.0059$\\
\object{W\,49} & 5 & $19^\mathrm{h}10^\mathrm{m}12\fs2$ & $+09\degr07\arcmin06\arcsec$ & $43.1772$  & $+$$0.0224$\\
\object{W\,49} & 6 & $19^\mathrm{h}10^\mathrm{m}11\fs6$ & $+09\degr05\arcmin21\arcsec$ & $43.1502$  & $+$$0.0112$\\
\object{W\,49} & P39 & $19^\mathrm{h}10^\mathrm{m}13\fs0$ & $+09\degr06\arcmin12\arcsec$ & $43.1654$  & $+$$0.0126$\\
\object{W\,51} & 1 & $19^\mathrm{h}23^\mathrm{m}50\fs9$ & $+14\degr32\arcmin52\arcsec$ & $49.5368$  & $-$$0.3944$\\
\object{W\,51} & 2 & $19^\mathrm{h}23^\mathrm{m}42\fs5$ & $+14\degr30\arcmin40\arcsec$ & $49.4883$  & $-$$0.3816$\\
\object{W\,51} & 3 & $19^\mathrm{h}23^\mathrm{m}40\fs2$ & $+14\degr31\arcmin07\arcsec$ & $49.4906$  & $-$$0.3700$\\
\object{W\,51} & 4 & $19^\mathrm{h}23^\mathrm{m}37\fs9$ & $+14\degr29\arcmin52\arcsec$ & $49.4679$  & $-$$0.3717$\\
\object{W\,51} & 5 & $19^\mathrm{h}23^\mathrm{m}33\fs2$ & $+14\degr29\arcmin58\arcsec$ & $49.4603$  & $-$$0.3541$\\
\object{W\,51} & 6 & $19^\mathrm{h}23^\mathrm{m}18\fs7$ & $+14\degr29\arcmin16\arcsec$ & $49.4225$  & $-$$0.3082$\\
\object{W\,51} & 7 & $19^\mathrm{h}23^\mathrm{m}14\fs0$ & $+14\degr27\arcmin16\arcsec$ & $49.3840$  & $-$$0.3070$\\
\object{W\,51} & 8 & $19^\mathrm{h}23^\mathrm{m}11\fs5$ & $+14\degr26\arcmin40\arcsec$ & $49.3705$  & $-$$0.3030$\\
\object{W\,51} & P39 & $19^\mathrm{h}23^\mathrm{m}43\fs0$ & $+14\degr30\arcmin31\arcsec$ & $49.4872$  & $-$$0.3847$\\
\object{DR\,21} & 1 & $20^\mathrm{h}39^\mathrm{m}01\fs8$ & $+42\degr19\arcmin41\arcsec$ & $81.6817$  & $+$$0.5375$\\
\hline
\end{tabular}
\end{center}
\end{table*}

In contrast to common practice, we did not subtract the continuum from the spectral-line data, neither before nor after imaging the visibilities. This has the advantage that planes in the produced data cube that are most affected by absorption need the least amount of cleaning, while the continuum contribution of the background emitter appears as a (strong) positive feature in absorption-free planes. If we had subtracted the continuum in the first place, it would be the opposite way. Then the results in the most interesting parts of the spectrum would need extremely careful cleaning. The JVLA data are corrected for the primary-beam response  before the relevant data are extracted.

We were unable to clean the continuum maps down to the thermal noise level. For the \object{Sgr\,B2}, \object{W\,49N}, and \object{W\,51} maps in particular, residual artifacts appear (this is reflected by the apparent high noise levels). Since much effort was invested in calibration and deconvolution of the data, we suspect that the poor uv-coverage of our snapshot data is simply not sufficient for the relatively complex (and very bright) sources. We can still use the derived continuum maps for our analysis, but the nominally higher noise level leads to larger uncertainties in the derived opacity and column density spectra.

For each of the reduced data sets we extracted one or more absorption spectra from the final data cube for sight lines containing sufficient background continuum to warrant good signal-to-noise ratio for further analyses. These sight lines are annotated in Figs.~\ref{fig:cont_maps1} and \ref{fig:cont_maps2}. For four targets (\object{G\,10.62$-$0.39}, \object{G\,34.3$+$0.1}, \object{W\,49N}, and \object{W\,51}), continuum intensities at the exact PRISMAS program positions are high enough to study the associated profiles in detail. These sight lines are labeled P39, and the PRISMAS 550-GHz beam size of $\sim$$39\arcsec$ is indicated by a red circle. For \object{SGR\,B2} two positions (M and N) were observed within the HEXOS project. The associated sight lines are annotated P39M and P39N, respectively.

Lacking sufficient 21~cm continuum emission, no meaningful $\ion{H}{i}$ column density could be determined toward the strong far-infrared source \object{W\,33\,A}, but only toward the \ion{H}{ii} region \object{W\,33\,(Main)}; compare Fig.~\ref{fig:cont_maps1} (we refer to \citealp{immer13} for a definition of the subregions of the \object{W\,33} complex). In addition, \object{DR\,21\,(OH)} shows only very weak radio continuum, and \ion{H}{i} absorption data are retrieved toward the strong radio emission of the compact \ion{H}{ii} region \object{DR\,21}. This is shown in Fig.~\ref{fig:cont_maps2}, where the centroid position is $\sim$$3\arcmin$ south of the position given in Table~\ref{tab:vla_sources}; the centroid position has also been a Herschel/HIFI target \citep[see, e.g., the first detection of H$_2$O$^+$ by][]{ossenkopf10}.

\subsection{Using archival data to infer emission spectra}\label{subsec:emissiondata}
While the JVLA data cubes have high angular and spectral resolution, they are unfortunately not suitable to obtain an \ion{H}{i} emission spectrum. This is because our JVLA data suffer severely from a lack of short spacings. In fact, there is little \ion{H}{i} emission visible in the data cubes.

Since all observed sight lines are located within the Galactic plane, emission line data can be extracted from the three Galactic plane surveys. These data sets are publicly available and have been combined with single-dish data to incorporate short spacings. In our case, the sight lines of \object{W\,51}, \object{W\,49N}, \object{G\,31.41$+$0.31}, and \object{G\,34.3$+$0.1} are contained in VGPS, \object{DR\,21} is from CGPS, while \object{W\,33} and \object{G\,10.62$-$0.39} sight lines were taken from SGPS. The Galactic center region needed for \object{Sgr\,B2} is included in the ATCA \ion{H}{i} Galactic Center Survey \citep{mcclure12}. In Table~\ref{tab:gpsurveys} basic survey parameters have been compiled.

\begin{figure*}[!t]
\centering%
\includegraphics[width=\textwidth,viewport=12 7 860 350,clip=]{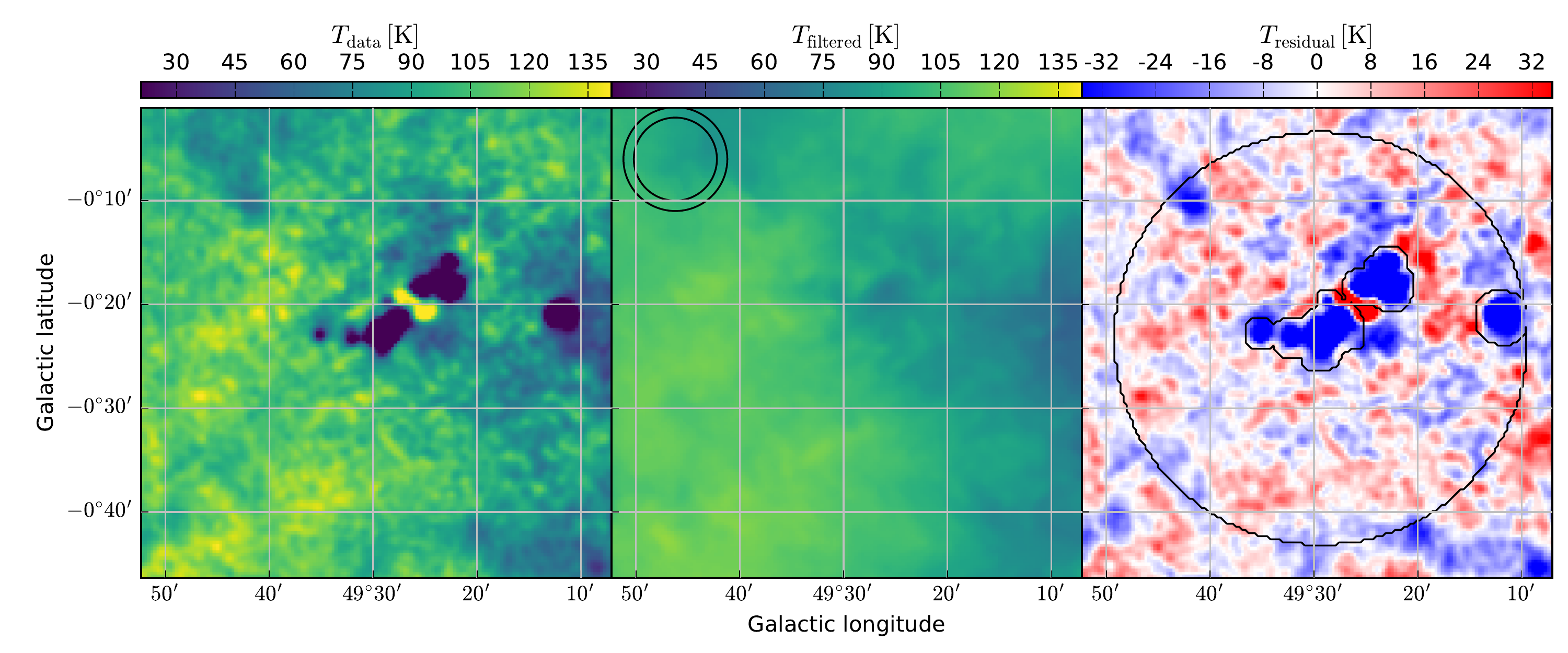}
\caption{Visualization of the moving-ring filter operation used to estimate the emission spectrum, $T_\mathrm{B}(\varv)$, at the positions affected by absorption features. The left panel shows a single plane ($\varv_\mathrm{lsr}=45.5~\mathrm{km\,s}^{-1}$) of the VGPS data cube of \object{W\,51} located in the center of the map. In the middle panel the filtered plane is shown. The size of the ring filter is marked with the two black circles in the upper left. The right panel contains the residual image. It is used to estimate the error of the emission spectrum by calculating the RMS. To improve the RMS calculation, a mask was applied (see text). The mask is indicated by the black contours.}%
\label{fig:spatial_filter}%
\end{figure*}

\begin{table*}[!tp]
\caption{Summary of the main observational parameters of the Galactic Plane surveys used to obtain emission spectra. Column~1: survey name, Col.~2: name of the interferometer used, Col.~3: angular resolution of the interferometric data (FWHM), Col.~4: spectral resolution, Col.~5: typical RMS noise level per spectral channel, Col.~6: single-dish telescope used to add-in the short-spacings information, Col.~7: angular resolution of the single-dish instrument, and Col.~8: the survey reference paper. Note that for the Galactic Center survey, $\sigma_\mathrm{rms}$ increases to about 2~K close to $(l,b)=(0\degr,0\degr)$. No meaningful $\ion{H}{i}$ column density could be determined toward the strong far-infrared source \object{DR\,21\,(OH)}. This was possible for the neighboring compact $\ion{H}{ii}$ region \object{DR\,21} (see text), however.}
\label{tab:gpsurveys}
\centering
\begin{tabular}{lllllllll}
\hline\hline
\rule{0ex}{3ex}Survey & Observatory  &  $\theta_\mathrm{maj,min}$ & $\delta \varv$ [$\mathrm{km\,s}^{-1}$] & $\sigma%_\mathrm{rms}
$ [$\mathrm{K}$] & SD &  $\theta_\mathrm{SD}$ & Reference\\
\hline
\rule{0ex}{3ex}VGPS & VLA  & $1\arcmin\times1\arcmin$ & $1.56$ & $2$ & GBT&  $9\arcmin$ & \citet{stil06}\\
CGPS & DRAO  & $1\arcmin\times1\arcmin\mathrm{csc}\delta$ & $1.32$ & $3$ & DRAO 26-m&  $36\arcmin$ & \citet{taylor03}\\
SGPS & ATCA  & $2\arcmin\times2\arcmin$ & $1.00$ & $1.6$ & Parkes&  $15\arcmin$ & \citet{mcclure05}\\
Galactic Center & ATCA & $145\arcsec\times145\arcsec$ & $1.00$ & $0.7$ & Parkes&  $15\arcmin$ & \citet{mcclure12}\\
\hline
\end{tabular}
\end{table*}

Of course, the Galactic plane surveys also exhibit \ion{H}{i} absorption features toward bright continuum sources, and in our case these can be quite extended. Therefore, special treatment is necessary to construct a suitable emission spectrum, $T_\mathrm{B}$. We applied a moving-ring filter to each plane in the data cube: Each pixel $p_{i,j}$ is replaced by the median value of all pixels in a ring around $p_{i,j}$. The inner and outer radii of this ring were set to $r_i=4\arcmin$ and $r_o=5\arcmin$ for all sources except for \object{Sgr\,B2}, where we used $r_i=7\arcmin$ and $r_o=9\arcmin$ to account for the larger extent of the background continuum source. The aim of the ring filter is to replace pixels affected by absorption with an average of the surrounding material. In principle, this would be necessary only for the pixels/sight lines of interest. However, applying the filter operation to the full map allowed us to estimate the residual scatter in regions around the sight lines, which is useful for subsequent error estimation. To further improve the filter operation we blanked pixels showing substantial absorption, $T<-25~\mathrm{K}\approx-10\sigma_\mathrm{rms}$, at any radial velocity prior to filtering. The unphysical negative brightness temperature values are merely a result of the Galactic plane survey data reduction, where the continuum levels were subtracted from the spectral line data cube. We note that the $r_i$ and $r_o$ values were made as low as possible to yield a minimal RMS level in the residual map, but under the constraint that spurious signs of the relevant absorption features must not be visible in the filtered maps.

The filtered cube was subsequently subtracted from the original data, yielding a residual data cube containing only small-scale fluctuations. For each plane in the residual cube, the RMS noise was calculated within a circle of radius $20\arcmin$ ($30\arcmin$ for \object{Sgr\,B2}) around the JVLA phase centers. This results in a ``noise spectrum'' $\Delta T_\mathrm{B}$ associated with each data set. The previously constructed blank mask was used again, after a dilation operator (five pixels in size) had been applied to it.

As an example, in Fig.~\ref{fig:spatial_filter} the result of such filtering is shown for one plane of the \object{W\,51} data cube ($\varv_\mathrm{lsr}=45.5~\mathrm{km\,s}^{-1}$). The left panel displays the original data, the middle panel shows the ring-filtered data, and the right panel contains the residual. Except for a few regions in the map, mostly those exhibiting strong absorption, the residual is relatively homogeneous.

\subsection{Correcting systematic errors}\label{subsec:systematicerrors}

For the derivation of the opacity and column densities formulas in Sect.~\ref{sec:theory}, we made several simplifying assumptions, which may in principle lead to systematic errors in our results. In the following, we briefly discuss the two main sources of potential uncertainty, and how to correct for them or lessen their impact.

\subsubsection{Short-spacings correction}\label{subsubsec:shortspacings}
\begin{figure}[!t]
\centering%
\includegraphics[width=0.48\textwidth,viewport=0 35 415 670,clip=]{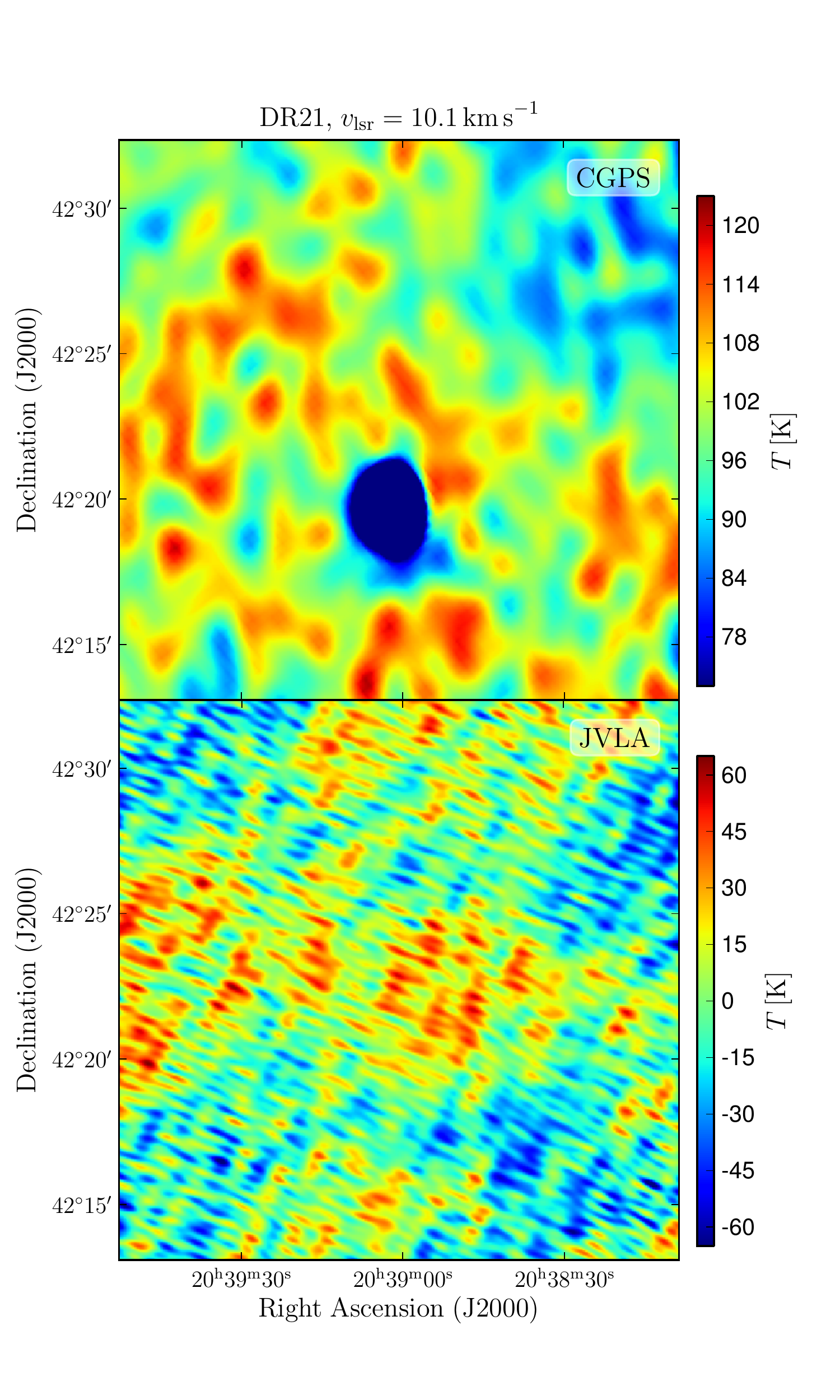}
\caption{Comparison between Galactic plane survey (top panel) and JVLA emission (bottom panel) data. The missing short-spacings in the latter cause a significant lack of (diffuse) emission. The example displays \object{DR\,21} data at $\varv_\mathrm{lsr}=10.1~\mathrm{km\,s}^{-1}$. Note the different intensity scales.}%
\label{fig:jvla_gps_comparison}%
\end{figure}

As mentioned at the beginning of Sect.~\ref{subsec:emissiondata}, the JVLA absorption data cubes are severely affected by the lack of short spacings. In Fig.~\ref{fig:jvla_gps_comparison} we show an example for the sight line to \object{DR\,21}. The top panel shows a plane of the CGPS data cube at $\varv_\mathrm{lsr}=10.1~\mathrm{km\,s}^{-1}$. The bottom panel contains the respective data from our JVLA data set. The former has a much higher base level of about 100~K, while in case of the JVLA, the mean is $\sim$0~K.

\begin{figure}[!t]
\centering%
\includegraphics[width=0.48\textwidth,clip=]{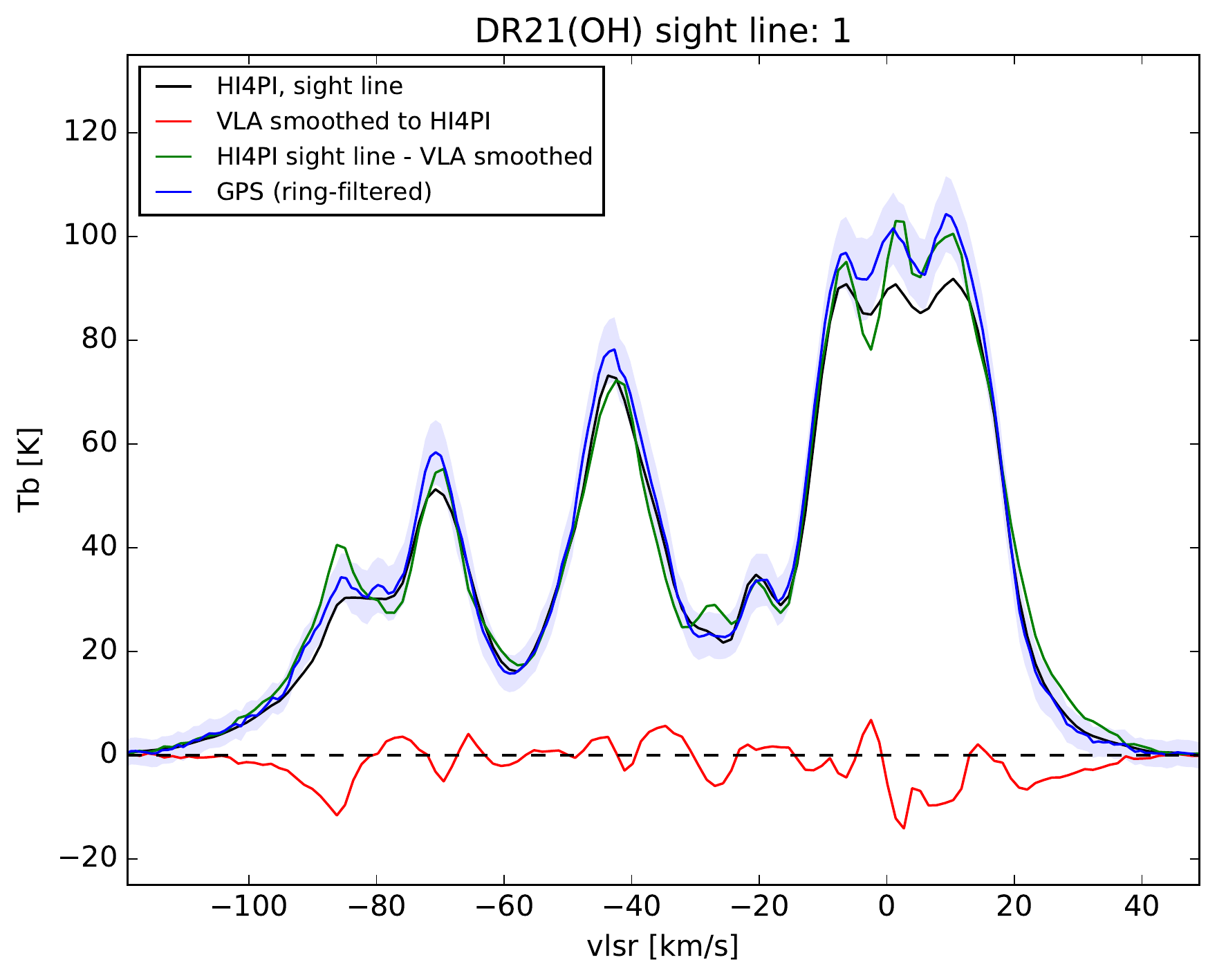}
\caption{Short-spacings correction for the JVLA absorption data. As an example, sight line~1 of the \object{DR\,21} source is shown. The black line is the original HI4PI spectrum, the red line is the JVLA data smoothed to HI4PI resolution, and the green line is the resulting short-spacings contribution. For comparison the result of the ring-filter operation, applied to the Galactic plane survey data cube, is displayed in blue (with $1\sigma$ uncertainties).}%
\label{fig:short_spacings}%
\end{figure}

To add the missing spacings, we made use of the recently published HI4PI data \citep{hi4pi16}. HI4PI is an all-sky single-dish \ion{H}{i} survey combined from observations with the 64 m Parkes/Australia and 100 m Effelsberg/Germany radio telescopes. In principle, we could also use the Galactic plane survey data sets, but HI4PI is more sensitive. There are various techniques to correct for missing spacings \citep[see, e.g.,][]{stanimirovic02}. Here, we chose combination in the image domain. First, we obtained suitable HI4PI data cubes of size $1\degr\times1\degr$ centered on the JVLA phase centers for each of our sources. Then, the JVLA data cubes were spatially and spectrally smoothed to the HI4PI resolution and resampled to the HI4PI voxel (volume pixel) grid, using the Python gridding software cygrid \citep{winkel16a}. This gives the brightness temperatures, $T_\mathrm{B}^\mathrm{JVLA,sm}=T_\mathrm{B}^\mathrm{small}$, that would have been observed by HI4PI if only the small angular scales, accessible to JVLA, were present. Of course, HI4PI, being a single-dish survey, observed all angular scales (i.e., $T_\mathrm{B}^\mathrm{HI4PI}=T_\mathrm{B}^\mathrm{all}$). Therefore, the larger angular scales, which are missing in the JVLA data, can be computed as $T_\mathrm{B}^\mathrm{large}=T_\mathrm{B}^\mathrm{all}-T_\mathrm{B}^\mathrm{small}$ and subsequently be used to correct the JVLA absorption spectra for each sight line.

As an example, we display the three quantities in Fig.~\ref{fig:short_spacings}. The black line is the original HI4PI spectrum, $T_\mathrm{B}^\mathrm{HI4PI}$, on the position of sight line~1 of the \object{DR\,21} source. From the smoothed JVLA data cube we extract $T_\mathrm{B}^\mathrm{JVLA,sm}$ (red line). The result, $T_\mathrm{B}^\mathrm{large}$, is plotted in green. For comparison, we also show the result from the ring-filtered Galactic plane survey data set (blue line, with $1\sigma$ errors). We also find for all other sources and sight lines that the short-spacings contribution is very similar to the GPS (ring-filtered), which means that no significant contribution stems from angular scales larger than the size of the ring-filter aperture. We note that this effectively means that the nominator in Eq.~(\ref{eq:tau}) reduces to the originally measured JVLA absorption spectrum, and the opacity spectrum is therefore almost exclusively determined by the JVLA absorption data alone. However, the interpolated brightness temperature spectrum, $T^\mathrm{off}$, is still mandatory for deriving spin temperatures and column densities.

\subsubsection{Beam-filling considerations}\label{subsubsec:beamfilling}
We have discussed in Sect.~\ref{sec:theory} that the effect of beam filling can usually not be neglected. Its consequence is a reduction of the observed intensity, $T_\mathrm{cont}^\mathrm{sou,obs}$, by a factor $f_\mathrm{B}$, that is, $T_\mathrm{cont}^\mathrm{sou,obs} = f_\mathrm{B} T_\mathrm{cont}^\mathrm{sou}$.
Likewise, the observed \textsc{On} spectrum is modified
\begin{equation}
T^\mathrm{on,obs} =\left(f_\mathrm{B}T_\mathrm{cont}^\mathrm{sou}+T_\mathrm{cont}^\mathrm{bg}\right) e^{-\tau} +T_\mathrm{B}\label{eq:t_on_obs}\,,
\end{equation}
and we can show that
\begin{equation}
e^{-\tau(\varv)}= \frac{T^\mathrm{on,obs}-T^\mathrm{off}}{T_\mathrm{cont}^\mathrm{sou,obs}}\label{eq:taumod}\,.
\end{equation}
Hereafter, we use the superscript `obs' to mark a quantity that is affected by beam filling (with respect to the continuum-emitting background source, only). Equation~(\ref{eq:taumod}) shows that it is not necessary to incorporate the beam-filling factor in the opacity calculations; we can just work with the observed quantities. However, the absorption profile itself is changed indeed: $f_\mathrm{B}$ effectively acts a weighting factor between emission and absorption in Eq.~(\ref{eq:t_on_obs}). This also explains why absorption features in low-angular resolution data are often much less pronounced. The relative strength of the absorption line (with respect to $T_\mathrm{B}$) is suppressed. If no other filling factor effects were to play a role, however,
we could infer the opacity equally well from such data. In practice, however, in lower-angular resolution single-dish observations, the structure of the absorbing source cannot be neglected, which makes an analysis much more challenging.

Although $f_\mathrm{B}$ does not play a role for estimating $\tau$, we still determined it for each sight line and quote the values (Sect.~\ref{sec:absorptionspectra}, Fig.~\ref{fig:results_dr21_1}, and Appendix~\ref{sec:othersightlines}) because it may be useful when comparing the absorption profiles.

The continuum maps (Figs.~\ref{fig:cont_maps1} and \ref{fig:cont_maps2}) show that most of the extracted sight lines are toward unresolved sources. Assuming the sources can be described by an elliptical Gaussian, the angular size of the beam and the intrinsic size of each source add quadratically. Thus, it is possible to obtain a rough estimate on $f_\mathrm{B}$ by fitting (elliptical) Gaussians to the sight lines. In several cases blending of features complicates the fitting, however. Stable results could be achieved by using an iterative approach, applying a clip mask (neglecting values above $3\sigma_\mathrm{rms}$) and increasing the fit aperture in steps from two to three beam widths. With the determined source sizes we calculated the beam-filling factors through numerical integration.

Unfortunately, we can only estimate the beam-filling effects for the background continuum emitter, but not for the absorbing material. As a consequence, a correction for the source covering factors is not possible. All our derived opacity and column density profiles will therefore include the implicit assumption that absorbing material is a constant sheet of gas (across the beam). However, we show below that in the derived opacity and column density spectra relatively strong spatial variations are detected, which casts doubts on the validity of this assumption.

\section{Computing \ion{H}{i} opacities and column densities}\label{sec:absorptionspectra}

In the following we describe how the two complementary data sets are used to compute optical depth, spin temperature, and column density spectra. To estimate reasonable error bars, we employed Bayesian inference using the \textit{pymc} framework (version 3), which is a library module available for the Python programming language. If provided with a list of stochastic variables and their relationship, the \textit{pymc} framework can automatically calculate the joint likelihood distribution. Given a start parameter vector, \textit{pymc} can then sample the complex parameter space using Markov-chain Monte Carlo (MCMC) calculations. From the resulting chains, it is possible to marginalize over individual parameters, which allows us to compute the expected value of each parameter and its error distribution.

In our case, we have four observed quantities (per spectral channel). For each we can estimate its uncertainty:

\begin{itemize}
\item For the absorption noise spectrum, we first smoothed the VLA data cube in velocity (using a Gaussian filter with a kernel width of eight spectral channels, FWHM) to approximately match the spectral resolution of the emission spectrum. For each sight line, we extracted the spectrum $T_\mathrm{VLA}^\mathrm{on,obs}$. Furthermore, in the resulting data cube we iteratively inferred the RMS, $\sigma_\mathrm{rms}$, per velocity plane in a robust manner, by blanking $4\sigma_\mathrm{rms}$ outliers in each step. This typically converged after three iterations and results in an estimate for $\sigma\left(T_\mathrm{VLA}^\mathrm{on,obs}\right)$.
\item For the emission spectrum, inferred from the Galactic plane surveys, the method described in Sect.~\ref{subsec:emissiondata} delivers not only the $T_\mathrm{B}\equiv T_\mathrm{GPS}^\mathrm{ring-filter}$ estimate, but also its frequency-dependent uncertainty spectrum, $\sigma\left(T_\mathrm{B}\right)$, for each target data set. For subsequent calculations, the resulting emission spectra were interpolated to the spectral grid defined by the VLA absorption data sets.
\item The continuum levels, $T_\mathrm{cont}^\mathrm{sou,obs}$, for each sight line were extracted from the continuum maps (Figs.~\ref{fig:cont_maps1} and \ref{fig:cont_maps2}). The associated errors $\sigma\left(T_\mathrm{cont}^\mathrm{sou,obs}\right)$, were already quoted in Sect.~\ref{subsec:jvla_data}, Table~\ref{tab:vla_sources}.
\item For the sky background, $T_\mathrm{cont}^\mathrm{bg}$, and its uncertainty, $\sigma\left(T_\mathrm{cont}^\mathrm{bg}\right)$, we took rough estimates from the Stockert 21 cm survey \citep{reich82,reich86}. However, the contribution of  $T_\mathrm{cont}^\mathrm{bg}$ to the JVLA continuum flux is smaller than 1\% in all cases and can safely be ignored, compared to the magnitude of all other uncertainties.
\end{itemize}

We note that all these estimates account for spatial variations and not only for the noise in one sight line. This means that potential errors caused by improper cleaning and short-spacings correction or the ring-filtering are included.

Before we processed the data in \textit{pymc}, we calculated the effectively observed \textsc{On} and \textsc{Off} profiles
\begin{align}
T^\mathrm{on,obs} &= T_\mathrm{VLA}^\mathrm{on,obs} + T_\mathrm{B}^\mathrm{large} + T_\mathrm{cont}^\mathrm{bg}\label{eq:model_ton_obs}\\
T^\mathrm{off} &= T_\mathrm{B} + T_\mathrm{cont}^\mathrm{bg}\label{eq:model_toff}\,,
\end{align}
where we added $T_\mathrm{B}^\mathrm{large}+ T_\mathrm{cont}^\mathrm{bg}$  to the VLA \ion{H}{i} absorption spectrum, $T_\mathrm{VLA}^\mathrm{obs}$, to account for short-spacings (see Sect.~\ref{subsubsec:shortspacings}). Because the errors of the inputs are approximately Normal-distributed, we can directly calculate the resulting uncertainties
\begin{align}
\sigma\left(T^\mathrm{on,obs}\right) &= \sqrt{\sigma\left(T_\mathrm{VLA}^\mathrm{on,obs}\right)^2 + \sigma\left(T_\mathrm{B}^\mathrm{large}\right)^2 + \sigma\left(T_\mathrm{cont}^\mathrm{bg}\right)^2}\\
\sigma\left(T^\mathrm{off}\right) &= \sqrt{\sigma\left(T_\mathrm{B}\right)^2 + \sigma\left(T_\mathrm{cont}^\mathrm{bg}\right)^2}\,.
\end{align}

For the four variables, $T^\mathrm{on,obs}$, $T^\mathrm{off}$, $T_\mathrm{cont}^\mathrm{sou,obs}$, and $T_\mathrm{cont}^\mathrm{bg}$ we assigned Normal distributions in \textit{pymc} with the according standard deviations. We note that technically we have to provide \textit{pymc} a Normal-distribution object with an \textit{observed} keyword argument along with the mean and standard deviation of the distribution if we wish to define observables. For $T^\mathrm{on,obs}$ and $T^\mathrm{off}$ we inserted the two model equations, Eqs.~(\ref{eq:t_on_obs})~and~(\ref{eq:t_off}), for the mean, and the inferred errors for the standard deviation. However, in case of $T_\mathrm{cont}^\mathrm{sou,obs}$ and $T_\mathrm{cont}^\mathrm{bg}$ we do not know the true mean of the distribution (our measurement is one draw from it, and \textit{pymc} cannot calculate it from other quantities). The solution is to define the mean as another random variable with a flat prior (i.e., using a uniform distribution).

Furthermore, we have two unknown variables, the opacity, $\tau$ and the spin temperature, $T_\mathrm{spin}$. For $\tau$, which certainly has an asymmetric distribution, we used a lognormal prior with $\mu=0$ and $\sigma=2$. Having no better prior information, we work with a flat prior for the spin temperature.

These six distributions are linked through Eqs.~(\ref{eq:t_on}) and~(\ref{eq:t_off}), and \textit{pymc} automatically computes the resulting likelihood function. For the posteriors of $\tau$ and $T_\mathrm{spin}$ we can then calculate distribution percentiles to quantify $1\sigma$ and $3\sigma$ confidence levels (CL). The $1\sigma$ interval is defined by the $15.87\%$ and $84.13\%$ percentiles, while the $3\sigma$ interval is given by the  $0.13\%$ and $99.87\%$ percentiles. Furthermore, we determine the $50\%$ percentile (i.e., the median); this is the most likely result.

\begin{figure}[!t]
\centering%
\includegraphics[width=0.48\textwidth,viewport=10 10 415 980,clip=]{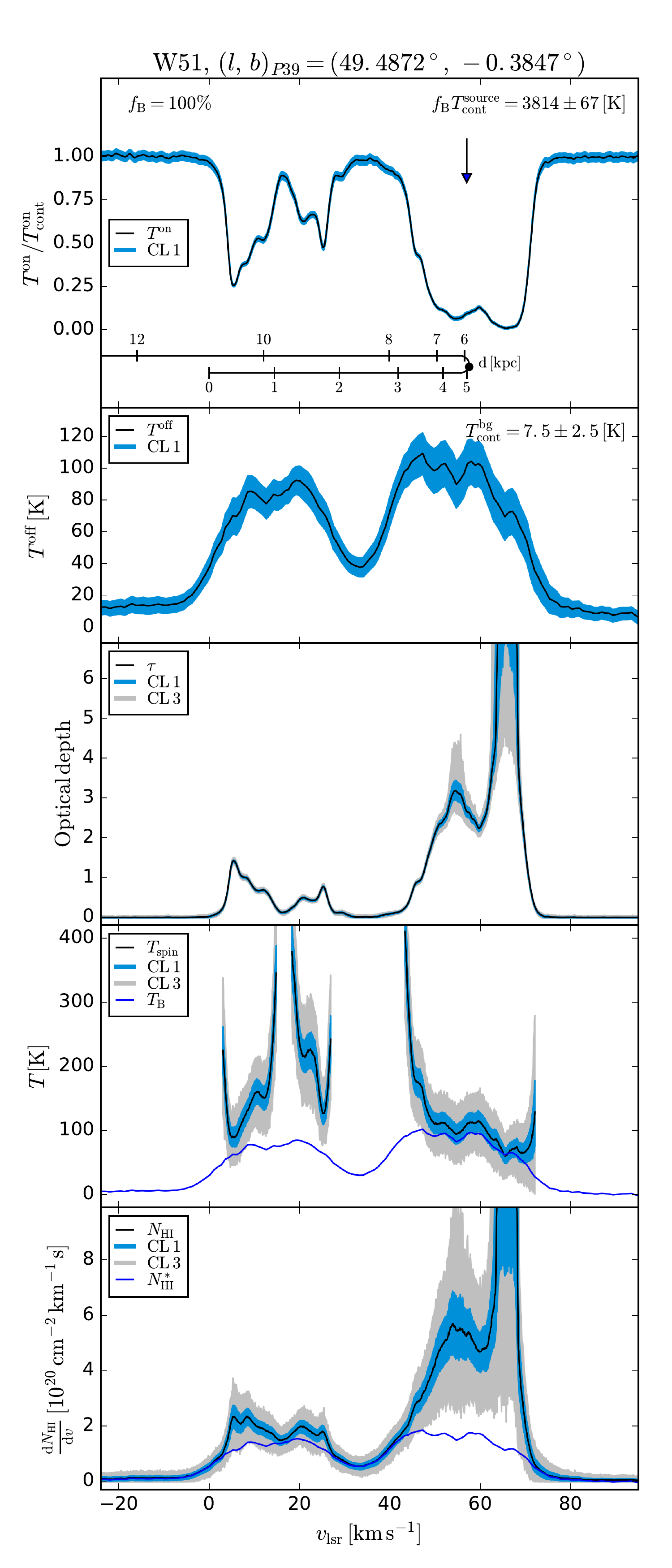}
\caption{Measured absorption and emission spectra, $T^\mathrm{on}$ and $T^\mathrm{off}$, with their respective errors, and derived quantities, optical depth, spin temperature, and \ion{H}{i} column density for the PRISMAS program source \object{W\,51}, sight line P39. For a detailed description see text. The labels CL\,1 and CL\,3 refer to the $1\sigma$ and $3\sigma$ confidence intervals, i.e., the 68\% and 99.7\% distribution percentiles (see also text).}%
\label{fig:results_w51_p1}%
\end{figure}

Figure~\ref{fig:results_w51_p1} shows the results of our absorption analysis for the example of the PRISMAS program source \object{W\,51} (sight line P39). Other PRISMAS/HEXOS sight lines are contained in Appendix~\ref{sec:prismassightlines}, Figs.~\ref{fig:results_prismas_1n2} to \ref{fig:results_prismas_5}, and all remaining spectra are compiled in Appendix~\ref{sec:othersightlines}, Figs.~\ref{fig:results_sgr_1n2} to \ref{fig:results_dr21_1} (compare Figs.~\ref{fig:cont_maps1} and \ref{fig:cont_maps2} for the spatial position of the sight lines). The top panel displays the normalized on-source spectrum, $T^\mathrm{on,obs}/T_\mathrm{cont}^\mathrm{sou,obs}$. The black solid line is the spectrally smoothed profile, which matches the resolution of the emission data. The blue interval visualizes the 1$\sigma$ interval. The continuum flux density, $T_\mathrm{cont}^\mathrm{sou,obs}$, of the background point source, as inferred from the continuum maps, is denoted in the upper panel.

The second panel contains the interpolated brightness temperature spectrum, $T^\mathrm{off}$ (black solid line), including $T_\mathrm{cont}^\mathrm{bg}$. The blue interval again shows the 1$\sigma$ interval. The value of $T_\mathrm{cont}^\mathrm{bg}$  is annotated in the plot.

The remaining three panels show the optical depth, spin temperature, and \ion{H}{i} column density. In each panel, the blue and gray intervals refer to the $1\sigma$ (68\% percentile) and $3\sigma$ (99.7\% percentile) confidence levels, as discussed above, while the black solid line is the median of the resulting distributions (50\% percentile). In the temperature panel, we also show the brightness temperature for comparison (blue solid line). By design, it provides a lower limit to the spin temperature because we added the emission spectrum to the \ion{H}{i} absorption profile for an approximate short-spacings correction. The bottom panel also displays the uncorrected column density spectrum, $N_\ion{H}{i}^\ast\propto T_\mathrm{B}$ (blue solid line).

\subsection{Discussion}\label{subsec:blending_discussion}

Similar to \citet{fish03}, we have plotted a distance indicator in each of the figures (upper panels). It is based on a very simple model of disk rotation of the Milky Way \citep[flat rotation curve with $R_0=8.34~\mathrm{kpc}$ and $\Theta_0=240~\mathrm{km\,s}^{-1}$ (A5 model); see][]{reid14} calculated for each of the sight lines. The bend in the bar refers to the terminal velocity (tangent point), and the black circle marks the distance to the PRISMAS/HEXOS program sources on the respective sight lines (compare Table~\ref{tab:vla_sources}). The associated LSR velocity of this black circle is in most cases inconsistent with observations; compare Table~\ref{tab:vla_sources}. This is either caused by peculiar velocities or by the applied MW rotation model being incorrect. Therefore, we also draw a black arrow in the top panel to indicate the approximate source velocity of the continuum-emitting PRISMAS/HEXOS source, as inferred from observations.  It is not possible to plot a distance indicator for the sight line to \object{Sgr\,B2 } since there is no radial velocity component toward the Galactic center in a simple disk-rotation model. According to \citet{levine06}, the \ion{H}{i} surface density has considerable values out to galactocentric radii of $R\sim20\mathrm{kpc}$. We can therefore calculate a maximum distance for each sight line out to which \ion{H}{i} emission can be expected. This is reflected by the length of the distance indicators in the plots. The emission spectrum covers the possible range of distances quite well in
all cases.

This also clearly illustrates our discussion of blending issues in Sect.~\ref{sec:theory}: the velocity--distance relation is degenerate. \ion{H}{i} features from the far end of the MW disk are folded into the velocity range covered by the absorbing material. In all sight lines the ``background'' continuum source is at much shorter distances than the full length of the sight line through the MW \ion{H}{i} gas. This additional contribution to the emission spectrum, $T^\mathrm{off}$, is unaccounted for when calculating spin temperatures and opacity-corrected column densities because we cannot separate the near- and far-side components. For the \textsc{On}--\textsc{Off} technique, only the near-side contribution would be relevant. This issue introduces a systematic error, which is not incorporated into the Bayesian uncertainty calculation, and leads to overestimation of $T_\mathrm{spin}$ and $N_\ion{H}{i}$. Because the received \ion{H}{i} emission from a thin sheet of gas at distance $[d, d+\Delta d]$ decreases with $d^2$ but at the same time the observed volume increases with $d^2$, the contribution of a near- and far-side component with identical radial velocity will be similar, given homogeneous density. In reality, however, the radial velocities and the \ion{H}{i} distribution along the sight lines are too complex for a simple estimation of the size of the overall effect.

For several of the targets we were able to extract more than one absorption profile. In some cases there is considerable small-scale variation in the absorbing material that is unrelated to the background objects: the absorption features differ significantly in the sight lines toward \object{G\,10.62$-$0.39} and \object{W\,51}. As an example, for \object{G\,10.62$-$0.39} the two analyzed sight lines are only 90\arcsec\ apart, and absorption features differ significantly on radial distances smaller than 2 to 3~kpc. This converts into spatial scales of about 1~pc. Fluctuations on even smaller scales (several AU) were detected with VLBI experiments \citep[e.g.,][]{faison98,brogan05,lazio09}, which used as background sources the bright continuum emission of the jets of the active galactic nuclei of \object{3C138}, \object{3C147} and other galaxies. For \object{W\,33} and \object{G\,34.3$+$0.1}, the JVLA profiles seem to have much less variation. The remaining targets do not permit such an analysis.

\subsection{Comparison with previous studies}\label{subsec:opacities_comparison}
As noted in the introduction, for several of our targets there exist other studies, which calculated \ion{H}{i} opacities and in some cases also $N_\ion{H}{i}$ values.

\subsubsection{Sgr\,B2}
\ion{H}{i} absorption in the Galactic center region has been studied in great detail in \citet{lang10}. Their \object{Sgr\,B2} sight line was also part of their analyses. The derived opacity spectrum is consistent with our findings, although the spectral resolution of $\delta\varv=2.5~\mathrm{km\,s}^{-1}$ is lower. Based on consistency arguments, \citet{lang10} assumed a spin temperature of 60~K, which is close to our findings for a good fraction of the absorbers along the sight line.

\subsubsection{G\,31.41$+$0.31 and G\,34.3$+$0.1}
For \object{G\,31.41$+$0.31} and \object{G\,34.3$+$0.1} we are not aware of dedicated \ion{H}{i} absorption measurements. However, \citet{dickey83} presented a spectrum toward the extragalactic source \object{1849$+$005}, $(l, b)=(33.5\degr,0.19\degr$, which is in close vicinity. The sensitivity and spectral resolution is much poorer, but nevertheless, the $\tau$ spectrum seems consistent, with one exception: in our \object{G\,34.3$+$0.1} data, the absorption is only visible below $\varv\lesssim70~\mathrm{km\,s}^{-1}$, while in \object{1849$+$005} and \object{G\,31.41$+$0.31} absorption occurs out to $\varv\lesssim120~\mathrm{km\,s}^{-1}$. This is not surprising because \object{G\,34.3$+$0.1} is located at much smaller distance.

\subsubsection{W\,49}
The source \object{W\,49} was also observed by \citet[][their sight line~L equals our sight line~4]{brogan01}. Their opacity spectrum only covers $-20<\varv_\mathrm{lsr}<20~\mathrm{km\,s}^{-1}$ and is, in this interval, consistent with our findings. They did not measure $T_\mathrm{spin}$ , but estimated it to be in the range of 20 to 150 K, which is compatible with our findings, if not a very strict limit.

\subsubsection{W\,51}
The whole \object{W\,51} complex has been studied by \citet{koo97}, while we only observed a smaller region in the northwestern part of \object{W\,51} (their sight line \object{G\,49.5$-$0.4\,e} corresponds to our PRISMAS sight line P39). Their spectra have relatively coarse resolution, but the opacities and integrated \ion{H}{i} column densities are very consistent, ours being only about 20\% higher. To derive $N_\ion{H}{i}$, \citet{koo97} assumed a constant spin temperature of 160~K, which is compatible with our findings at local velocities, but for $40\lesssim\varv_\mathrm{lsr}\lesssim70~\mathrm{km\,s}^{-1}$ we find significantly lower values of $T_\mathrm{spin}\lesssim100~K$ (which is probably even overestimated; see Sect.~\ref{sec:theory}).

\subsubsection{DR\,21}
The source \object{DR\,21} is interesting because it features a molecular outflow \citep{roberts97} that is visible as a broad wing in the absorption  spectrum ($-30<\varv_\mathrm{lsr}<-6~\mathrm{km\,s}^{-1}$). The opacities, derived in \citet{roberts97}, are slightly higher than our values, and their assumed $T_\mathrm{spin}$ of 20~K is significantly below the $\sim$100~K from our calculation.

\section{Molecular hydrogen fraction on the observed sight lines}\label{sec:molfraction}
With the \ion{H}{i} column densities at hand, we can derive the molecular hydrogen fraction
$f_{\mathrm{H}_2}$ $\equiv 2 N_{\mathrm{H}_2}/(N_{\ion{H}{i}}+ 2 N_{\mathrm{H}_2})$ in a relatively straightforward way. As already stated in the introduction, the direct measurement of $N_{\mathrm{H}_2}$ is excluded on sight lines that are mainly populated with translucent clouds \citep{rachford02}, and where possible, it has a limited spectral resolution (typically $20~\mathrm{km\,s}^{-1}$). We therefore used the HF absorption spectra from HIFI (key projects PRISMAS, \citealp{gerin12}, and for \object{Sgr\,B2}, HEXOS, \citealp{bergin10}) as a proxy for H$_2$. Data analysis was performed with the CLASS software\footnote{http://www.iram.fr/IRAMFR/GILDAS}. We did not consider the sight lines to \object{DR\,21\,(OH)} and \object{W\,33\,A} because their radio continuum is too weak to allow for a reliable estimate of $f_{\mathrm{H}_2}$.

\subsection{Computing HF column densities}\label{sec:molecularHydrogen}
As described in the introduction, HF and CH can be used as surrogates for H$_2$. The rotational ground-state line of CH has a relatively large hyperfine splitting of $\sim$$20~\mathrm{km\,s}^{-1}$. At the radial velocities of the hot cores, illuminating the diffuse gas on the sight line, the CH lines easily turn into emission. Both circumstances lead to complications that we prefer to avoid here. The correspondence between HF and CH column densities has been discussed by \citet[][and references therein]{godard12}. \citet{wiesemeyer16} have confirmed this correspondence on the sight line to \object{W\,49N}. Within 15\%, the conversion factor deduced from the HIFI/PRISMAS absorption spectrum is in agreement with an abundance of HF of $1.4\cdot 10^{-8}$, which agrees with recent findings by \citet{sonnertrucker15}, who in turn confirmed the measurements of \citet{indriolo13} in the $v = 1$$-$$0, R(0)$ ro-vibrational transition of HF using recent
measurements of the HF+H$_2$ reaction rate \citep{tizniti14}.

The HF molecule, which was first discovered on the sight line to \object{Sgr\,B2} \citep{neufeld97}, owes its usefulness as proxy for H$_2$ to the reaction F(H$_2$,H)HF (fluorine is the only element capable of such an exothermic reaction with H$_2$). Light hydrides, such as HF and CH, have large Einstein A-coefficients, which is why it is difficult to excite them collisionally in diffuse clouds. In the radiative equilibrium with the Galactic interstellar radiation field, we can safely assume that basically all HF molecules are in the ground state.

\begin{figure*}[!t]
\centering%
\includegraphics[width=0.48\textwidth,viewport=10 10 415 719,clip=]{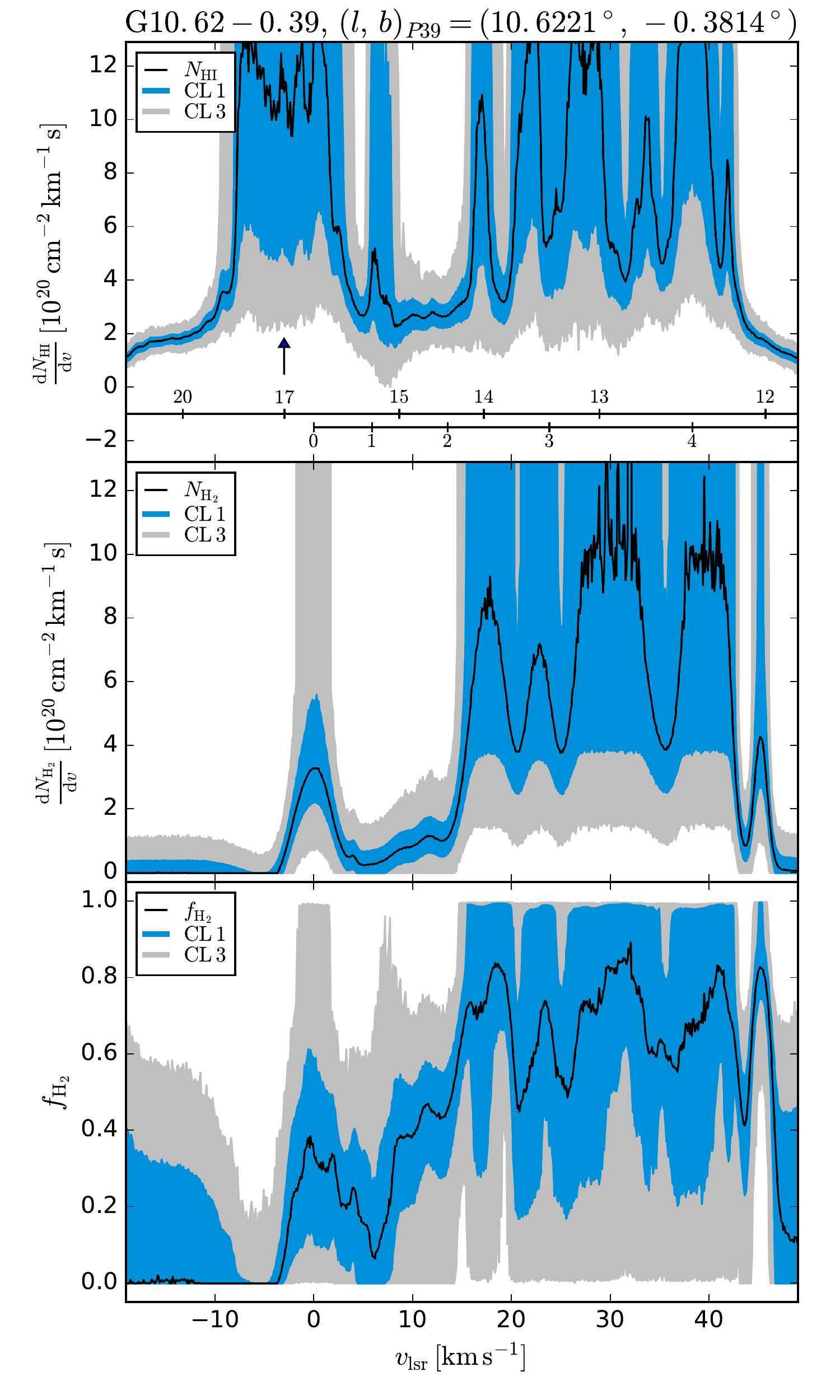}\hfill
\includegraphics[width=0.48\textwidth,viewport=10 10 415 719,clip=]{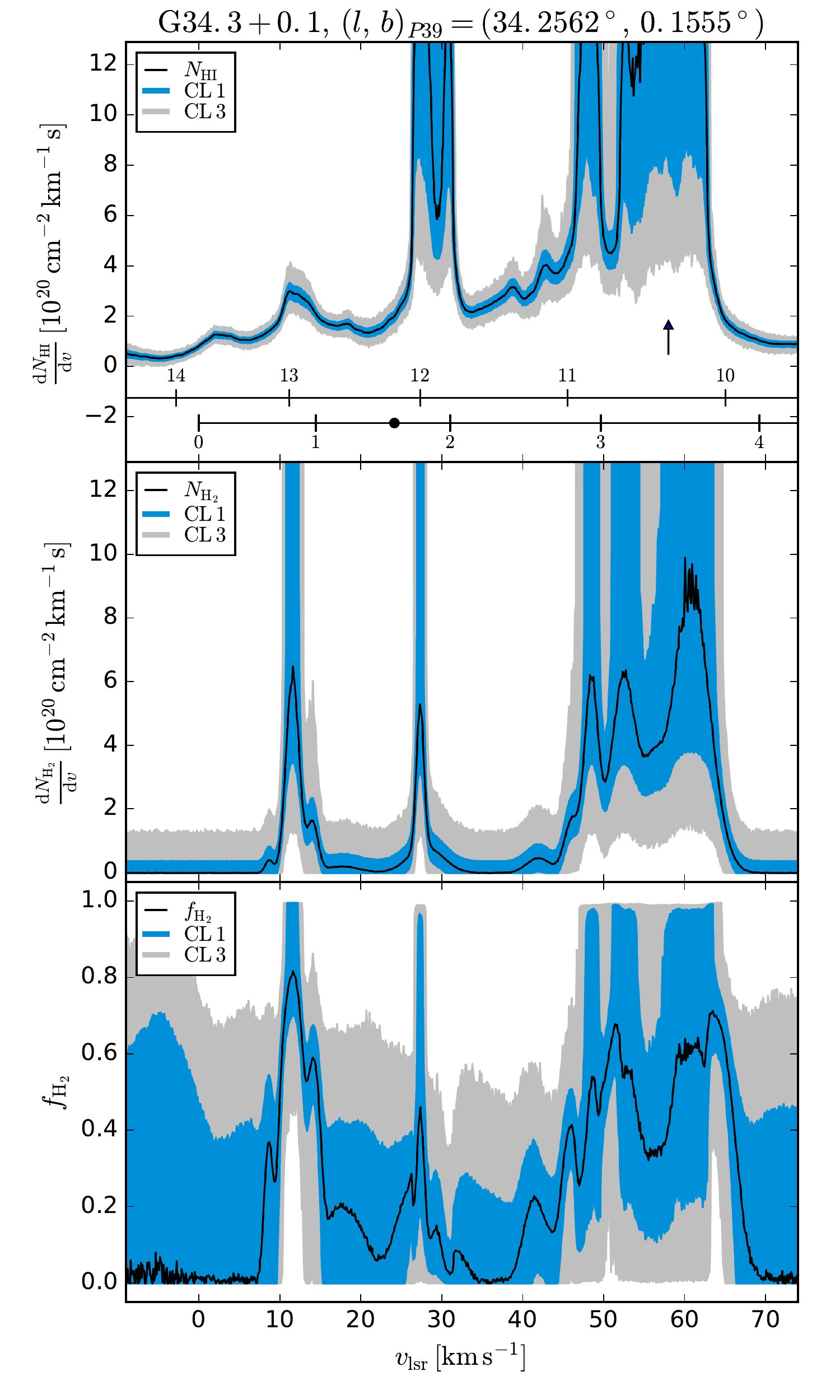}
\caption{Molecular hydrogen fraction $f_{\mathrm{H}_2}$ $\equiv 2N_{\mathrm{H}_2}/(N_{\ion{H}{i}}+ 2 N_{\mathrm{H}_2})$ on the sight lines toward \object{G\,10.62$-$0.39} (\object{W\,31C}) and \object{G\,34.3$+$0.1}.}
\label{fig:results_nh2a}%
\end{figure*}

\begin{figure*}[!t]
\centering%
\includegraphics[width=0.48\textwidth,viewport=10 10 415 719,clip=]{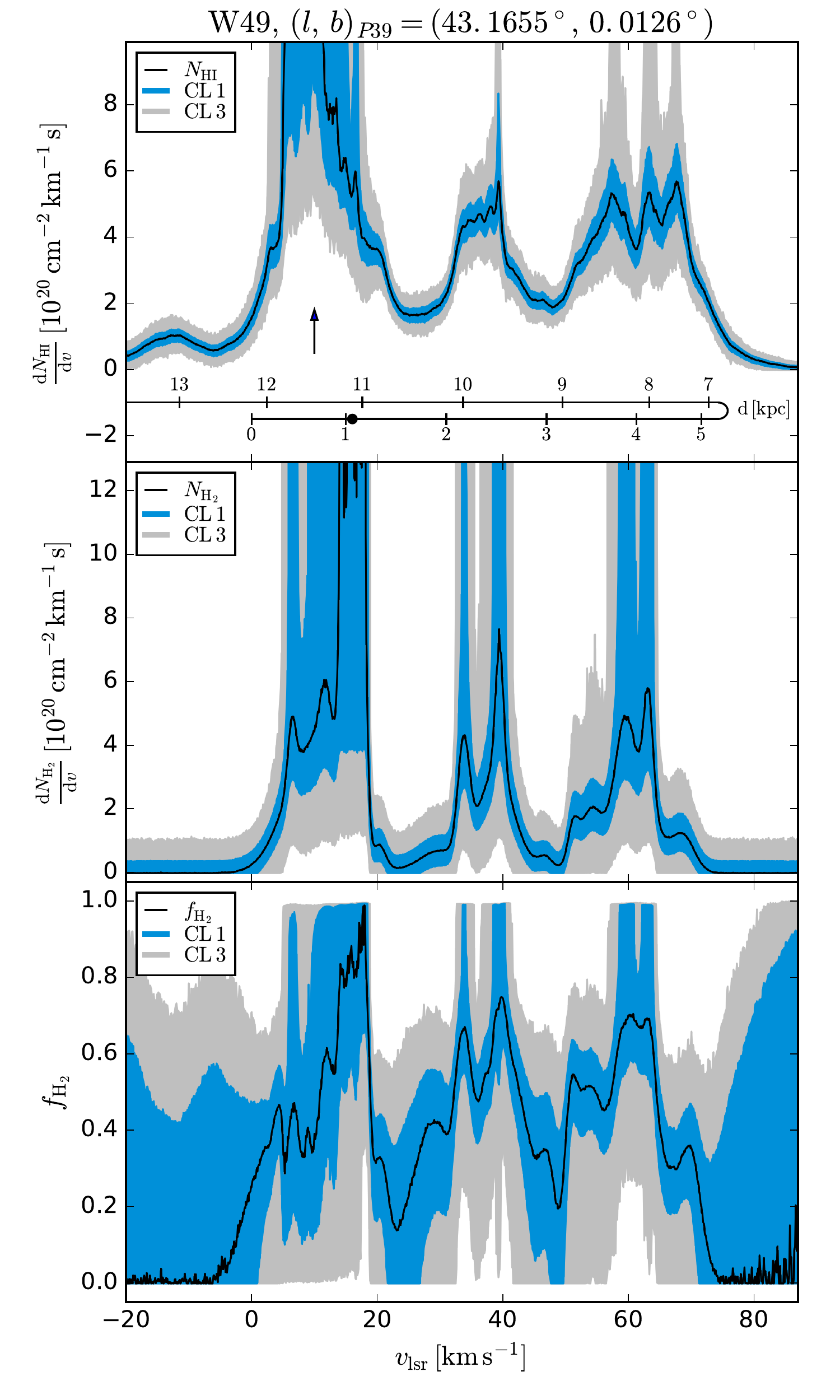}\hfill
\includegraphics[width=0.48\textwidth,viewport=10 10 415 719,clip=]{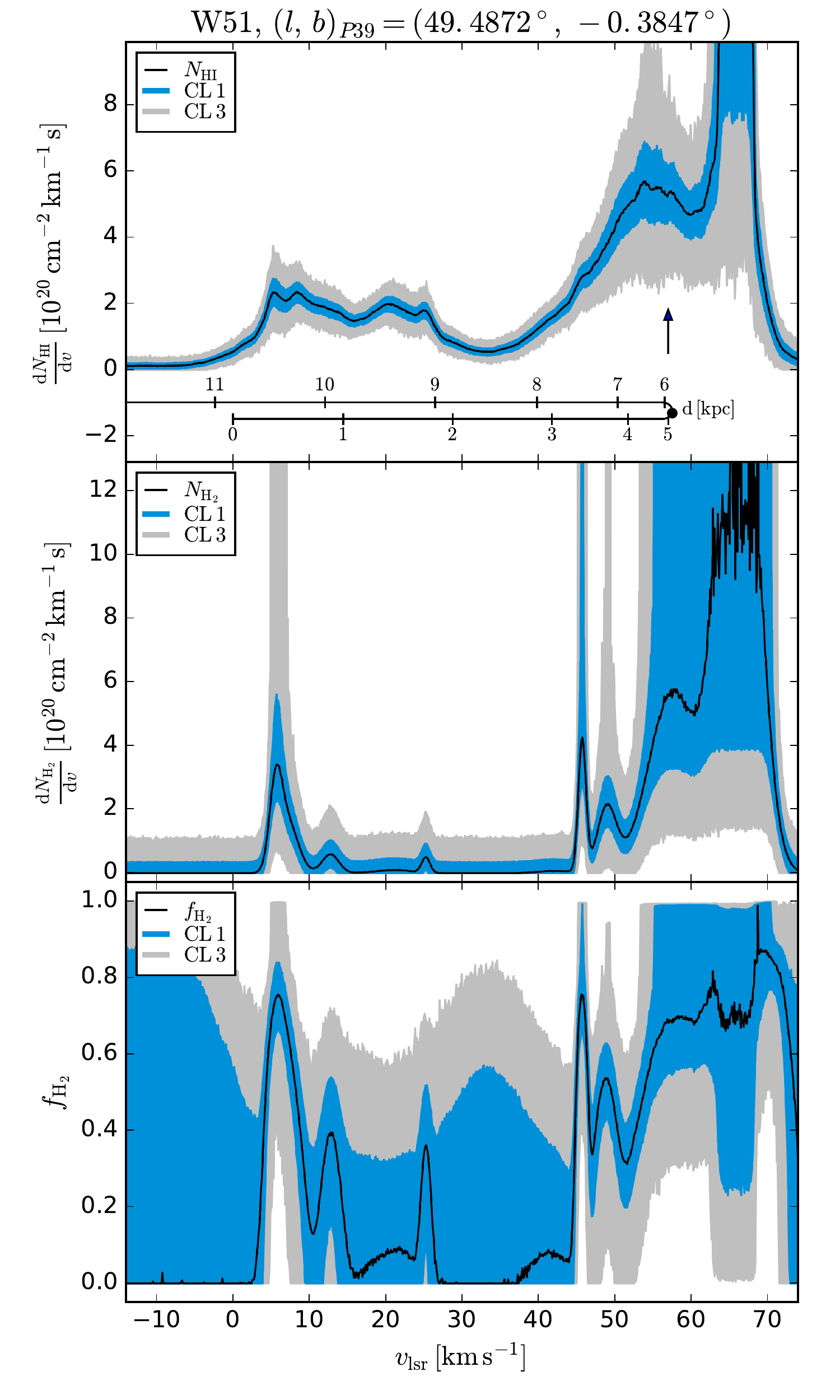}
\caption{As Fig.~\ref{fig:results_nh2a} for the sight lines toward \object{W\,49N} and \object{W\,51e2}.}
\label{fig:results_nh2b}%
\end{figure*}

\begin{figure*}[!t]
\centering%
\includegraphics[width=0.48\textwidth,viewport=10 10 415 719,clip=]{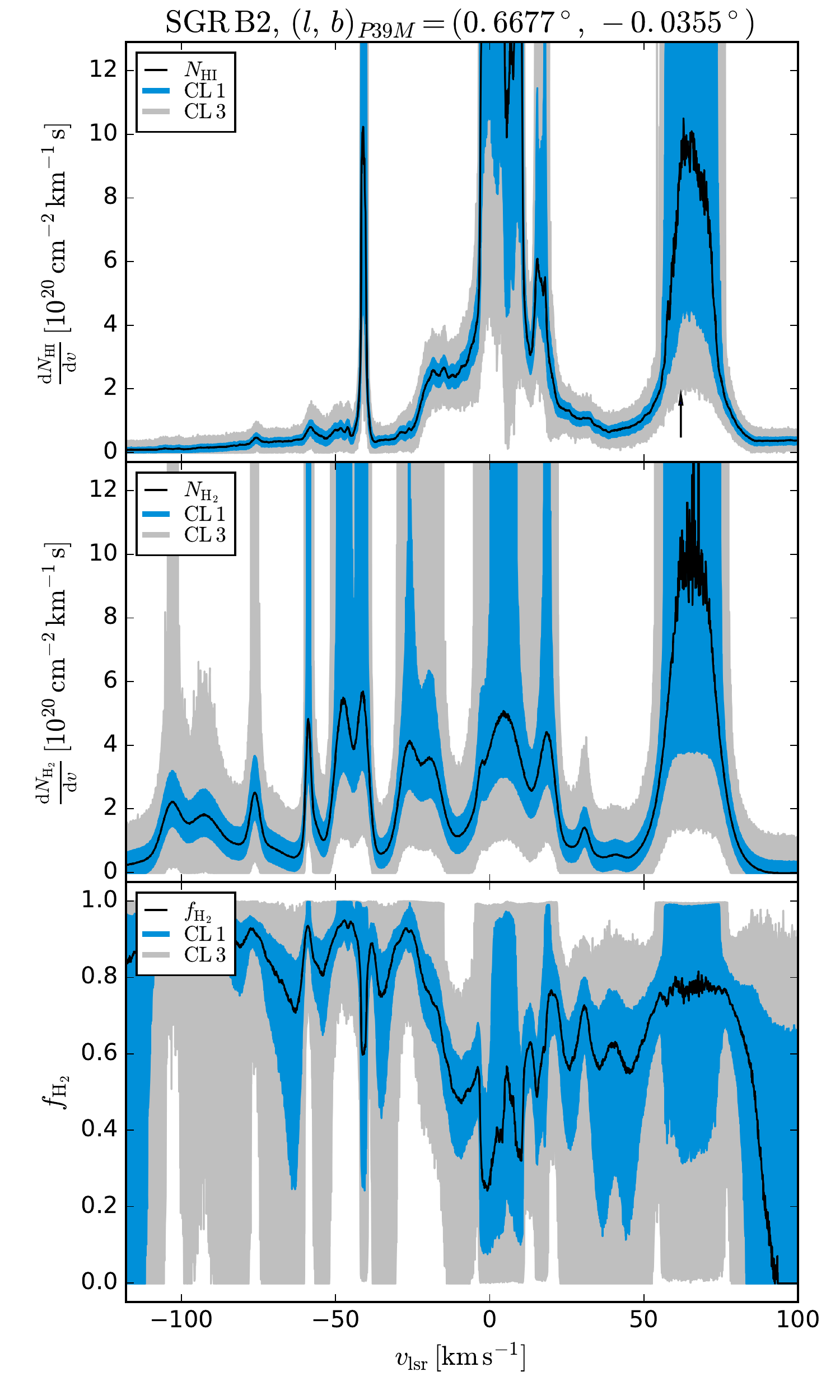}
\includegraphics[width=0.48\textwidth,viewport=10 10 415 719,clip=]{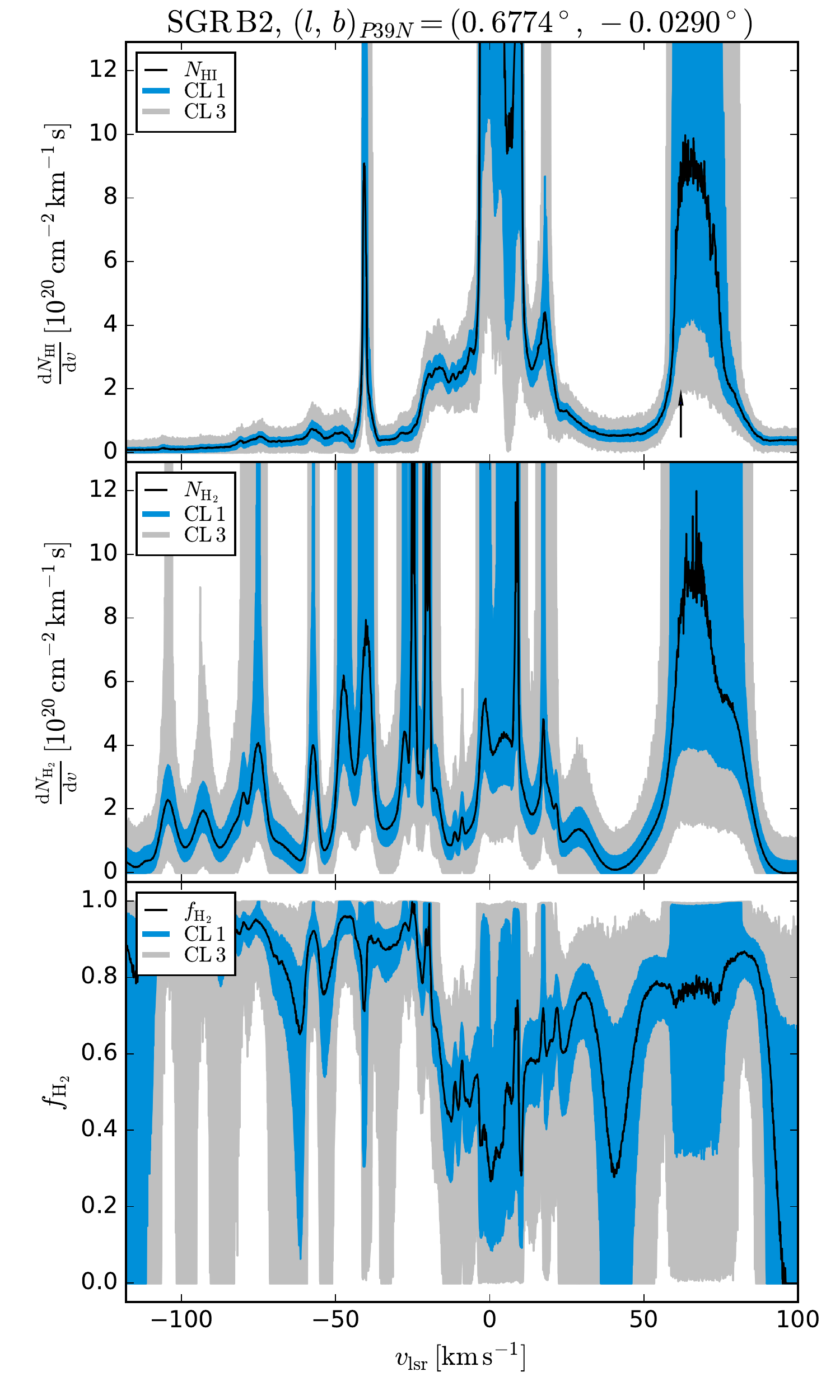}
\caption{As Fig.~\ref{fig:results_nh2a} for the sight lines toward \object{Sgr\,B2}.}
\label{fig:results_nh2c}%
\end{figure*}

This allows us to directly determine the H$_2$ column density, $dN/d\varv$, from the measured optical depth $\tau$ by means of the equation \citep[e.g.,][]{cannon12}
\begin{equation}
\tau = \frac{A_\mathrm{E}c^3}{8\pi \nu^3} \left[1-\exp{\left(-\frac{E_\mathrm{u}}{k_\mathrm{b}T}\right)}\right]
\frac{g_{\mathrm{u}}}{Q(T)}\frac{dN}{d\varv}\,,
\end{equation}
where $A_\mathrm{E}$ is the spontaneous emission Einstein coefficient of the transition of frequency $\nu$, $E_\mathrm{u}$ is the energy of the upper level with degeneracy $g_\mathrm{u}$, and $Q(T)$ is the partition function at temperature $T$. For thermalization of the level population at temperatures of up to $\sim 10$~K (e.g., for radiative equilibrium with the cosmic microwave background, $T_\mathrm{cmb} = 2.728$~K), the factor $\left[1-\exp\left(-\frac{E_\mathrm{u}}{k_\mathrm{b}T}\right)\right]/Q(T) \simeq 0.99$, and the total column density per unit velocity can be directly determined from the measured opacity. For the HF $J=1-0$ transition, the underlying parameters are $\nu=1232.47627$~GHz, $A_\mathrm{E}=0.024234$~s$^{-1}$, $E_\mathrm{u}=59.14919$~K, $g_\mathrm{u}=2J_{\rm u}+1=3$, and $Q(9.375$~K$) = 1.006$ \citep[][further references therein]{pickett98}. The resulting column density profiles are shown in Fig.~\ref{fig:results_nh2a} for sight lines to \object{G\,10.62$-$0.39} (W\,31C) and \object{G\,34.3$+$0.1}, in Fig.~\ref{fig:results_nh2b} for sight lines to \object{W\,49N} and \object{W\,51e4}, and in Fig.~\ref{fig:results_nh2c} for the sight line to \object{Sgr\,B2}.

Associated opacity and spin temperature profiles are presented in Appendix~\ref{sec:prismassightlines}, Figs.~\ref{fig:results_prismas_1n2} to \ref{fig:results_prismas_5}. The evaluation of confidence levels follows the method described in Sect.~\ref{sec:absorptionspectra}. Again, large errors are apparent where the sight line absorption turns to saturation either in \ion{H}{i} or in HF, or both. This is systematically the case toward the illuminating background sources, where our assumption of a complete ground-state occupation breaks down, and where the chemistry is very different from that in diffuse clouds (for the chemistry of light hydrides in hot cores see, e.g., \citealp{bruderer10}). Another reason for the large errors is that we derive $f_{\mathrm{H}_2}$ from ratios of column density velocity profiles, not from quantities averaged across a given velocity interval (allowing for a cutoff at velocities where $f_{\mathrm{H}_2}$ is fraught with large uncertainties), thought to represent a spiral arm crossing, for instance. Both approaches have their conveniences and caveats, which are briefly discussed in \citet{wiesemeyer16} (their Appendix B). Table~\ref{tab:fh2} summarizes the molecular hydrogen fraction toward the five investigated sight lines. The reported velocity intervals correspond to peaks in the column density distributions of \ion{H}{i} and/or HF. The intervals are taken large enough so that the uncertainties in $f_{\mathrm{H}_2}$ are not dominated by velocity components with large errors, and at the same time small enough to avoid a blend of spiral arm crossings.

\begin{table}[!t]
\caption{Molecular hydrogen fraction $f_{\mathrm{H}_2}$ in selected velocity intervals on the sight lines toward four PRISMAS sources and \object{Sgr\,B2}.}
\label{tab:fh2}
\centering
\begin{tabular}{llllll}
\hline\hline
\rule{0ex}{3ex}Source & Location & $\varv_\mathrm{lsr}$ & $f_{\mathrm{H}_2}$\\
                      &          & $\mathrm{km\,s}^{-1}$ &    \\
\hline
\rule{0ex}{2.5ex}Sgr\,B2\,(M)    & GC (3~kpc arm)       & $(-50,-13)$      & $0.78_{-0.15}^{+0.21}$ \\
\rule{0ex}{2.5ex}                & GC                   & $(-9,+8)$      & $0.21_{-0.08}^{+0.66}$ \\
\rule{0ex}{2.5ex}                & Sgr                  & $(+12,+22)$      & $0.54_{-0.14}^{+0.39}$ \\
\rule{0ex}{2.5ex}                & Scutum/Crux          & $(+25,+39)$      & $0.63_{-0.21}^{+0.10}$ \\
\rule{0ex}{2.5ex}Sgr\,B2\,(N)    & GC (3~kpc arm)       & $(-50,-13)$      & $0.94_{-0.25}^{+0.05}$ \\
\rule{0ex}{2.5ex}                & GC                   & $(-9,+8)$      & $0.25_{-0.13}^{+0.62}$ \\
\rule{0ex}{2.5ex}                & Sgr                  & $(+12,+22)$      & $0.59_{-0.13}^{+0.33}$ \\
\rule{0ex}{2.5ex}                & Scutum/Crux          & $(+25,+39)$      & $0.69_{-0.14}^{+0.10}$ \\
\rule{0ex}{2.5ex}G\,10.62$-$0.39 & local gas            & $(-3,+5)$      & $0.14_{-0.06}^{+0.10}$ \\
\rule{0ex}{2.5ex}                &                      & $(+5,+14)$     & $0.24_{-0.13}^{+0.12}$ \\
\rule{0ex}{2.5ex}G\,34.3$+$0.1   & Sgr                  & $(+7,+16)$     & $0.81_{-0.31}^{+0.17}$ \\
\rule{0ex}{2.5ex}                &                      & $(+16,+22)$    & $0.16_{-0.16}^{+0.25}$ \\
\rule{0ex}{2.5ex}                &                      & $(+22,+34)$    & $0.11_{-0.06}^{+0.53}$ \\
\rule{0ex}{2.5ex}W\,49N          & local gas            & $(-3,+5)$      & $0.32_{-0.14}^{+0.14}$ \\
\rule{0ex}{2.5ex}                &                      & $(+5,+20)$     & $0.78_{-0.57}^{+0.16}$ \\
\rule{0ex}{2.5ex}                & interarm             & $(+20,+30)$    & $0.31_{-0.22}^{+0.16}$ \\
\rule{0ex}{2.5ex}                & Sgr$^{\mathrm{(a)}}$ & $(+30,+45)$    & $0.76_{-0.30}^{+0.20}$ \\
\rule{0ex}{2.5ex}                & Sgr$^{\mathrm{(b)}}$ & $(+45,+70)$    & $0.54_{-0.14}^{+0.42}$ \\
\rule{0ex}{2.5ex}W\,51e4         & local gas            & $(+3,+10)$     & $0.59_{-0.12}^{+0.10}$ \\
\rule{0ex}{2.5ex}                &                      & $(+10,+15)$    & $0.27_{-0.22}^{+0.17}$ \\
\rule{0ex}{2.5ex}                &                      & $(+15,+30)$    & $0.10_{-0.10}^{+0.26}$ \\
\rule{0ex}{2.5ex}                &                      & $(+30,+45)$    & $0.13_{-0.08}^{+0.31}$ \\[0.5ex]
\hline
\end{tabular}
\tablefoot{(a) Near-side crossing of the Sagittarius spiral arm, (b) far-side crossing.}
\end{table}

\subsection{Discussion}
Despite the relatively large errors in $f_{\mathrm{H}_2}$, it is possible to identify some spiral arm crossings. In the following we briefly discuss their fingerprints, which are detected in the velocity profiles of the molecular hydrogen fraction. This overview is based on the kinematic model of the Galaxy of \citet[][for further references see Table~\ref{tab:vla_sources}]{reid14}. The velocity-averaged molecular hydrogen densities mentioned in the following are listed in Table~\ref{tab:fh2}. We recall that even at a given velocity (i.e., spiral arm crossing) the derived $f_{\mathrm{H}_2}$ represents an average of individual clouds with varying physical and chemical properties.

\subsubsection{Sgr\,B2}
The sight line toward \object{Sgr\,B2} (M) and (N) crosses the Sagittarius spiral arm, the Scutum/Crux arm and the 3~kpc expanding ring before it enters the Galactic center. While the absorption off the HF ground state is saturated in the environment of \object{Sgr\,B2}, near $\varv_\mathrm{lsr} \sim 62\mathrm{km\,s}^{-1}$, we obtained the $f_{\mathrm{H}_2}$ profile for the intervening spiral arms and the 3~kpc ring. The result shows that the diffuse gas is mainly molecular, which does not preclude entities of diffuse, mainly atomic gas, such as clouds or envelopes of cloud cores. Our results confirm another analysis of this sightline by \citet{menten11}, who studied the absorption observed in $^{13}\mathrm{CH}^+$ and $\mathrm{SH}^+$. Moreover, in the inner Galaxy the molecular fraction in the interarm gas contrasts only moderately with that in the spiral arms \citep[$\sim$20\%][]{koda16}. While the detailed analysis of this sight line will be postponed to future work, here we comment on the slight differences in the molecular hydrogen fraction $f_{\mathrm{H}_2}$ toward \object{Sgr\,B2\,M} and \object{Sgr\,B2\,N} that appear in Fig.~\ref{fig:results_nh2c}. The two components are separated by an apparent distance of $45\arcsec$, which at a distance of 7.8~kpc \citep{reid14} corresponds to a length scale of 1.7~pc. The varying molecular hydrogen fractions are therefore due to diffuse or translucent cloud entities of at most such a size scale. This upper limit holds for the local environment of Sgr\,B2. For the near side of the expanding 3~kpc arm, the upper limit to the length scale of the absorbing cloud entities would drop to 1~pc and correspondingly less for the Scutum/Crux spiral arm and the near side of the Sagittarius spiral arm.

\subsubsection{G\,10.62$-$0.39}
The background source is located in the 3 kpc arm, at $4.95\pm 0.47~\mathrm{kpc}$ distance \citep{sanna14}. The crossings of the Scutum and Sagittarius arms are clearly visible at around $40$ and $0~\mathrm{km\,s}^{-1}$, respectively, while the velocity of \object{G\,10.62$-$0.39} is identical to that of the Sagittarius arm crossing (i.e., coincides in velocity with local gas). A further interpretation of the corresponding velocity interval is therefore impossible. At the spiral arm crossings, the molecular hydrogen fractions $f_{\mathrm{H}_2}$ are close to unity. \cite{liszt16} estimate a typical clump size of 5.5~pc. At the distance of the Scutum arm (4~kpc), this corresponds to an angular extent of $4\farcm7$, which is well above the expected far-infrared size of \object{G\,10.62$-$0.39}.

\subsubsection{G\,34.3$+$0.1}
With $1.56\pm 0.12~\mathrm{kpc}$ \citep{kurayama11}, this is a relatively short sight line, grazing the Sagittarius arm almost tangentially. The HF column densities show a stronger arm-to-interarm contrast than those seen in \ion{H}{i}. Although it is short, this sight line contains clouds of markedly different $f_{\mathrm{H}_2}$, which may have several reasons: The variations may be of truly intrinsic origin, since on a short sight line we have a better chance to separate individual foreground clouds with different $f_{\mathrm{H}_2}$, whereas on longer distances the chances for blends are higher, simply owing to sight-line crowding. The other reason could be the misinterpretation of the data. As we have stated before, \ion{H}{i} emission from the rear part of the sight line may lead to a significant contamination of the derived column density profile. This caveat holds for all derived $f_{\mathrm{H}_2}$ profiles, not only for this one.

\subsubsection{W\,49N}
This $11.11\pm 0.86~\mathrm{kpc}$ long sight line \citep{zhang13} contains three groups of diffuse clouds, both in \ion{H}{i} and in HF. Their velocity intervals are $0$ to $20~\mathrm{km\,s}^{-1}$, $30$ to $45~\mathrm{km\,s}^{-1}$, and $45$ to $70~\mathrm{km\,s}^{-1}$, corresponding to the Perseus arm (where \object{W\,49N} is located), and the near- and far-side crossings of the Sagittarius arm, respectively \citep[cf.][]{vallee08}. In the recent literature, the far-side crossing of the Sagittarius arm is instead assigned to a crossing of the molecular ring, as modeled by \citet{dobbs12a}. Different from \citet{vallee08}, the authors assume that the Scutum-Centaurus arm starts immediately at the bar, not at the Sagittarius–Carina arm. \citet{li16} modeled the molecular ring as an ensemble formed by two bar-driven spiral arms and the Scutum arm. In the framework of either model it seems fair to say that the velocities around $50~\mathrm{km\,s}^{-1}$ are also assigned to interarm gas. By consequence, the corresponding molecular hydrogen fractions are significantly different: typically 20\% in the interarm gas, and above 50\% in the spiral arms. Such an arm-to-interarm contrast compares to that found in the outer Galaxy \citep{koda16}. The sightline to \object{W\,49}, located at a galactocentric distance of 7.6~kpc (Table~\ref{tab:vla_sources}), crosses a larger fraction of atomic gas than the sightlines toward targets in the inner Galaxy.

\subsubsection{W\,51e4}
The sight line to \object{W\,51e4}, which is at $5.41\pm 0.21~\mathrm{kpc}$ distance \citep{sato10}, follows the Sagittarius spiral arm. As in the case of \object{W\,49N}, the contrast between the interarm region and the near-side crossing of the Sagittarius spiral arm is higher in HF than in \ion{H}{i}.

\section{Summary}\label{sec:summary}
Based on JVLA 21~cm absorption profiles combined with \ion{H}{i} emission line data of the interferometric galactic plane surveys, we were able to infer the optical depth, spin temperatures, and \ion{H}{i} column densities toward eight continuum-bright galactic targets, which allowed us to study the total hydrogen content of absorbing clouds in the foreground.

We employed Bayesian inference using the \textit{pymc} Python module to estimate the error distributions of the resulting profiles, which are highly asymmetric. Despite a careful study of the statistical uncertainties, there are several systematic effects that we were unable to account for with the observational data and the tools available to us. Especially the blending of \ion{H}{i} emission originating from behind the continuum source into the velocity range of interest can potentially cause a strong bias to the inferred spin temperatures and \ion{H}{i} column densities.

Using archival Herschel data from the PRISMAS and HEXOS key projects \citep[][respectively]{gerin12, bergin10}, we have determined the molecular hydrogen fraction $f_{\mathrm{H}_2}$ in the spiral arm crossings located on the sight lines toward the background sources \object{Sgr\,B2}, \object{G\,10.62$-$0.39} (W\,31C), \object{G\,34.3$+$0.1}, \object{W\,49N,} and \object{W\,51e4}. Despite the large errors in the $dN/d\varv$ profile, the data strongly suggest that the majority of the diffuse clouds crossed in the spiral arms is predominantly molecular and not mainly atomic. By consequence, the contrast between the column densities in the spiral arms and interarm regions is significantly higher in \ion{H}{i} than in HF. Toward \object{W\,49N} a similar contrast was found between CO and \ion{H}{i} \citep{heyer98} and between OH$^+$ and OH \citep{wiesemeyer16}. With the results derived here, we expect to provide a good basis for a broad range of investigations studying the chemical and physical conditions along prominent sight lines through the Milky Way disk.

\begin{acknowledgements}
We are grateful to Arnaud Belloche for careful proof-reading of the manuscript. We also thank the anonymous referee for a very constructive report and Lars Fl\"{o}er for useful hints regarding \textit{pymc}.

The research presented in this paper has used data from the Canadian Galactic Plane Survey, a Canadian project with international partners, supported by the Natural Sciences and Engineering Research Council. The National Radio Astronomy Observatory is a facility of the National Science Foundation operated under cooperative agreement by Associated Universities, Inc.

HI4PI is based on observations with the 100 m telescope of the MPIfR (Max-Planck-Institut für Radioastronomie) at Effelsberg and the Parkes Radio Telescope, which is part of the Australia Telescope, funded by the Commonwealth of Australia for operation as a National Facility managed by CSIRO.

MG would like to thank CNES for financial support.

We would like to express our gratitude to the developers of the many C/C++ and Python libraries, made available as open-source software, that we have used: most importantly, NumPy \citep{NumPy}, SciPy \citep{SciPy}, and Astropy \citep{Astropy}. The Bayesian inferences were derived using the pymc3 package \citep{pymc}. Figures have been prepared using matplotlib \citep{Matplotlib} and in part using the Kapteyn Package \citep{KapteynPackage}.
\end{acknowledgements}

\bibliographystyle{aa} % style aa.bst
\bibliography{references} % your references Yourfile.bib

\begin{thebibliography}{114}
\expandafter\ifx\csname natexlab\endcsname\relax\def\natexlab#1{#1}\fi

\bibitem[{{Astropy Collaboration} {et~al.}(2013){Astropy Collaboration},
  {Robitaille}, {Tollerud}, {Greenfield}, {Droettboom}, {Bray}, {Aldcroft},
  {Davis}, {Ginsburg}, {Price-Whelan}, {Kerzendorf}, {Conley}, {Crighton},
  {Barbary}, {Muna}, {Ferguson}, {Grollier}, {Parikh}, {Nair}, {Unther},
  {Deil}, {Woillez}, {Conseil}, {Kramer}, {Turner}, {Singer}, {Fox}, {Weaver},
  {Zabalza}, {Edwards}, {Azalee Bostroem}, {Burke}, {Casey}, {Crawford},
  {Dencheva}, {Ely}, {Jenness}, {Labrie}, {Lim}, {Pierfederici}, {Pontzen},
  {Ptak}, {Refsdal}, {Servillat}, \& {Streicher}}]{Astropy}
{Astropy Collaboration}, {Robitaille}, T.~P., {Tollerud}, E.~J., {et~al.} 2013,
  \aap, 558, A33

\bibitem[{{Belloche} {et~al.}(2013){Belloche}, {M{\"u}ller}, {Menten},
  {Schilke}, \& {Comito}}]{belloche13}
{Belloche}, A., {M{\"u}ller}, H.~S.~P., {Menten}, K.~M., {Schilke}, P., \&
  {Comito}, C. 2013, \aap, 559, A47

\bibitem[{{Bergin} {et~al.}(2010){Bergin}, {Phillips}, {Comito}, {Crockett},
  {Lis}, {Schilke}, {Wang}, {Bell}, {Blake}, {Bumble}, {Caux}, {Cabrit},
  {Ceccarelli}, {Cernicharo}, {Daniel}, {de Graauw}, {Dubernet},
  {Emprechtinger}, {Encrenaz}, {Falgarone}, {Gerin}, {Giesen}, {Goicoechea},
  {Goldsmith}, {Gupta}, {Hartogh}, {Helmich}, {Herbst}, {Joblin}, {Johnstone},
  {Kawamura}, {Langer}, {Latter}, {Lord}, {Maret}, {Martin}, {Melnick},
  {Menten}, {Morris}, {M{\"u}ller}, {Murphy}, {Neufeld}, {Ossenkopf}, {Pagani},
  {Pearson}, {P{\'e}rault}, {Plume}, {Roelfsema}, {Qin}, {Salez}, {Schlemmer},
  {Stutzki}, {Tielens}, {Trappe}, {van der Tak}, {Vastel}, {Yorke}, {Yu}, \&
  {Zmuidzinas}}]{bergin10}
{Bergin}, E.~A., {Phillips}, T.~G., {Comito}, C., {et~al.} 2010, \aap, 521, L20

\bibitem[{{Beuther} {et~al.}(2016){Beuther}, {Bihr}, {Rugel}, {Johnston},
  {Wang}, {Walter}, {Brunthaler}, {Walsh}, {Ott}, {Stil}, {Henning},
  {Schierhuber}, {Kainulainen}, {Heyer}, {Goldsmith}, {Anderson}, {Longmore},
  {Klessem}, {Glover}, {Urquhart}, {Plume}, {Ragan}, {Schneider},
  {McClure-Griffiths}, {Menten}, {Smith}, {Roy}, {Shanahan}, {Nguyen-Luong}, \&
  {Bigiel}}]{beuther16}
{Beuther}, H., {Bihr}, S., {Rugel}, M., {et~al.} 2016, ArXiv e-prints

\bibitem[{{Bihr} {et~al.}(2015){Bihr}, {Beuther}, {Ott}, {Johnston},
  {Brunthaler}, {Anderson}, {Bigiel}, {Carlhoff}, {Churchwell}, {Glover},
  {Goldsmith}, {Heitsch}, {Henning}, {Heyer}, {Hill}, {Hughes}, {Klessen},
  {Linz}, {Longmore}, {McClure-Griffiths}, {Menten}, {Motte}, {Nguyen-Luong},
  {Plume}, {Ragan}, {Roy}, {Schilke}, {Schneider}, {Smith}, {Stil}, {Urquhart},
  {Walsh}, \& {Walter}}]{bihr15}
{Bihr}, S., {Beuther}, H., {Ott}, J., {et~al.} 2015, \aap, 580, A112

\bibitem[{{Braun} {et~al.}(2009){Braun}, {Thilker}, {Walterbos}, \&
  {Corbelli}}]{braun09}
{Braun}, R., {Thilker}, D.~A., {Walterbos}, R.~A.~M., \& {Corbelli}, E. 2009,
  \apj, 695, 937

\bibitem[{{Brogan} \& {Troland}(2001)}]{brogan01}
{Brogan}, C.~L. \& {Troland}, T.~H. 2001, \apj, 550, 799

\bibitem[{{Brogan} {et~al.}(2005){Brogan}, {Zauderer}, {Lazio}, {Goss},
  {DePree}, \& {Faison}}]{brogan05}
{Brogan}, C.~L., {Zauderer}, B.~A., {Lazio}, T.~J., {et~al.} 2005, \aj, 130,
  698

\bibitem[{{Bronfman} {et~al.}(1996){Bronfman}, {Nyman}, \& {May}}]{bronfman96}
{Bronfman}, L., {Nyman}, L.-A., \& {May}, J. 1996, \aaps, 115, 81

\bibitem[{{Bruderer} {et~al.}(2010){Bruderer}, {Benz}, {St{\"a}uber}, \&
  {Doty}}]{bruderer10}
{Bruderer}, S., {Benz}, A.~O., {St{\"a}uber}, P., \& {Doty}, S.~D. 2010, \apj,
  720, 1432

\bibitem[{{Cannon}(2012)}]{cannon12}
{Cannon}, C.~J. 2012, {The Transfer of Spectral Line Radiation} (Cambridge
  University Press)

\bibitem[{{Carruthers}(1970)}]{carruthers70}
{Carruthers}, G.~R. 1970, \apjl, 161, L81

\bibitem[{{Chengalur} {et~al.}(2013){Chengalur}, {Kanekar}, \&
  {Roy}}]{chengalur13}
{Chengalur}, J.~N., {Kanekar}, N., \& {Roy}, N. 2013, \mnras, 432, 3074

\bibitem[{{de Graauw} {et~al.}(2010){de Graauw}, {Helmich}, {Phillips},
  {Stutzki}, {Caux}, {Whyborn}, {Dieleman}, {Roelfsema}, {Aarts}, {Assendorp},
  {Bachiller}, {Baechtold}, {Barcia}, {Beintema}, {Belitsky}, {Benz}, {Bieber},
  {Boogert}, {Borys}, {Bumble}, {Ca{\"i}s}, {Caris}, {Cerulli-Irelli},
  {Chattopadhyay}, {Cherednichenko}, {Ciechanowicz}, {Coeur-Joly}, {Comito},
  {Cros}, {de Jonge}, {de Lange}, {Delforges}, {Delorme}, {den Boggende},
  {Desbat}, {Diez-Gonz{\'a}lez}, {di Giorgio}, {Dubbeldam}, {Edwards},
  {Eggens}, {Erickson}, {Evers}, {Fich}, {Finn}, {Franke}, {Gaier}, {Gal},
  {Gao}, {Gallego}, {Gauffre}, {Gill}, {Glenz}, {Golstein}, {Goulooze},
  {Gunsing}, {G{\"u}sten}, {Hartogh}, {Hatch}, {Higgins}, {Honingh}, {Huisman},
  {Jackson}, {Jacobs}, {Jacobs}, {Jarchow}, {Javadi}, {Jellema}, {Justen},
  {Karpov}, {Kasemann}, {Kawamura}, {Keizer}, {Kester}, {Klapwijk}, {Klein},
  {Kollberg}, {Kooi}, {Kooiman}, {Kopf}, {Krause}, {Krieg}, {Kramer},
  {Kruizenga}, {Kuhn}, {Laauwen}, {Lai}, {Larsson}, {Leduc}, {Leinz}, {Lin},
  {Liseau}, {Liu}, {Loose}, {L{\'o}pez-Fernandez}, {Lord}, {Luinge}, {Marston},
  {Mart{\'{\i}}n-Pintado}, {Maestrini}, {Maiwald}, {McCoey}, {Mehdi}, {Megej},
  {Melchior}, {Meinsma}, {Merkel}, {Michalska}, {Monstein}, {Moratschke},
  {Morris}, {Muller}, {Murphy}, {Naber}, {Natale}, {Nowosielski}, {Nuzzolo},
  {Olberg}, {Olbrich}, {Orfei}, {Orleanski}, {Ossenkopf}, {Peacock}, {Pearson},
  {Peron}, {Phillip-May}, {Piazzo}, {Planesas}, {Rataj}, {Ravera}, {Risacher},
  {Salez}, {Samoska}, {Saraceno}, {Schieder}, {Schlecht}, {Schl{\"o}der},
  {Schm{\"u}lling}, {Schultz}, {Schuster}, {Siebertz}, {Smit}, {Szczerba},
  {Shipman}, {Steinmetz}, {Stern}, {Stokroos}, {Teipen}, {Teyssier}, {Tils},
  {Trappe}, {van Baaren}, {van Leeuwen}, {van de Stadt}, {Visser}, {Wildeman},
  {Wafelbakker}, {Ward}, {Wesselius}, {Wild}, {Wulff}, {Wunsch}, {Tielens},
  {Zaal}, {Zirath}, {Zmuidzinas}, \& {Zwart}}]{degraauw10}
{de Graauw}, T., {Helmich}, F.~P., {Phillips}, T.~G., {et~al.} 2010, \aap, 518,
  L6

\bibitem[{{Dickey} {et~al.}(1983){Dickey}, {Kulkarni}, {van Gorkom}, \&
  {Heiles}}]{dickey83}
{Dickey}, J.~M., {Kulkarni}, S.~R., {van Gorkom}, J.~H., \& {Heiles}, C.~E.
  1983, \apjs, 53, 591

\bibitem[{{Dickey} {et~al.}(2003){Dickey}, {McClure-Griffiths}, {Gaensler}, \&
  {Green}}]{dickey03}
{Dickey}, J.~M., {McClure-Griffiths}, N.~M., {Gaensler}, B.~M., \& {Green},
  A.~J. 2003, \apj, 585, 801

\bibitem[{{Dickey} {et~al.}(2000){Dickey}, {Mebold}, {Stanimirovic}, \&
  {Staveley-Smith}}]{dickey00}
{Dickey}, J.~M., {Mebold}, U., {Stanimirovic}, S., \& {Staveley-Smith}, L.
  2000, \apj, 536, 756

\bibitem[{{Dickey} {et~al.}(2009){Dickey}, {Strasser}, {Gaensler}, {Haverkorn},
  {Kavars}, {McClure-Griffiths}, {Stil}, \& {Taylor}}]{dickey09}
{Dickey}, J.~M., {Strasser}, S., {Gaensler}, B.~M., {et~al.} 2009, \apj, 693,
  1250

\bibitem[{{Dobbs} \& {Burkert}(2012)}]{dobbs12a}
{Dobbs}, C.~L. \& {Burkert}, A. 2012, \mnras, 421, 2940

\bibitem[{{Dobbs} {et~al.}(2012){Dobbs}, {Pringle}, \& {Burkert}}]{dobbs12b}
{Dobbs}, C.~L., {Pringle}, J.~E., \& {Burkert}, A. 2012, \mnras, 425, 2157

\bibitem[{{Faison} {et~al.}(1998){Faison}, {Goss}, {Diamond}, \&
  {Taylor}}]{faison98}
{Faison}, M.~D., {Goss}, W.~M., {Diamond}, P.~J., \& {Taylor}, G.~B. 1998, \aj,
  116, 2916

\bibitem[{{Falgarone} {et~al.}(2010){Falgarone}, {Godard}, {Cernicharo}, {de
  Luca}, {Gerin}, {Phillips}, {Black}, {Lis}, {Bell}, {Boulanger}, {Coutens},
  {Dartois}, {Encrenaz}, {Giesen}, {Goicoechea}, {Goldsmith}, {Gupta}, {Gry},
  {Hennebelle}, {Herbst}, {Hily-Blant}, {Joblin}, {Ka{\'z}mierczak},
  {Ko{\l}os}, {Kre{\l}owski}, {Martin-Pintado}, {Monje}, {Mookerjea},
  {Neufeld}, {Perault}, {Pearson}, {Persson}, {Plume}, {Salez}, {Schmidt},
  {Sonnentrucker}, {Stutzki}, {Teyssier}, {Vastel}, {Yu}, {Menten}, {Geballe},
  {Schlemmer}, {Shipman}, {Tielens}, {Philipp}, {Cros}, {Zmuidzinas},
  {Samoska}, {Klein}, {Lorenzani}, {Szczerba}, {P{\'e}ron}, {Cais}, {Gaufre},
  {Cros}, {Ravera}, {Morris}, {Lord}, \& {Planesas}}]{falgarone10}
{Falgarone}, E., {Godard}, B., {Cernicharo}, J., {et~al.} 2010, \aap, 521, L15

\bibitem[{{Fish} {et~al.}(2003){Fish}, {Reid}, {Wilner}, \&
  {Churchwell}}]{fish03}
{Fish}, V.~L., {Reid}, M.~J., {Wilner}, D.~J., \& {Churchwell}, E. 2003, \apj,
  587, 701

\bibitem[{{Flagey} {et~al.}(2013){Flagey}, {Goldsmith}, {Lis}, {Gerin},
  {Neufeld}, {Sonnentrucker}, {De Luca}, {Godard}, {Goicoechea}, {Monje}, \&
  {Phillips}}]{flagey13}
{Flagey}, N., {Goldsmith}, P.~F., {Lis}, D.~C., {et~al.} 2013, \apj, 762, 11

\bibitem[{{Gerin} {et~al.}(2010{\natexlab{a}}){Gerin}, {de Luca}, {Black},
  {Goicoechea}, {Herbst}, {Neufeld}, {Falgarone}, {Godard}, {Pearson}, {Lis},
  {Phillips}, {Bell}, {Sonnentrucker}, {Boulanger}, {Cernicharo}, {Coutens},
  {Dartois}, {Encrenaz}, {Giesen}, {Goldsmith}, {Gupta}, {Gry}, {Hennebelle},
  {Hily-Blant}, {Joblin}, {Kazmierczak}, {Kolos}, {Krelowski},
  {Martin-Pintado}, {Monje}, {Mookerjea}, {Perault}, {Persson}, {Plume},
  {Rimmer}, {Salez}, {Schmidt}, {Stutzki}, {Teyssier}, {Vastel}, {Yu},
  {Contursi}, {Menten}, {Geballe}, {Schlemmer}, {Shipman}, {Tielens},
  {Philipp-May}, {Cros}, {Zmuidzinas}, {Samoska}, {Klein}, \&
  {Lorenzani}}]{gerin10a}
{Gerin}, M., {de Luca}, M., {Black}, J., {et~al.} 2010{\natexlab{a}}, \aap,
  518, L110

\bibitem[{{Gerin} {et~al.}(2010{\natexlab{b}}){Gerin}, {de Luca}, {Goicoechea},
  {Herbst}, {Falgarone}, {Godard}, {Bell}, {Coutens}, {Ka{\'z}mierczak},
  {Sonnentrucker}, {Black}, {Neufeld}, {Phillips}, {Pearson}, {Rimmer},
  {Hassel}, {Lis}, {Vastel}, {Boulanger}, {Cernicharo}, {Dartois}, {Encrenaz},
  {Giesen}, {Goldsmith}, {Gupta}, {Gry}, {Hennebelle}, {Hily-Blant}, {Joblin},
  {Ko{\l}os}, {Kre{\l}owski}, {Mart{\'{\i}}n-Pintado}, {Monje}, {Mookerjea},
  {Perault}, {Persson}, {Plume}, {Salez}, {Schmidt}, {Stutzki}, {Teyssier},
  {Yu}, {Contursi}, {Menten}, {Geballe}, {Schlemmer}, {Morris}, {Hatch},
  {Imram}, {Ward}, {Caux}, {G{\"u}sten}, {Klein}, {Roelfsema}, {Dieleman},
  {Schieder}, {Honingh}, \& {Zmuidzinas}}]{gerin10b}
{Gerin}, M., {de Luca}, M., {Goicoechea}, J.~R., {et~al.} 2010{\natexlab{b}},
  \aap, 521, L16

\bibitem[{{Gerin} {et~al.}(2012){Gerin}, {Levrier}, {Falgarone}, {Godard},
  {Hennebelle}, {Le Petit}, {De Luca}, {Neufeld}, {Sonnentrucker}, {Goldsmith},
  {Flagey}, {Lis}, {Persson}, {Black}, {Goicoechea}, \& {Menten}}]{gerin12}
{Gerin}, M., {Levrier}, F., {Falgarone}, E., {et~al.} 2012, Philosophical
  Transactions of the Royal Society of London Series A, 370, 5174

\bibitem[{{Gerin} {et~al.}(2016){Gerin}, {Neufeld}, \& {Goicoechea}}]{gerin16}
{Gerin}, M., {Neufeld}, D.~A., \& {Goicoechea}, J.~R. 2016, ArXiv e-prints

\bibitem[{{Gerin} {et~al.}(2015){Gerin}, {Ruaud}, {Goicoechea}, {Gusdorf},
  {Godard}, {de Luca}, {Falgarone}, {Goldsmith}, {Lis}, {Menten}, {Neufeld},
  {Phillips}, \& {Liszt}}]{gerin15}
{Gerin}, M., {Ruaud}, M., {Goicoechea}, J.~R., {et~al.} 2015, \aap, 573, A30

\bibitem[{{Gibson} {et~al.}(2005{\natexlab{a}}){Gibson}, {Taylor}, {Higgs},
  {Brunt}, \& {Dewdney}}]{gibson05a}
{Gibson}, S.~J., {Taylor}, A.~R., {Higgs}, L.~A., {Brunt}, C.~M., \& {Dewdney},
  P.~E. 2005{\natexlab{a}}, \apj, 626, 195

\bibitem[{{Gibson} {et~al.}(2005{\natexlab{b}}){Gibson}, {Taylor}, {Higgs},
  {Brunt}, \& {Dewdney}}]{gibson05b}
{Gibson}, S.~J., {Taylor}, A.~R., {Higgs}, L.~A., {Brunt}, C.~M., \& {Dewdney},
  P.~E. 2005{\natexlab{b}}, \apj, 626, 214

\bibitem[{{Gibson} {et~al.}(2000){Gibson}, {Taylor}, {Higgs}, \&
  {Dewdney}}]{gibson00}
{Gibson}, S.~J., {Taylor}, A.~R., {Higgs}, L.~A., \& {Dewdney}, P.~E. 2000,
  \apj, 540, 851

\bibitem[{{Godard} {et~al.}(2012){Godard}, {Falgarone}, {Gerin}, {Lis}, {De
  Luca}, {Black}, {Goicoechea}, {Cernicharo}, {Neufeld}, {Menten}, \&
  {Emprechtinger}}]{godard12}
{Godard}, B., {Falgarone}, E., {Gerin}, M., {et~al.} 2012, \aap, 540, A87

\bibitem[{{Heiles}(1997)}]{heiles97}
{Heiles}, C. 1997, \apj, 481, 193

\bibitem[{{Heiles} \& {Troland}(2003{\natexlab{a}})}]{heiles03a}
{Heiles}, C. \& {Troland}, T.~H. 2003{\natexlab{a}}, \apjs, 145, 329

\bibitem[{{Heiles} \& {Troland}(2003{\natexlab{b}})}]{heiles03b}
{Heiles}, C. \& {Troland}, T.~H. 2003{\natexlab{b}}, \apj, 586, 1067

\bibitem[{{Heyer} \& {Terebey}(1998)}]{heyer98}
{Heyer}, M.~H. \& {Terebey}, S. 1998, \apj, 502, 265

\bibitem[{{Heyminck} {et~al.}(2012){Heyminck}, {Graf}, {G{\"u}sten}, {Stutzki},
  {H{\"u}bers}, \& {Hartogh}}]{heyminck12}
{Heyminck}, S., {Graf}, U.~U., {G{\"u}sten}, R., {et~al.} 2012, \aap, 542, L1

\bibitem[{{HI4PI Collaboration} {et~al.}(2016){HI4PI Collaboration}, {Ben
  Bekhti}, {Fl{\"o}er}, {Keller}, {Kerp}, {Lenz}, {Winkel}, {Bailin},
  {Calabretta}, {Dedes}, {Ford}, {Gibson}, {Haud}, {Janowiecki}, {Kalberla},
  {Lockman}, {McClure-Griffiths}, {Murphy}, {Nakanishi}, {Pisano}, \&
  {Staveley-Smith}}]{hi4pi16}
{HI4PI Collaboration}, {Ben Bekhti}, N., {Fl{\"o}er}, L., {et~al.} 2016, \aap,
  594, A116

\bibitem[{Hunter(2007)}]{Matplotlib}
Hunter, J. 2007, Computing in Science Engineering, 9, 90

\bibitem[{{Immer} {et~al.}(2013){Immer}, {Reid}, {Menten}, {Brunthaler}, \&
  {Dame}}]{immer13}
{Immer}, K., {Reid}, M.~J., {Menten}, K.~M., {Brunthaler}, A., \& {Dame}, T.~M.
  2013, \aap, 553, A117

\bibitem[{{Indriolo} {et~al.}(2015){Indriolo}, {Neufeld}, {Gerin}, {Schilke},
  {Benz}, {Winkel}, {Menten}, {Chambers}, {Black}, {Bruderer}, {Falgarone},
  {Godard}, {Goicoechea}, {Gupta}, {Lis}, {Ossenkopf}, {Persson},
  {Sonnentrucker}, {van der Tak}, {van Dishoeck}, {Wolfire}, \&
  {Wyrowski}}]{indriolo15}
{Indriolo}, N., {Neufeld}, D.~A., {Gerin}, M., {et~al.} 2015, \apj, 800, 40

\bibitem[{{Indriolo} {et~al.}(2013){Indriolo}, {Neufeld}, {Seifahrt}, \&
  {Richter}}]{indriolo13}
{Indriolo}, N., {Neufeld}, D.~A., {Seifahrt}, A., \& {Richter}, M.~J. 2013,
  \apj, 764, 188

\bibitem[{Jones {et~al.}(2001)Jones, Oliphant, Peterson, {et~al.}}]{SciPy}
Jones, E., Oliphant, T., Peterson, P., {et~al.} 2001, {SciPy}: Open source
  scientific tools for {Python}, [Online; accessed 2015-06-27]

\bibitem[{{Koda} {et~al.}(2016){Koda}, {Scoville}, \& {Heyer}}]{koda16}
{Koda}, J., {Scoville}, N., \& {Heyer}, M. 2016, \apj, 823, 76

\bibitem[{{Koo}(1997)}]{koo97}
{Koo}, B.-C. 1997, \apjs, 108, 489

\bibitem[{{Kurayama} {et~al.}(2011){Kurayama}, {Nakagawa}, {Sawada-Satoh},
  {Sato}, {Honma}, {Sunada}, {Hirota}, \& {Imai}}]{kurayama11}
{Kurayama}, T., {Nakagawa}, A., {Sawada-Satoh}, S., {et~al.} 2011, \pasj, 63,
  513

\bibitem[{{Lang} {et~al.}(2010){Lang}, {Goss}, {Cyganowski}, \&
  {Clubb}}]{lang10}
{Lang}, C.~C., {Goss}, W.~M., {Cyganowski}, C., \& {Clubb}, K.~I. 2010, \apjs,
  191, 275

\bibitem[{{Lazareff}(1975)}]{lazareff75}
{Lazareff}, B. 1975, \aap, 42, 25

\bibitem[{{Lazio} {et~al.}(2009){Lazio}, {Brogan}, {Goss}, \&
  {Stanimirovi{\'c}}}]{lazio09}
{Lazio}, T.~J.~W., {Brogan}, C.~L., {Goss}, W.~M., \& {Stanimirovi{\'c}}, S.
  2009, \aj, 137, 4526

\bibitem[{{Lee} {et~al.}(2015){Lee}, {Stanimirovi{\'c}}, {Murray}, {Heiles}, \&
  {Miller}}]{lee15}
{Lee}, M.-Y., {Stanimirovi{\'c}}, S., {Murray}, C.~E., {Heiles}, C., \&
  {Miller}, J. 2015, \apj, 809, 56

\bibitem[{{Levine} {et~al.}(2006){Levine}, {Blitz}, \& {Heiles}}]{levine06}
{Levine}, E.~S., {Blitz}, L., \& {Heiles}, C. 2006, Science, 312, 1773

\bibitem[{{Li} {et~al.}(2016){Li}, {Gerhard}, {Shen}, {Portail}, \&
  {Wegg}}]{li16}
{Li}, Z., {Gerhard}, O., {Shen}, J., {Portail}, M., \& {Wegg}, C. 2016, \apj,
  824, 13

\bibitem[{{Lis} {et~al.}(2010){Lis}, {Phillips}, {Goldsmith}, {Neufeld},
  {Herbst}, {Comito}, {Schilke}, {M{\"u}ller}, {Bergin}, {Gerin}, {Bell},
  {Emprechtinger}, {Black}, {Blake}, {Boulanger}, {Caux}, {Ceccarelli},
  {Cernicharo}, {Coutens}, {Crockett}, {Daniel}, {Dartois}, {de Luca},
  {Dubernet}, {Encrenaz}, {Falgarone}, {Geballe}, {Godard}, {Giesen},
  {Goicoechea}, {Gry}, {Gupta}, {Hennebelle}, {Hily-Blant}, {Ko{\l}os},
  {Kre{\l}owski}, {Joblin}, {Johnstone}, {Ka{\'z}mierczak}, {Lord}, {Maret},
  {Martin}, {Mart{\'{\i}}n-Pintado}, {Melnick}, {Menten}, {Monje}, {Mookerjea},
  {Morris}, {Murphy}, {Ossenkopf}, {Pearson}, {P{\'e}rault}, {Persson},
  {Plume}, {Qin}, {Salez}, {Schlemmer}, {Schmidt}, {Sonnentrucker}, {Stutzki},
  {Teyssier}, {Trappe}, {van der Tak}, {Vastel}, {Wang}, {Yorke}, {Yu},
  {Zmuidzinas}, {Boogert}, {Erickson}, {Karpov}, {Kooi}, {Maiwald}, {Schieder},
  \& {Zaal}}]{lis10}
{Lis}, D.~C., {Phillips}, T.~G., {Goldsmith}, P.~F., {et~al.} 2010, \aap, 521,
  L26

\bibitem[{{Liszt} \& {Gerin}(2016)}]{liszt16}
{Liszt}, H.~S. \& {Gerin}, M. 2016, \aap, 585, A80

\bibitem[{{Liszt} {et~al.}(1983){Liszt}, {van der Hulst}, {Burton}, \&
  {Ondrechen}}]{liszt83}
{Liszt}, H.~S., {van der Hulst}, J.~M., {Burton}, W.~B., \& {Ondrechen}, M.~P.
  1983, \aap, 126, 341

\bibitem[{{Martin} {et~al.}(2015){Martin}, {Blagrave}, {Pinheiro Goncalves},
  {Lockman}, {Boothroyd}, {Miville-Deschenes}, {Joncas}, \&
  {Stephan}}]{martin15}
{Martin}, P.~G., {Blagrave}, K.~P.~M., {Pinheiro Goncalves}, D., {et~al.} 2015,
  ArXiv e-prints

\bibitem[{{McClure-Griffiths} {et~al.}(2012){McClure-Griffiths}, {Dickey},
  {Gaensler}, {Green}, {Green}, \& {Haverkorn}}]{mcclure12}
{McClure-Griffiths}, N.~M., {Dickey}, J.~M., {Gaensler}, B.~M., {et~al.} 2012,
  \apjs, 199, 12

\bibitem[{{McClure-Griffiths} {et~al.}(2006){McClure-Griffiths}, {Dickey},
  {Gaensler}, {Green}, \& {Haverkorn}}]{mcclure06}
{McClure-Griffiths}, N.~M., {Dickey}, J.~M., {Gaensler}, B.~M., {Green}, A.~J.,
  \& {Haverkorn}, M. 2006, \apj, 652, 1339

\bibitem[{{McClure-Griffiths} {et~al.}(2005){McClure-Griffiths}, {Dickey},
  {Gaensler}, {Green}, {Haverkorn}, \& {Strasser}}]{mcclure05}
{McClure-Griffiths}, N.~M., {Dickey}, J.~M., {Gaensler}, B.~M., {et~al.} 2005,
  \apjs, 158, 178

\bibitem[{{Mebold} {et~al.}(1997){Mebold}, {D{\"u}sterberg}, {Dickey},
  {Staveley-Smith}, \& {Kalberla}}]{mebold97}
{Mebold}, U., {D{\"u}sterberg}, C., {Dickey}, J.~M., {Staveley-Smith}, L., \&
  {Kalberla}, P. 1997, \apjl, 490, L65

\bibitem[{{Menten} {et~al.}(2011){Menten}, {Wyrowski}, {Belloche},
  {G{\"u}sten}, {Dedes}, \& {M{\"u}ller}}]{menten11}
{Menten}, K.~M., {Wyrowski}, F., {Belloche}, A., {et~al.} 2011, \aap, 525, A77

\bibitem[{{Monje} {et~al.}(2011){Monje}, {Emprechtinger}, {Phillips}, {Lis},
  {Goldsmith}, {Bergin}, {Bell}, {Neufeld}, \& {Sonnentrucker}}]{monje11}
{Monje}, R.~R., {Emprechtinger}, M., {Phillips}, T.~G., {et~al.} 2011, \apjl,
  734, L23

\bibitem[{{Mookerjea} {et~al.}(2012){Mookerjea}, {Hassel}, {Gerin}, {Giesen},
  {Stutzki}, {Herbst}, {Black}, {Goldsmith}, {Menten}, {Kre{\l}owski}, {De
  Luca}, {Csengeri}, {Joblin}, {Ka{\'z}mierczak}, {Schmidt}, {Goicoechea}, \&
  {Cernicharo}}]{mookerjea12}
{Mookerjea}, B., {Hassel}, G.~E., {Gerin}, M., {et~al.} 2012, \aap, 546, A75

\bibitem[{{Neufeld} {et~al.}(2015{\natexlab{a}}){Neufeld}, {Black}, {Gerin},
  {Goicoechea}, {Goldsmith}, {Gry}, {Gupta}, {Herbst}, {Indriolo}, {Lis},
  {Menten}, {Monje}, {Mookerjea}, {Persson}, {Schilke}, {Sonnentrucker}, \&
  {Wolfire}}]{neufeld15b}
{Neufeld}, D.~A., {Black}, J.~H., {Gerin}, M., {et~al.} 2015{\natexlab{a}},
  \apj, 807, 54

\bibitem[{{Neufeld} {et~al.}(2015{\natexlab{b}}){Neufeld}, {Godard}, {Gerin},
  {Pineau des For{\^e}ts}, {Bernier}, {Falgarone}, {Graf}, {G{\"u}sten},
  {Herbst}, {Lesaffre}, {Schilke}, {Sonnentrucker}, \&
  {Wiesemeyer}}]{neufeld15a}
{Neufeld}, D.~A., {Godard}, B., {Gerin}, M., {et~al.} 2015{\natexlab{b}}, \aap,
  577, A49

\bibitem[{{Neufeld} \& {Wolfire}(2016)}]{neufeld16}
{Neufeld}, D.~A. \& {Wolfire}, M.~G. 2016, \apj, 826, 183

\bibitem[{{Neufeld} {et~al.}(2005){Neufeld}, {Wolfire}, \&
  {Schilke}}]{neufeld05}
{Neufeld}, D.~A., {Wolfire}, M.~G., \& {Schilke}, P. 2005, \apj, 628, 260

\bibitem[{{Neufeld} {et~al.}(1997){Neufeld}, {Zmuidzinas}, {Schilke}, \&
  {Phillips}}]{neufeld97}
{Neufeld}, D.~A., {Zmuidzinas}, J., {Schilke}, P., \& {Phillips}, T.~G. 1997,
  \apjl, 488, L141

\bibitem[{{Ossenkopf} {et~al.}(2010){Ossenkopf}, {M{\"u}ller}, {Lis},
  {Schilke}, {Bell}, {Bruderer}, {Bergin}, {Ceccarelli}, {Comito}, {Stutzki},
  {Bacman}, {Baudry}, {Benz}, {Benedettini}, {Berne}, {Blake}, {Boogert},
  {Bottinelli}, {Boulanger}, {Cabrit}, {Caselli}, {Caux}, {Cernicharo},
  {Codella}, {Coutens}, {Crimier}, {Crockett}, {Daniel}, {Demyk}, {Dieleman},
  {Dominik}, {Dubernet}, {Emprechtinger}, {Encrenaz}, {Falgarone}, {France},
  {Fuente}, {Gerin}, {Giesen}, {di Giorgio}, {Goicoechea}, {Goldsmith},
  {G{\"u}sten}, {Harris}, {Helmich}, {Herbst}, {Hily-Blant}, {Jacobs}, {Jacq},
  {Joblin}, {Johnstone}, {Kahane}, {Kama}, {Klein}, {Klotz}, {Kramer},
  {Langer}, {Lefloch}, {Leinz}, {Lorenzani}, {Lord}, {Maret}, {Martin},
  {Martin-Pintado}, {McCoey}, {Melchior}, {Melnick}, {Menten}, {Mookerjea},
  {Morris}, {Murphy}, {Neufeld}, {Nisini}, {Pacheco}, {Pagani}, {Parise},
  {Pearson}, {P{\'e}rault}, {Phillips}, {Plume}, {Quin}, {Rizzo}, {R{\"o}llig},
  {Salez}, {Saraceno}, {Schlemmer}, {Simon}, {Schuster}, {van der Tak},
  {Tielens}, {Teyssier}, {Trappe}, {Vastel}, {Viti}, {Wakelam}, {Walters},
  {Wang}, {Whyborn}, {van der Wiel}, {Yorke}, {Yu}, \&
  {Zmuidzinas}}]{ossenkopf10}
{Ossenkopf}, V., {M{\"u}ller}, H.~S.~P., {Lis}, D.~C., {et~al.} 2010, \aap,
  518, L111

\bibitem[{Patil {et~al.}(2010)Patil, Huard, \& Fonnesbeck}]{pymc}
Patil, A., Huard, D., \& Fonnesbeck, C. 2010, Journal of Statistical Software,
  35, 1

\bibitem[{{Persson} {et~al.}(2010){Persson}, {Black}, {Cernicharo},
  {Goicoechea}, {Hassel}, {Herbst}, {Gerin}, {de Luca}, {Bell}, {Coutens},
  {Falgarone}, {Goldsmith}, {Gupta}, {Ka{\'z}mierczak}, {Lis}, {Mookerjea},
  {Neufeld}, {Pearson}, {Phillips}, {Sonnentrucker}, {Stutzki}, {Vastel}, {Yu},
  {Boulanger}, {Dartois}, {Encrenaz}, {Geballe}, {Giesen}, {Godard}, {Gry},
  {Hennebelle}, {Hily-Blant}, {Joblin}, {Ko{\l}os}, {Kre{\l}owski},
  {Mart{\'{\i}}n-Pintado}, {Menten}, {Monje}, {Perault}, {Plume}, {Salez},
  {Schlemmer}, {Schmidt}, {Teyssier}, {P{\'e}ron}, {Cais}, {Gaufre}, {Cros},
  {Ravera}, {Morris}, {Lord}, \& {Planesas}}]{persson10}
{Persson}, C.~M., {Black}, J.~H., {Cernicharo}, J., {et~al.} 2010, \aap, 521,
  L45

\bibitem[{{Persson} {et~al.}(2012){Persson}, {De Luca}, {Mookerjea},
  {Olofsson}, {Black}, {Gerin}, {Herbst}, {Bell}, {Coutens}, {Godard},
  {Goicoechea}, {Hassel}, {Hily-Blant}, {Menten}, {M{\"u}ller}, {Pearson}, \&
  {Yu}}]{persson12}
{Persson}, C.~M., {De Luca}, M., {Mookerjea}, B., {et~al.} 2012, \aap, 543,
  A145

\bibitem[{{Pickett} {et~al.}(1998){Pickett}, {Poynter}, {Cohen}, {Delitsky},
  {Pearson}, \& {M{\"u}ller}}]{pickett98}
{Pickett}, H.~M., {Poynter}, R.~L., {Cohen}, E.~A., {et~al.} 1998, \jqsrt, 60,
  883

\bibitem[{{Qin} {et~al.}(2010){Qin}, {Schilke}, {Comito}, {M{\"o}ller},
  {Rolffs}, {M{\"u}ller}, {Belloche}, {Menten}, {Lis}, {Phillips}, {Bergin},
  {Bell}, {Crockett}, {Blake}, {Cabrit}, {Caux}, {Ceccarelli}, {Cernicharo},
  {Daniel}, {Dubernet}, {Emprechtinger}, {Encrenaz}, {Falgarone}, {Gerin},
  {Giesen}, {Goicoechea}, {Goldsmith}, {Gupta}, {Herbst}, {Joblin},
  {Johnstone}, {Langer}, {Lord}, {Maret}, {Martin}, {Melnick}, {Morris},
  {Murphy}, {Neufeld}, {Ossenkopf}, {Pagani}, {Pearson}, {P{\'e}rault},
  {Plume}, {Salez}, {Schlemmer}, {Stutzki}, {Trappe}, {van der Tak}, {Vastel},
  {Wang}, {Yorke}, {Yu}, {Zmuidzinas}, {Boogert}, {G{\"u}sten}, {Hartogh},
  {Honingh}, {Karpov}, {Kooi}, {Krieg}, {Schieder}, {Diez-Gonzalez},
  {Bachiller}, {Martin-Pintado}, {Baechtold}, {Olberg}, {Nordh}, {Gill}, \&
  {Chattopadhyay}}]{qin10}
{Qin}, S.-L., {Schilke}, P., {Comito}, C., {et~al.} 2010, \aap, 521, L14

\bibitem[{{Rachford} {et~al.}(2009){Rachford}, {Snow}, {Destree}, {Ross},
  {Ferlet}, {Friedman}, {Gry}, {Jenkins}, {Morton}, {Savage}, {Shull},
  {Sonnentrucker}, {Tumlinson}, {Vidal-Madjar}, {Welty}, \&
  {York}}]{rachford09}
{Rachford}, B.~L., {Snow}, T.~P., {Destree}, J.~D., {et~al.} 2009, \apjs, 180,
  125

\bibitem[{{Rachford} {et~al.}(2002){Rachford}, {Snow}, {Tumlinson}, {Shull},
  {Blair}, {Ferlet}, {Friedman}, {Gry}, {Jenkins}, {Morton}, {Savage},
  {Sonnentrucker}, {Vidal-Madjar}, {Welty}, \& {York}}]{rachford02}
{Rachford}, B.~L., {Snow}, T.~P., {Tumlinson}, J., {et~al.} 2002, \apj, 577,
  221

\bibitem[{{Radhakrishnan}(1960)}]{radhakrishnan60}
{Radhakrishnan}, V. 1960, \pasp, 72, 296

\bibitem[{{Reich} \& {Reich}(1986)}]{reich86}
{Reich}, P. \& {Reich}, W. 1986, \aaps, 63, 205

\bibitem[{{Reich}(1982)}]{reich82}
{Reich}, W. 1982, \aaps, 48, 219

\bibitem[{{Reid} {et~al.}(2016){Reid}, {Dame}, {Menten}, \&
  {Brunthaler}}]{raid16}
{Reid}, M.~J., {Dame}, T.~M., {Menten}, K.~M., \& {Brunthaler}, A. 2016, \apj\
  {\it submitted}

\bibitem[{{Reid} {et~al.}(2014){Reid}, {Menten}, {Brunthaler}, {Zheng}, {Dame},
  {Xu}, {Wu}, {Zhang}, {Sanna}, {Sato}, {Hachisuka}, {Choi}, {Immer},
  {Moscadelli}, {Rygl}, \& {Bartkiewicz}}]{reid14}
{Reid}, M.~J., {Menten}, K.~M., {Brunthaler}, A., {et~al.} 2014, \apj, 783, 130

\bibitem[{{Roberts} {et~al.}(1997){Roberts}, {Dickel}, \& {Goss}}]{roberts97}
{Roberts}, D.~A., {Dickel}, H.~R., \& {Goss}, W.~M. 1997, \apj, 476, 209

\bibitem[{{Rygl} {et~al.}(2012){Rygl}, {Brunthaler}, {Sanna}, {Menten}, {Reid},
  {van Langevelde}, {Honma}, {Torstensson}, \& {Fujisawa}}]{rygl12}
{Rygl}, K.~L.~J., {Brunthaler}, A., {Sanna}, A., {et~al.} 2012, \aap, 539, A79

\bibitem[{{Sanna} {et~al.}(2014){Sanna}, {Reid}, {Menten}, {Dame}, {Zhang},
  {Sato}, {Brunthaler}, {Moscadelli}, \& {Immer}}]{sanna14}
{Sanna}, A., {Reid}, M.~J., {Menten}, K.~M., {et~al.} 2014, \apj, 781, 108

\bibitem[{{Sato} {et~al.}(2010){Sato}, {Reid}, {Brunthaler}, \&
  {Menten}}]{sato10}
{Sato}, M., {Reid}, M.~J., {Brunthaler}, A., \& {Menten}, K.~M. 2010, \apj,
  720, 1055

\bibitem[{{Savage} {et~al.}(1977){Savage}, {Bohlin}, {Drake}, \&
  {Budich}}]{savage77}
{Savage}, B.~D., {Bohlin}, R.~C., {Drake}, J.~F., \& {Budich}, W. 1977, \apj,
  216, 291

\bibitem[{{Schilke} {et~al.}(2014){Schilke}, {Neufeld}, {M{\"u}ller}, {Comito},
  {Bergin}, {Lis}, {Gerin}, {Black}, {Wolfire}, {Indriolo}, {Pearson},
  {Menten}, {Winkel}, {S{\'a}nchez-Monge}, {M{\"o}ller}, {Godard}, \&
  {Falgarone}}]{schilke14}
{Schilke}, P., {Neufeld}, D.~A., {M{\"u}ller}, H.~S.~P., {et~al.} 2014, \aap,
  566, A29

\bibitem[{{Shull} {et~al.}(2000){Shull}, {Tumlinson}, {Jenkins}, {Moos},
  {Rachford}, {Savage}, {Sembach}, {Snow}, {Sonneborn}, {York}, {Blair},
  {Green}, {Friedman}, \& {Sahnow}}]{shull00}
{Shull}, J.~M., {Tumlinson}, J., {Jenkins}, E.~B., {et~al.} 2000, \apjl, 538,
  L73

\bibitem[{{Snow} \& {McCall}(2006)}]{snow06}
{Snow}, T.~P. \& {McCall}, B.~J. 2006, \araa, 44, 367

\bibitem[{{Sonnentrucker} {et~al.}(2010){Sonnentrucker}, {Neufeld}, {Phillips},
  {Gerin}, {Lis}, {de Luca}, {Goicoechea}, {Black}, {Bell}, {Boulanger},
  {Cernicharo}, {Coutens}, {Dartois}, {Ka{\'z}mierczak}, {Encrenaz},
  {Falgarone}, {Geballe}, {Giesen}, {Godard}, {Goldsmith}, {Gry}, {Gupta},
  {Hennebelle}, {Herbst}, {Hily-Blant}, {Joblin}, {Ko{\l}os}, {Kre{\l}owski},
  {Mart{\'{\i}}n-Pintado}, {Menten}, {Monje}, {Mookerjea}, {Pearson},
  {Perault}, {Persson}, {Plume}, {Salez}, {Schlemmer}, {Schmidt}, {Stutzki},
  {Teyssier}, {Vastel}, {Yu}, {Caux}, {G{\"u}sten}, {Hatch}, {Klein}, {Mehdi},
  {Morris}, \& {Ward}}]{sonnentrucker10}
{Sonnentrucker}, P., {Neufeld}, D.~A., {Phillips}, T.~G., {et~al.} 2010, \aap,
  521, L12

\bibitem[{{Sonnentrucker} {et~al.}(2015{\natexlab{a}}){Sonnentrucker},
  {Wolfire}, {Neufeld}, {Flagey}, {Gerin}, {Goldsmith}, {Lis}, \&
  {Monje}}]{sonnentrucker15}
{Sonnentrucker}, P., {Wolfire}, M., {Neufeld}, D.~A., {et~al.}
  2015{\natexlab{a}}, \apj, 806, 49

\bibitem[{{Sonnentrucker} {et~al.}(2015{\natexlab{b}}){Sonnentrucker},
  {Wolfire}, {Neufeld}, {Flagey}, {Gerin}, {Goldsmith}, {Lis}, \&
  {Monje}}]{sonnertrucker15}
{Sonnentrucker}, P., {Wolfire}, M., {Neufeld}, D.~A., {et~al.}
  2015{\natexlab{b}}, \apj, 806, 49

\bibitem[{{Stanimirovic}(2002)}]{stanimirovic02}
{Stanimirovic}, S. 2002, in Astronomical Society of the Pacific Conference
  Series, Vol. 278, Single-Dish Radio Astronomy: Techniques and Applications,
  ed. S.~{Stanimirovic}, D.~{Altschuler}, P.~{Goldsmith}, \& C.~{Salter},
  375--396

\bibitem[{{Stanimirovi{\'c}} {et~al.}(2014){Stanimirovi{\'c}}, {Murray}, {Lee},
  {Heiles}, \& {Miller}}]{stanimirovic14}
{Stanimirovi{\'c}}, S., {Murray}, C.~E., {Lee}, M.-Y., {Heiles}, C., \&
  {Miller}, J. 2014, \apj, 793, 132

\bibitem[{{Stil} {et~al.}(2006){Stil}, {Taylor}, {Dickey}, {Kavars}, {Martin},
  {Rothwell}, {Boothroyd}, {Lockman}, \& {McClure-Griffiths}}]{stil06}
{Stil}, J.~M., {Taylor}, A.~R., {Dickey}, J.~M., {et~al.} 2006, \aj, 132, 1158

\bibitem[{{Strasser} \& {Taylor}(2004)}]{strasser04}
{Strasser}, S. \& {Taylor}, A.~R. 2004, \apj, 603, 560

\bibitem[{{Strasser} {et~al.}(2007){Strasser}, {Dickey}, {Taylor}, {Boothroyd},
  {Gaensler}, {Green}, {Kavars}, {Lockman}, {Martin}, {McClure-Griffiths},
  {Rothwell}, \& {Stil}}]{strasser07}
{Strasser}, S.~T., {Dickey}, J.~M., {Taylor}, A.~R., {et~al.} 2007, \aj, 134,
  2252

\bibitem[{{Taylor} {et~al.}(2003){Taylor}, {Gibson}, {Peracaula}, {Martin},
  {Landecker}, {Brunt}, {Dewdney}, {Dougherty}, {Gray}, {Higgs}, {Kerton},
  {Knee}, {Kothes}, {Purton}, {Uyaniker}, {Wallace}, {Willis}, \&
  {Durand}}]{taylor03}
{Taylor}, A.~R., {Gibson}, S.~J., {Peracaula}, M., {et~al.} 2003, \aj, 125,
  3145

\bibitem[{{Terlouw} \& {Vogelaar}(2015)}]{KapteynPackage}
{Terlouw}, J.~P. \& {Vogelaar}, M.~G.~R. 2015, {Kapteyn Package, version 2.3},
  {Kapteyn Astronomical Institute}, Groningen, available from
  \url{http://www.astro.rug.nl/software/kapteyn/}

\bibitem[{{Tizniti} {et~al.}(2014){Tizniti}, {Le Picard}, {Lique},
  {Berteloite}, {Canosa}, {Alexander}, \& {Sims}}]{tizniti14}
{Tizniti}, M., {Le Picard}, S.~D., {Lique}, F., {et~al.} 2014, Nature
  Chemistry, 6, 141

\bibitem[{{Valdivia} {et~al.}(2016){Valdivia}, {Hennebelle}, {G{\'e}rin}, \&
  {Lesaffre}}]{valdivia16}
{Valdivia}, V., {Hennebelle}, P., {G{\'e}rin}, M., \& {Lesaffre}, P. 2016,
  \aap, 587, A76

\bibitem[{{Vall{\'e}e}(2008)}]{vallee08}
{Vall{\'e}e}, J.~P. 2008, \aj, 135, 1301

\bibitem[{van~der Walt {et~al.}(2011)van~der Walt, Colbert, \&
  Varoquaux}]{NumPy}
van~der Walt, S., Colbert, S., \& Varoquaux, G. 2011, Computing in Science
  Engineering, 13, 22

\bibitem[{{van Dishoeck} {et~al.}(2013){van Dishoeck}, {Herbst}, \&
  {Neufeld}}]{vandishoeck13}
{van Dishoeck}, E.~F., {Herbst}, E., \& {Neufeld}, D.~A. 2013, Chemical
  Reviews, 113, 9043

\bibitem[{{van Dishoeck} {et~al.}(1993){van Dishoeck}, {Jansen}, {Schilke}, \&
  {Phillips}}]{vandishoeck93}
{van Dishoeck}, E.~F., {Jansen}, D.~J., {Schilke}, P., \& {Phillips}, T.~G.
  1993, \apjl, 416, L83

\bibitem[{{Wiesemeyer} {et~al.}(2016){Wiesemeyer}, {G{\"u}sten}, {Heyminck},
  {H{\"u}bers}, {Menten}, {Neufeld}, {Richter}, {Simon}, {Stutzki}, {Winkel},
  \& {Wyrowski}}]{wiesemeyer16}
{Wiesemeyer}, H., {G{\"u}sten}, R., {Heyminck}, S., {et~al.} 2016, \aap, 585,
  A76

\bibitem[{{Wiesemeyer} {et~al.}(2012){Wiesemeyer}, {G{\"u}sten}, {Heyminck},
  {Jacobs}, {Menten}, {Neufeld}, {Requena-Torres}, \& {Stutzki}}]{wiesemeyer12}
{Wiesemeyer}, H., {G{\"u}sten}, R., {Heyminck}, S., {et~al.} 2012, \aap, 542,
  L7

\bibitem[{{Wilson} {et~al.}(2013){Wilson}, {Rohlfs}, \&
  {H{\"u}ttemeister}}]{wilson13}
{Wilson}, T.~L., {Rohlfs}, K., \& {H{\"u}ttemeister}, S. 2013, {Tools of Radio
  Astronomy} (Springer-Verlag Berlin Heidelberg)

\bibitem[{{Winkel} {et~al.}(2016){Winkel}, {Lenz}, \& {Fl{\"o}er}}]{winkel16a}
{Winkel}, B., {Lenz}, D., \& {Fl{\"o}er}, L. 2016, \aap, 591, A12

\bibitem[{{Wyrowski} {et~al.}(2012){Wyrowski}, {G{\"u}sten}, {Menten},
  {Wiesemeyer}, \& {Klein}}]{wyrowski12}
{Wyrowski}, F., {G{\"u}sten}, R., {Menten}, K.~M., {Wiesemeyer}, H., \&
  {Klein}, B. 2012, \aap, 542, L15

\bibitem[{{Wyrowski} {et~al.}(2010){Wyrowski}, {Menten}, {G{\"u}sten}, \&
  {Belloche}}]{wyrowski10}
{Wyrowski}, F., {Menten}, K.~M., {G{\"u}sten}, R., \& {Belloche}, A. 2010,
  \aap, 518, A26

\bibitem[{{Zhang} {et~al.}(2014){Zhang}, {Moscadelli}, {Sato}, {Reid},
  {Menten}, {Zheng}, {Brunthaler}, {Dame}, {Xu}, \& {Immer}}]{zhang14}
{Zhang}, B., {Moscadelli}, L., {Sato}, M., {et~al.} 2014, \apj, 781, 89

\bibitem[{{Zhang} {et~al.}(2013){Zhang}, {Reid}, {Menten}, {Zheng},
  {Brunthaler}, {Dame}, \& {Xu}}]{zhang13}
{Zhang}, B., {Reid}, M.~J., {Menten}, K.~M., {et~al.} 2013, \apj, 775, 79

\end{thebibliography}

\appendix
\section{Prismas sight lines}\label{sec:prismassightlines}

\begin{figure*}[!t]
\centering%
\includegraphics[width=0.48\textwidth,viewport=10 10 415 980,clip=]{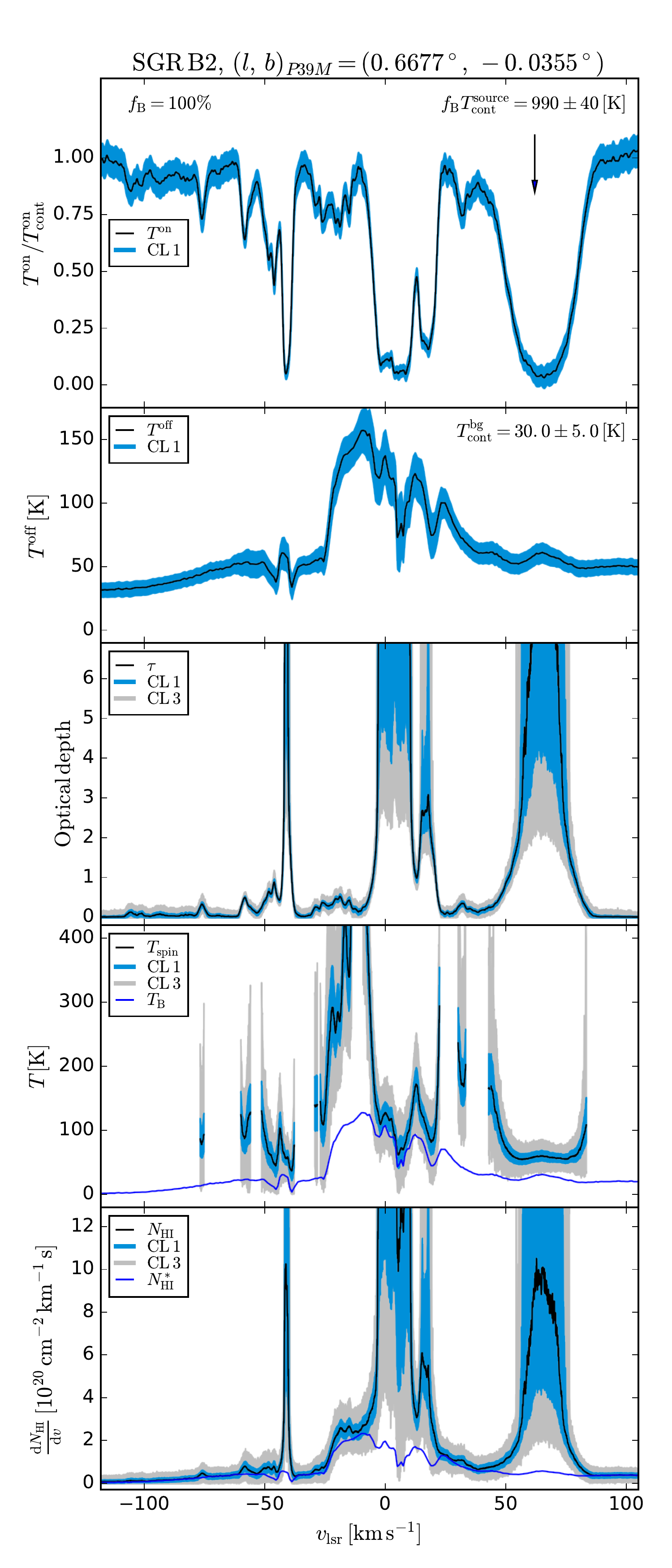}\hfill
\includegraphics[width=0.48\textwidth,viewport=10 10 415 980,clip=]{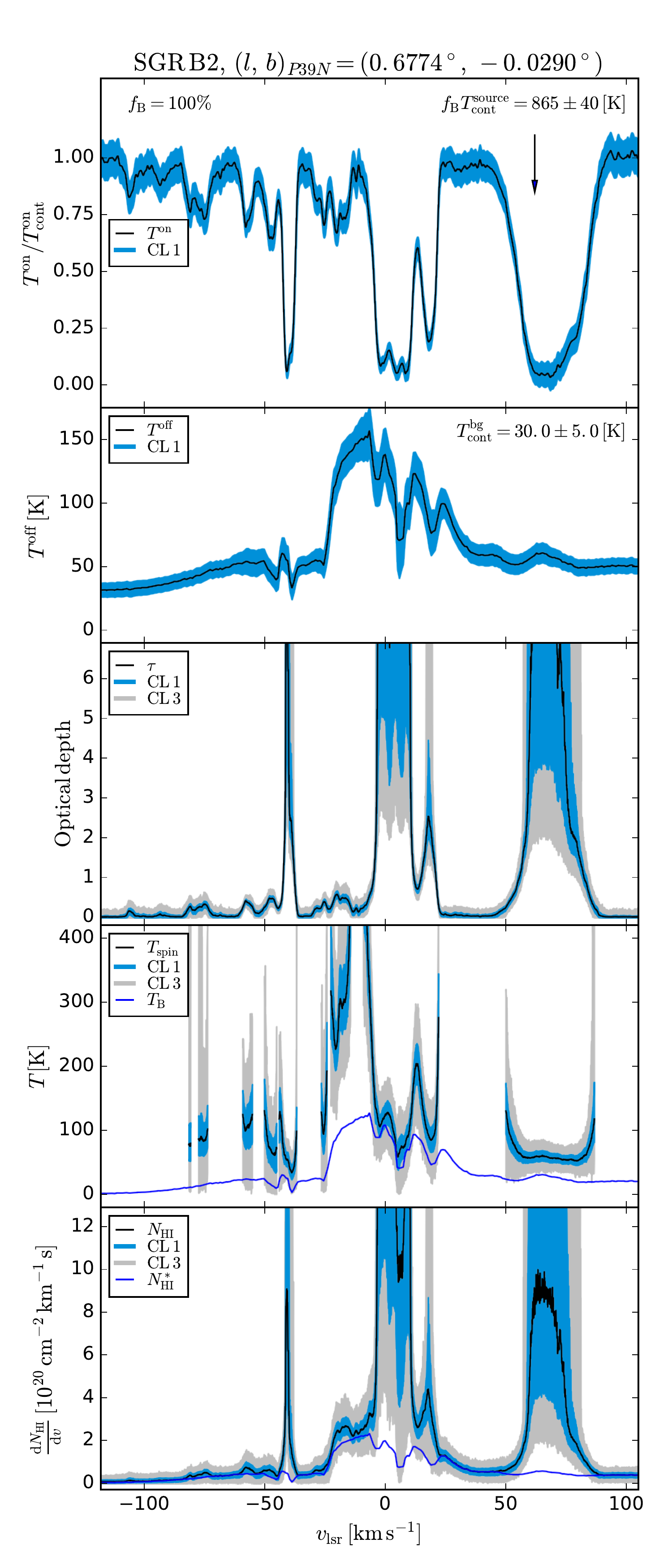}
\caption{Measured absorption and emission spectra, $T^\mathrm{on}$ and $T^\mathrm{off}$, with their respective errors, and derived quantities, optical depth, spin temperature, and \ion{H}{i} column density for the HEXOS program sources \object{Sgr\,B2\,(M)} and \object{Sgr\,B2\,(N)}. For a detailed description see text.}%
\label{fig:results_prismas_1n2}%
\end{figure*}

\begin{figure*}[!t]
\centering%
\includegraphics[width=0.48\textwidth,viewport=10 10 415 980,clip=]{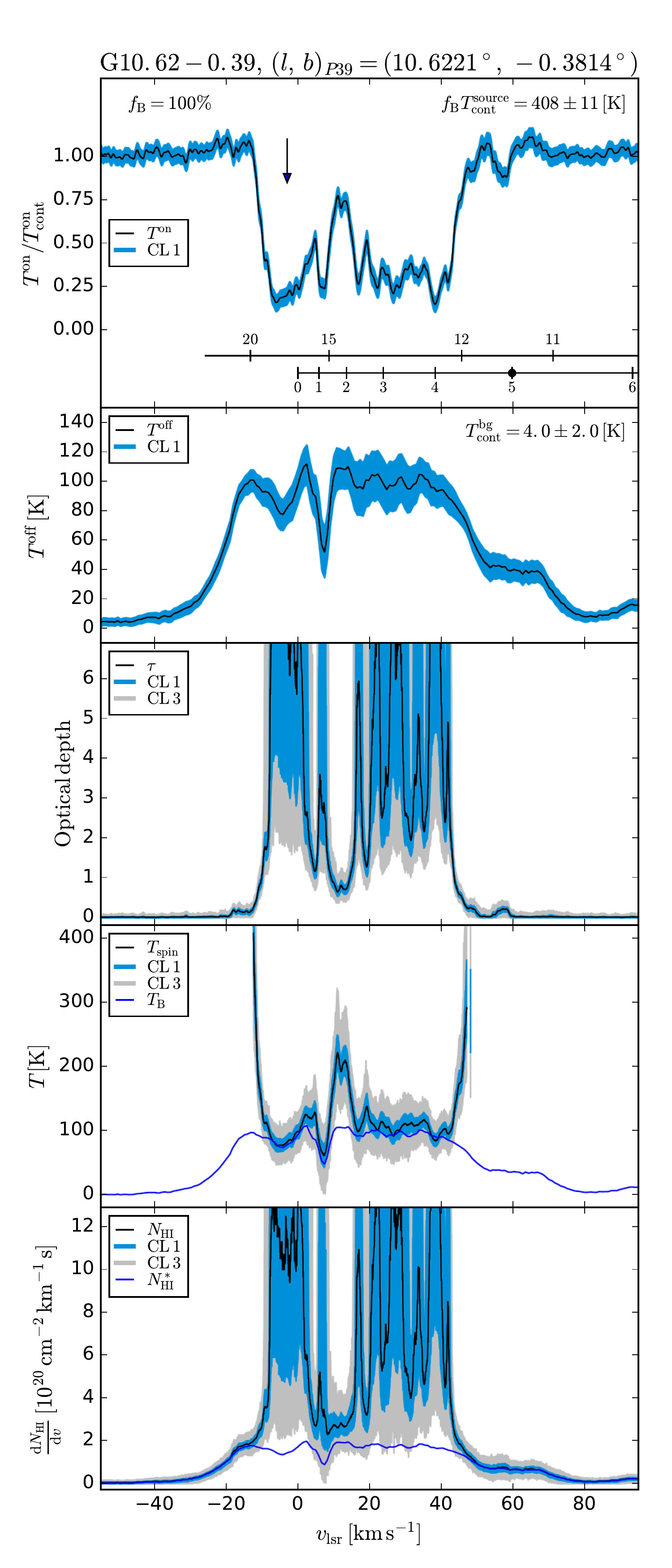}\hfill
\includegraphics[width=0.48\textwidth,viewport=10 10 415 980,clip=]{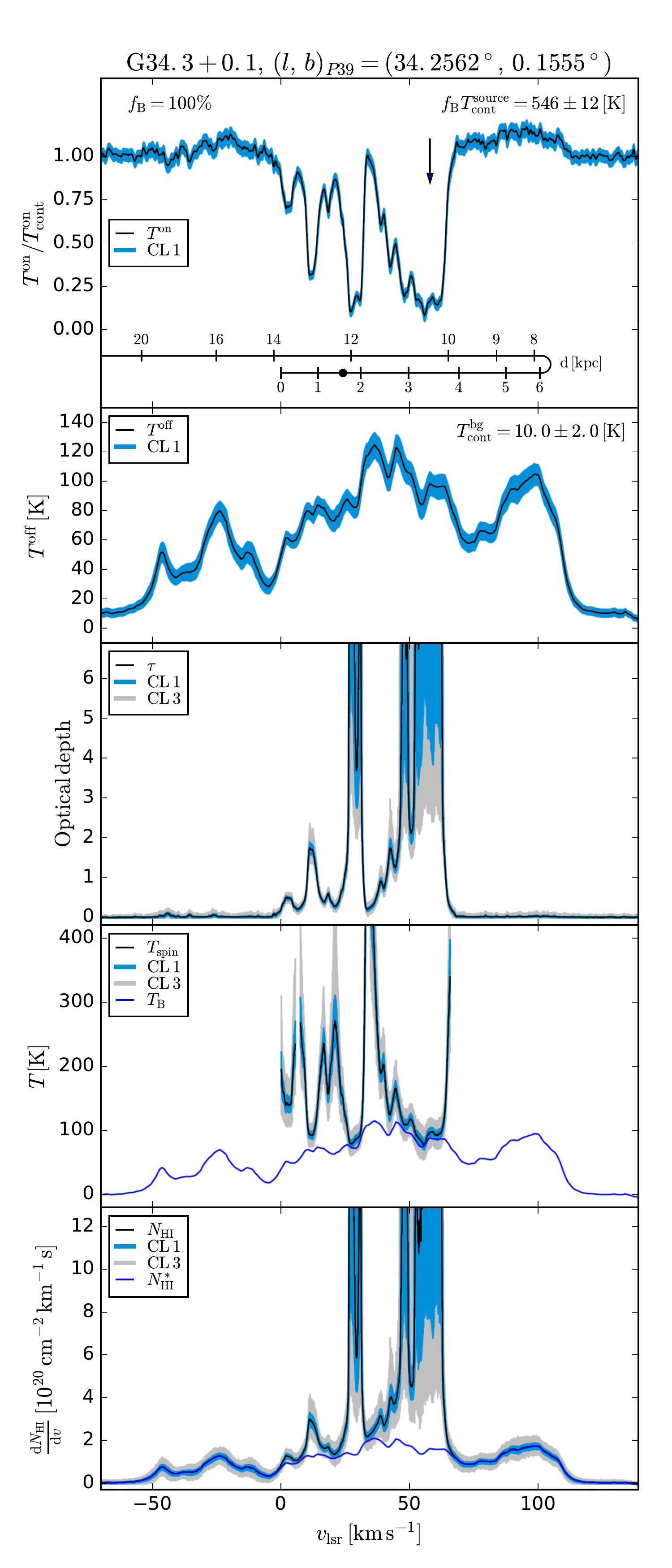}
\caption{As Fig.~\ref{fig:results_prismas_1n2} for the PRISMAS program sources \object{G\,10.62$-$0.39} and \object{G\,34.3$+$0.1}.}%
\label{fig:results_prismas_3n4}%
\end{figure*}

\begin{figure}[!t]
\centering%
\includegraphics[width=0.48\textwidth,viewport=10 10 415 980,clip=]{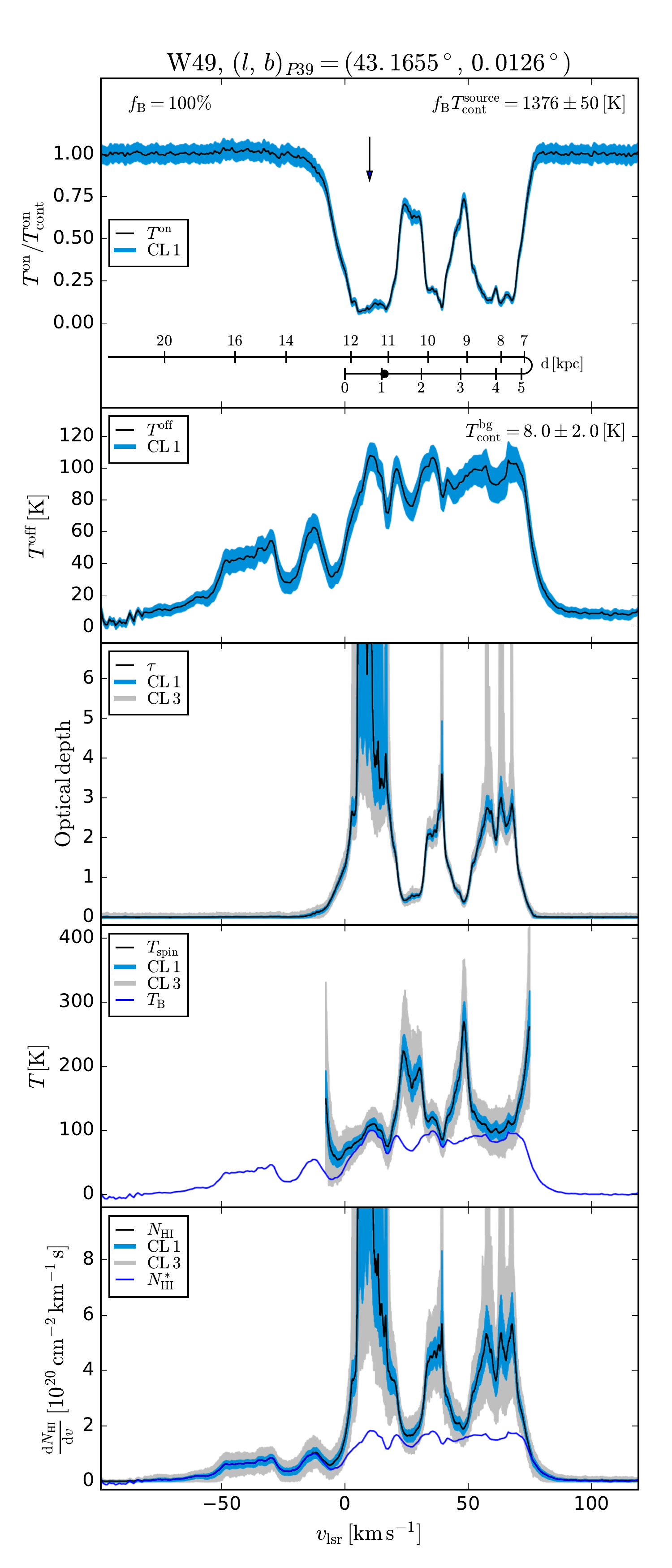}
\caption{As Fig.~\ref{fig:results_prismas_1n2} for the PRISMAS program source \object{W\,49}.}%
\label{fig:results_prismas_5}%
\end{figure}

\clearpage
\section{Other sight lines}\label{sec:othersightlines}

\begin{figure*}[!t]
\centering%
\includegraphics[width=0.48\textwidth,viewport=10 10 415 980,clip=]{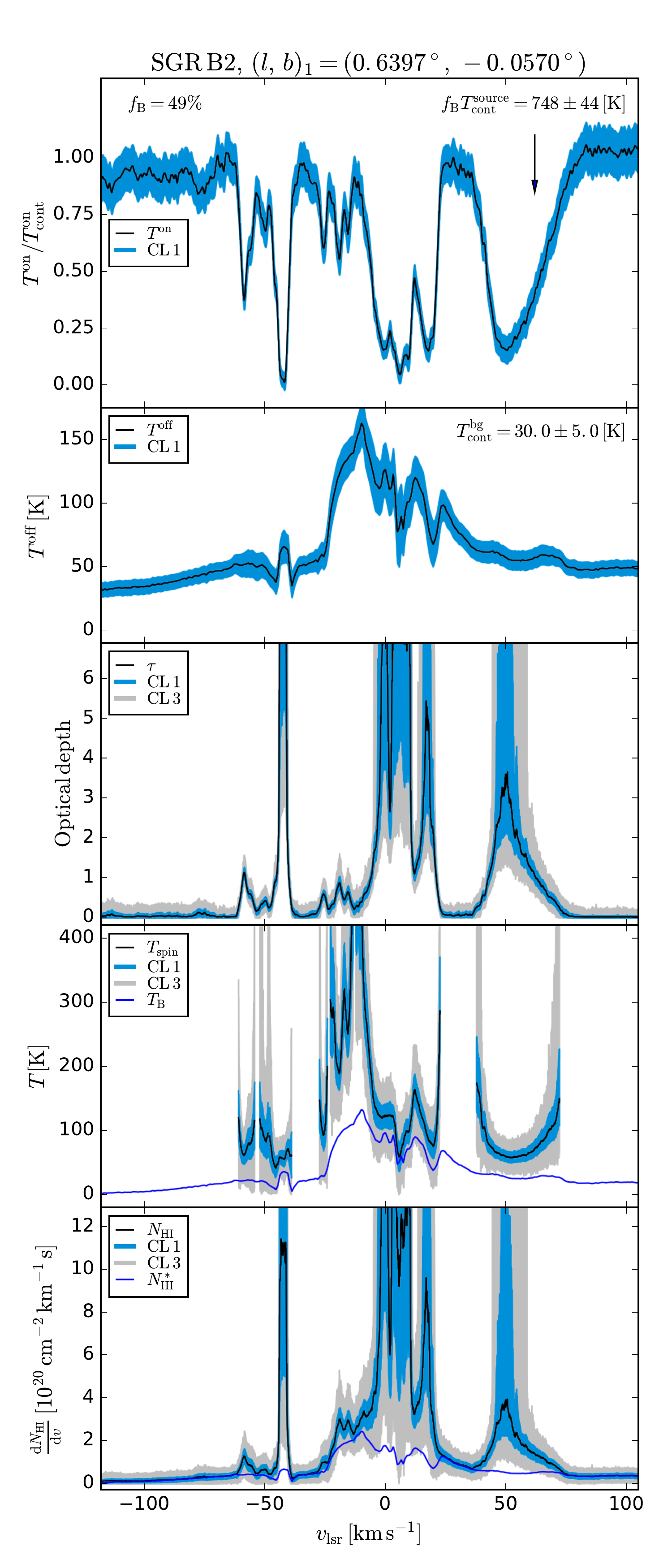}\hfill
\includegraphics[width=0.48\textwidth,viewport=10 10 415 980,clip=]{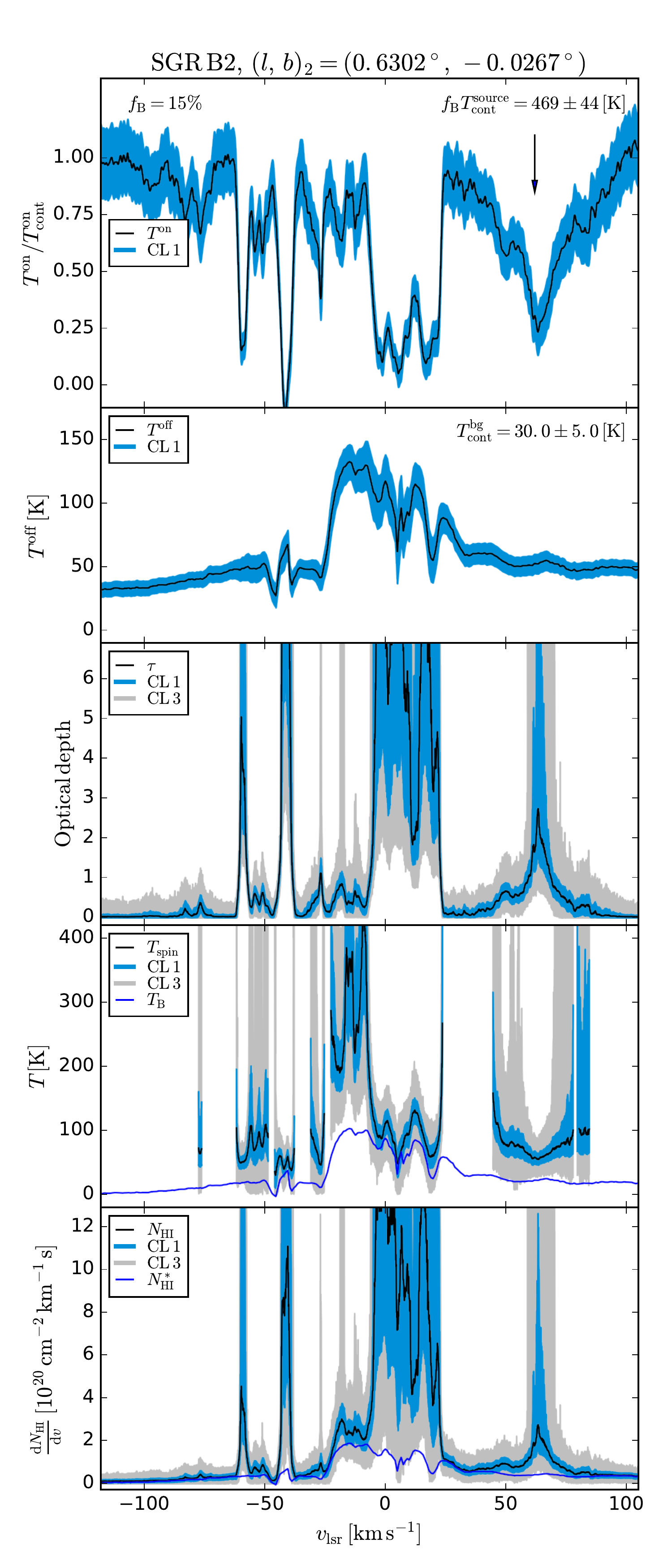}
\caption{Measured absorption and emission spectra, $T^\mathrm{on}$ and $T^\mathrm{off}$, with their respective errors, and derived quantities, optical depth, spin temperature, and \ion{H}{i} column density for the source \object{Sgr\,B2}, sight lines (1) and (2). For a detailed description see text.}%
\label{fig:results_sgr_1n2}%
\end{figure*}

\begin{figure*}[!t]
\centering%
\includegraphics[width=0.48\textwidth,viewport=10 10 415 980,clip=]{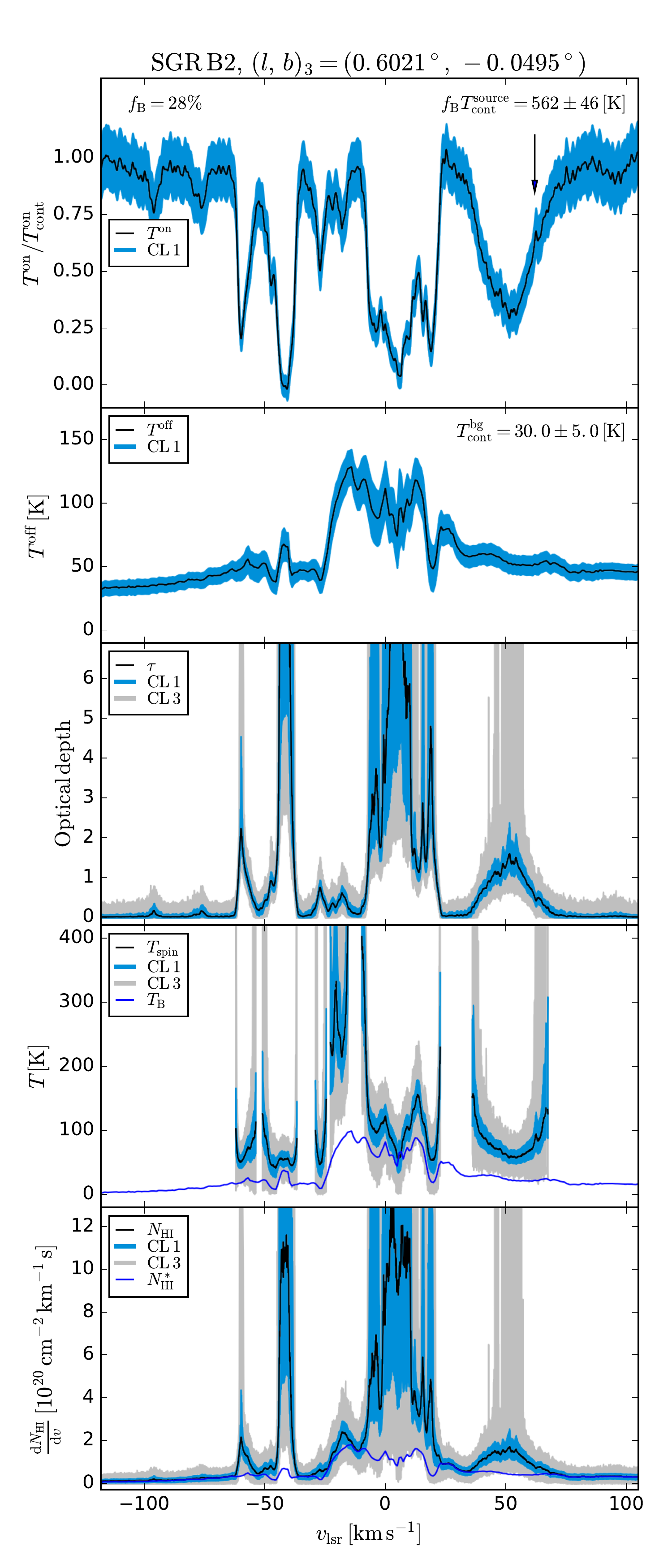}
\caption{As Fig.~\ref{fig:results_sgr_1n2} for sight line (3).}%
\label{fig:results_sgr_3}%
\end{figure*}

\begin{figure*}[!t]
\centering%
\includegraphics[width=0.48\textwidth,viewport=10 10 415 980,clip=]{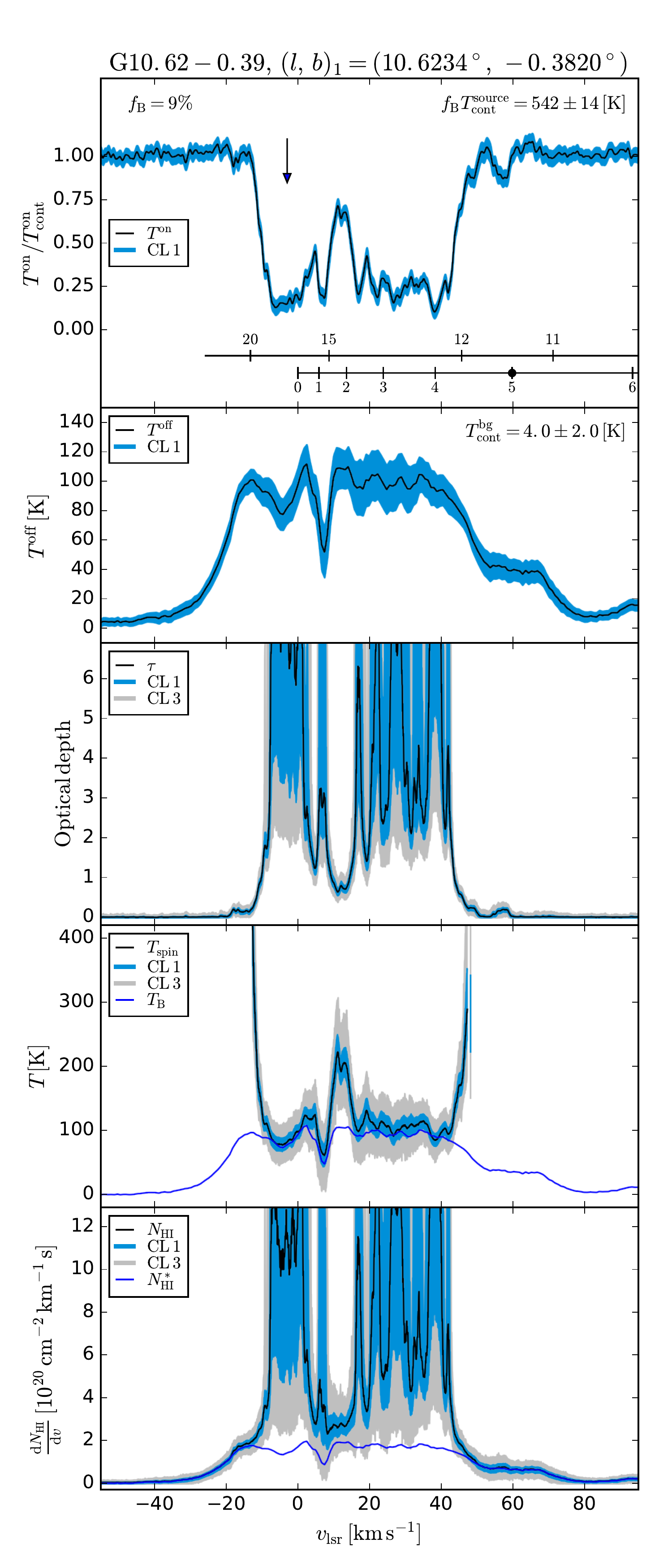}\hfill
\includegraphics[width=0.48\textwidth,viewport=10 10 415 980,clip=]{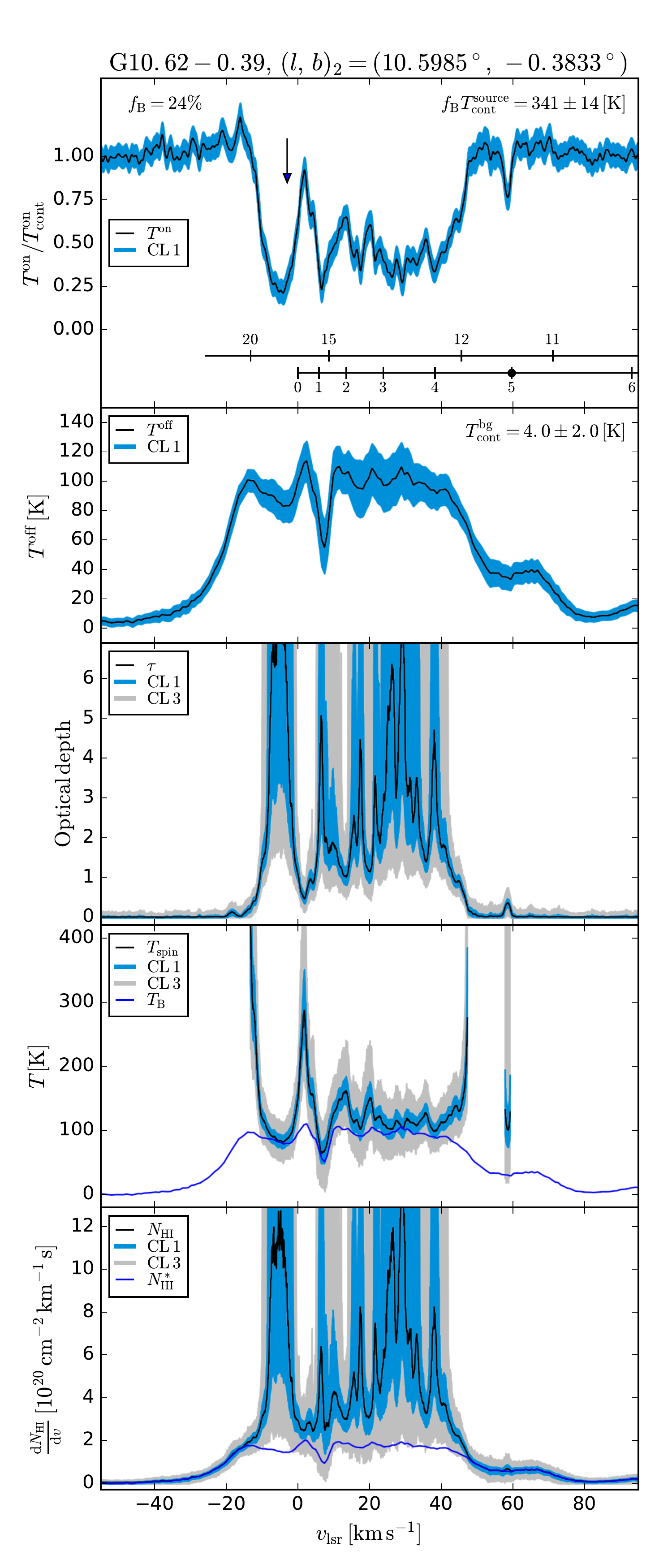}
\caption{As Fig.~\ref{fig:results_sgr_1n2} for source \object{G\,10.62$-$0.39}, sight lines (1) and (2).}%
\label{fig:results_g10_1n2}%
\end{figure*}

\begin{figure*}[!t]
\centering%
\includegraphics[width=0.48\textwidth,viewport=10 10 415 980,clip=]{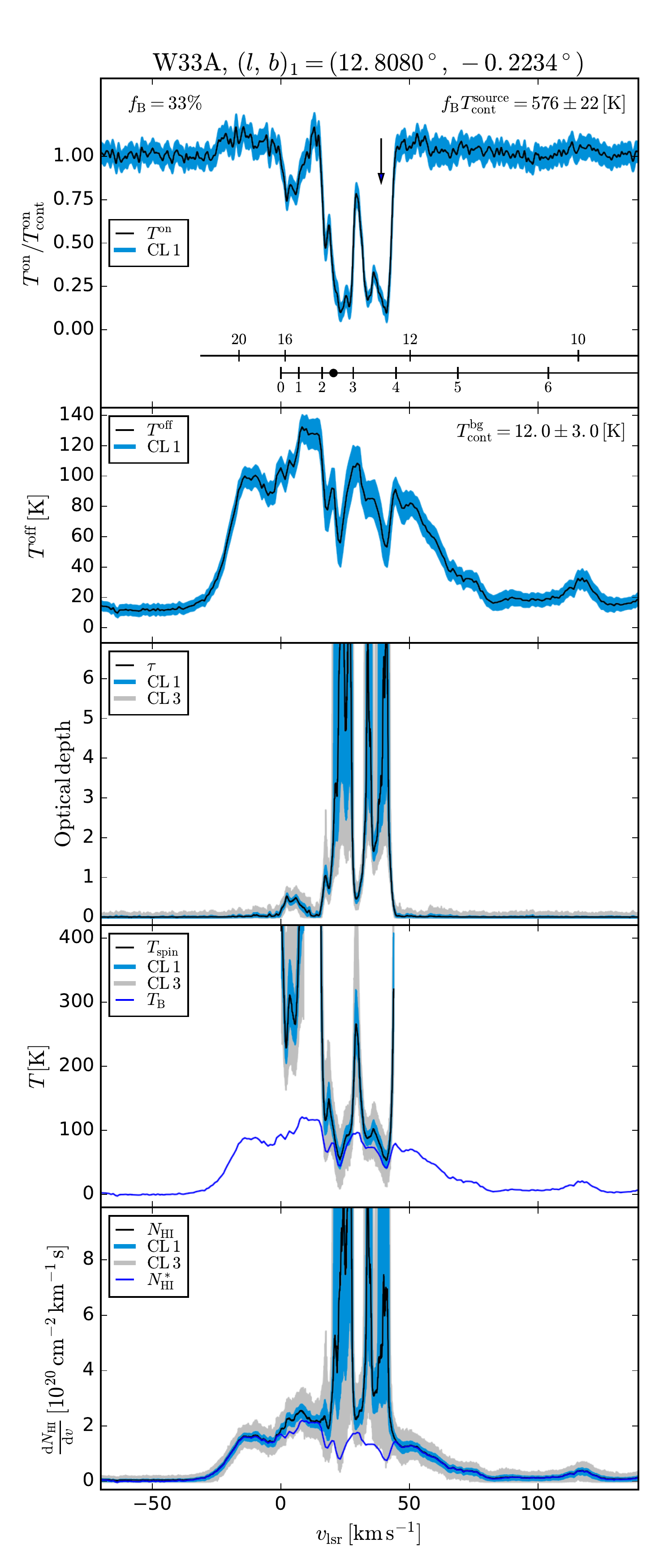}\hfill
\includegraphics[width=0.48\textwidth,viewport=10 10 415 980,clip=]{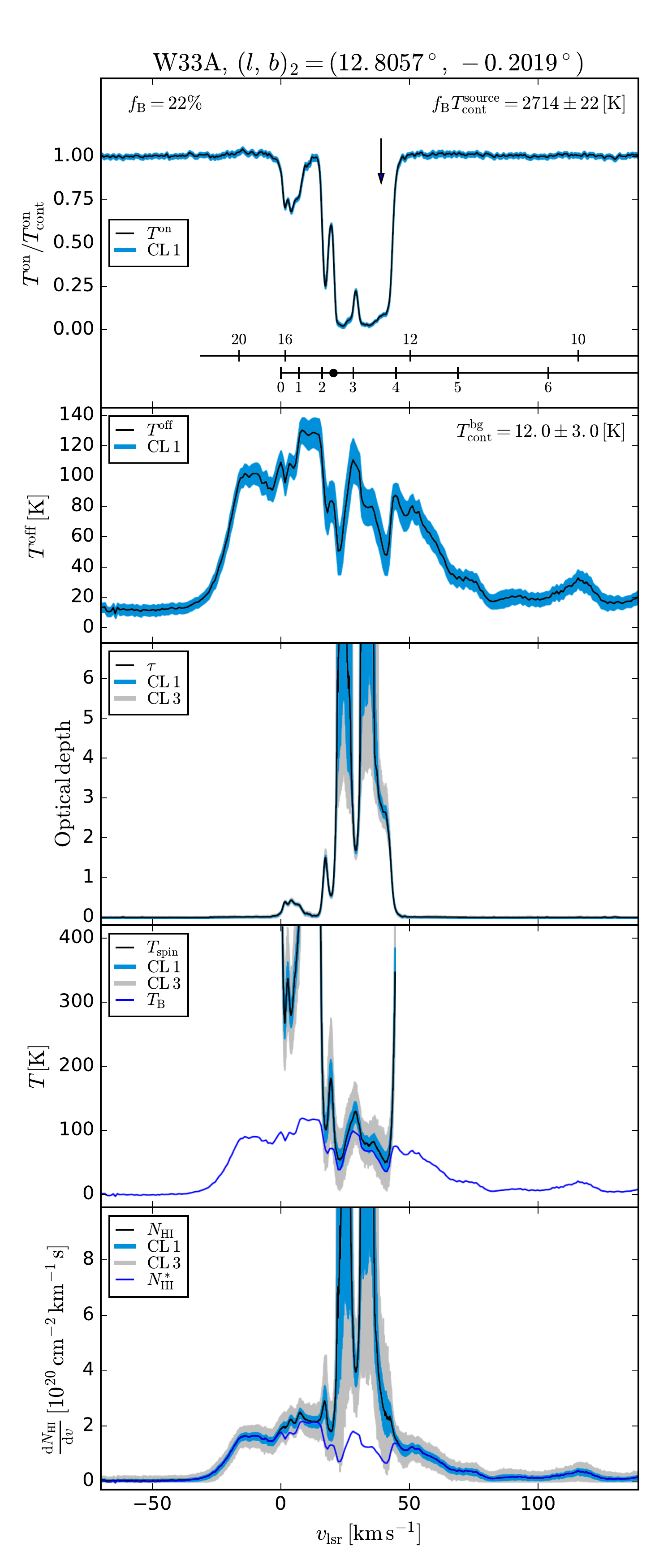}
\caption{As Fig.~\ref{fig:results_sgr_1n2} for source \object{W\,33}, sight lines (1) and (2).}%
\label{fig:results_w33_1n2}%
\end{figure*}

\begin{figure}[!t]
\centering%
\includegraphics[width=0.48\textwidth,viewport=10 10 415 980,clip=]{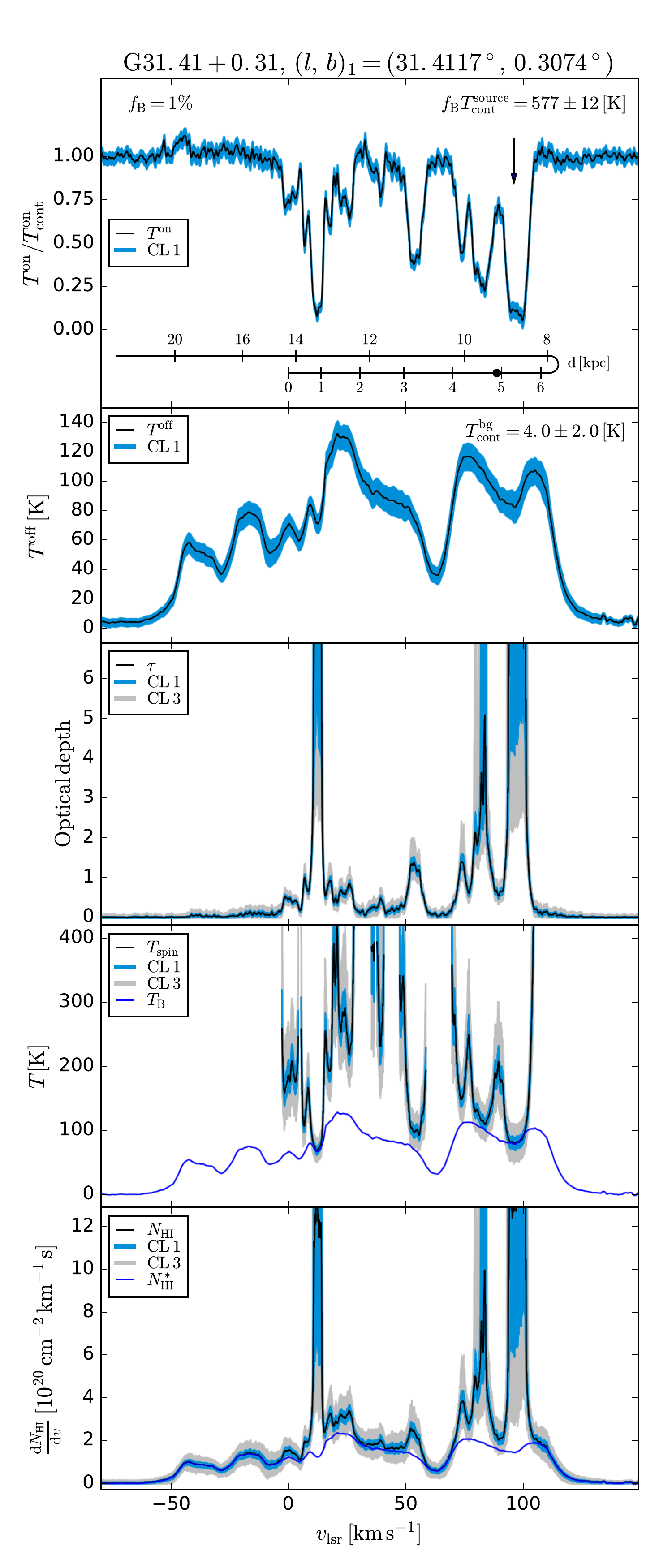}
\caption{As Fig.~\ref{fig:results_sgr_1n2} for source \object{G\,31.41$+$0.31}, sight line (1).}%
\label{fig:results_g31_1}%
\end{figure}

\begin{figure*}[!t]
\centering%
\includegraphics[width=0.48\textwidth,viewport=10 10 415 980,clip=]{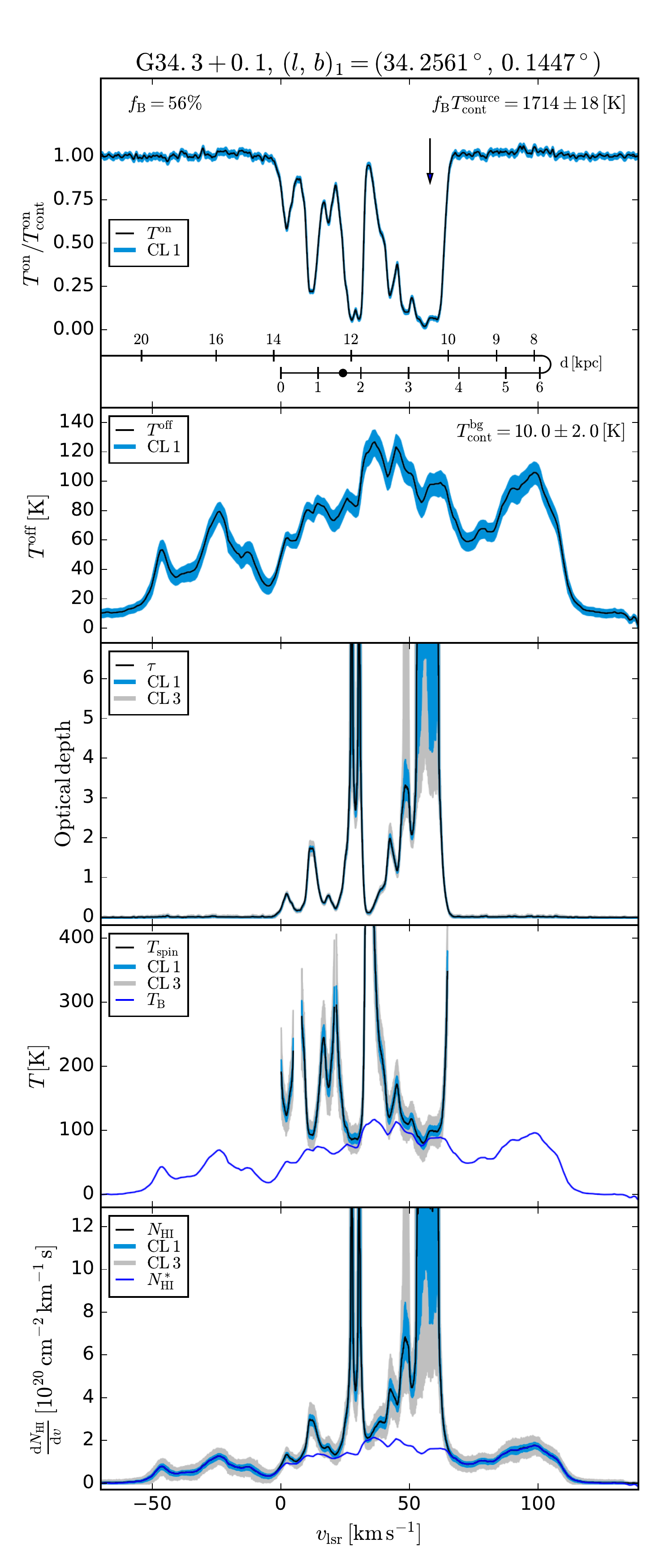}\hfill
\includegraphics[width=0.48\textwidth,viewport=10 10 415 980,clip=]{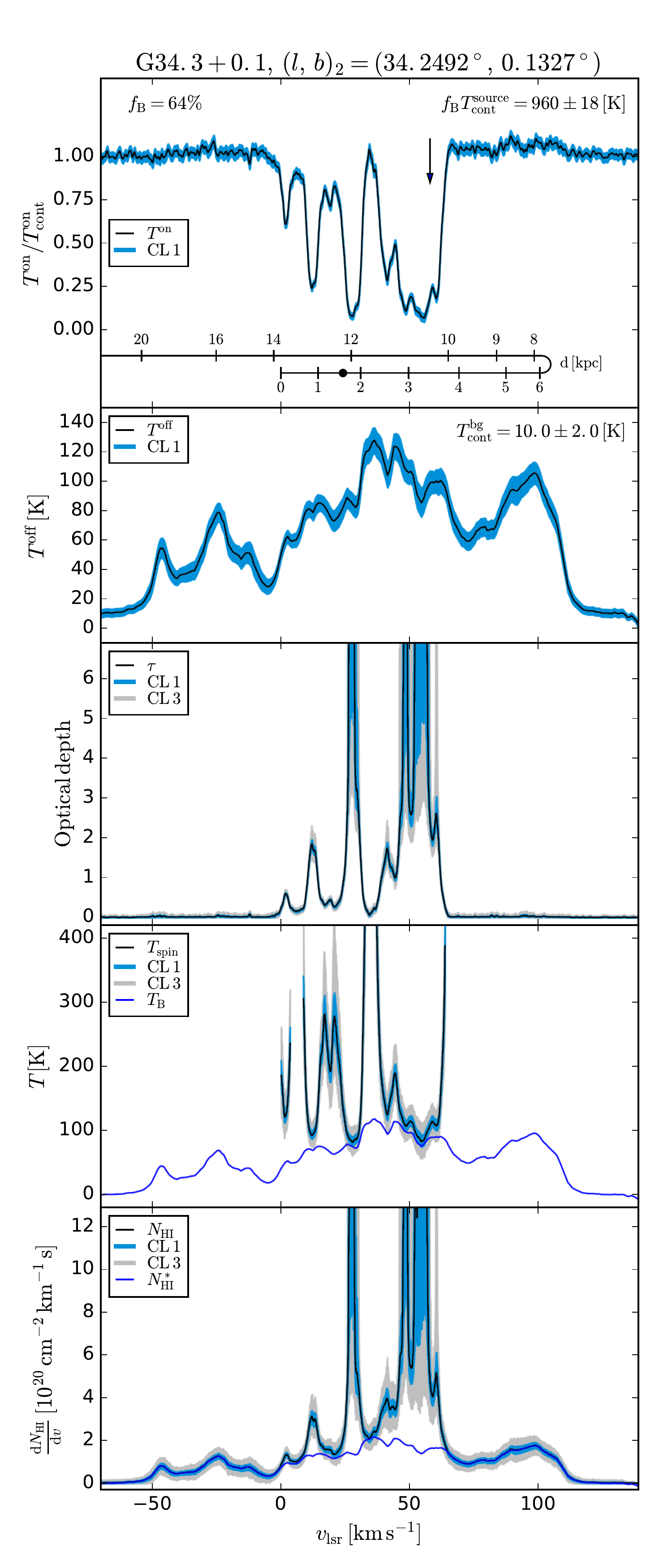}
\caption{As Fig.~\ref{fig:results_sgr_1n2} for source \object{G\,34.3$+$0.1}, sight lines (1) and (2).}%
\label{fig:results_g34_1n2}%
\end{figure*}

\begin{figure*}[!t]
\centering%
\includegraphics[width=0.48\textwidth,viewport=10 10 415 980,clip=]{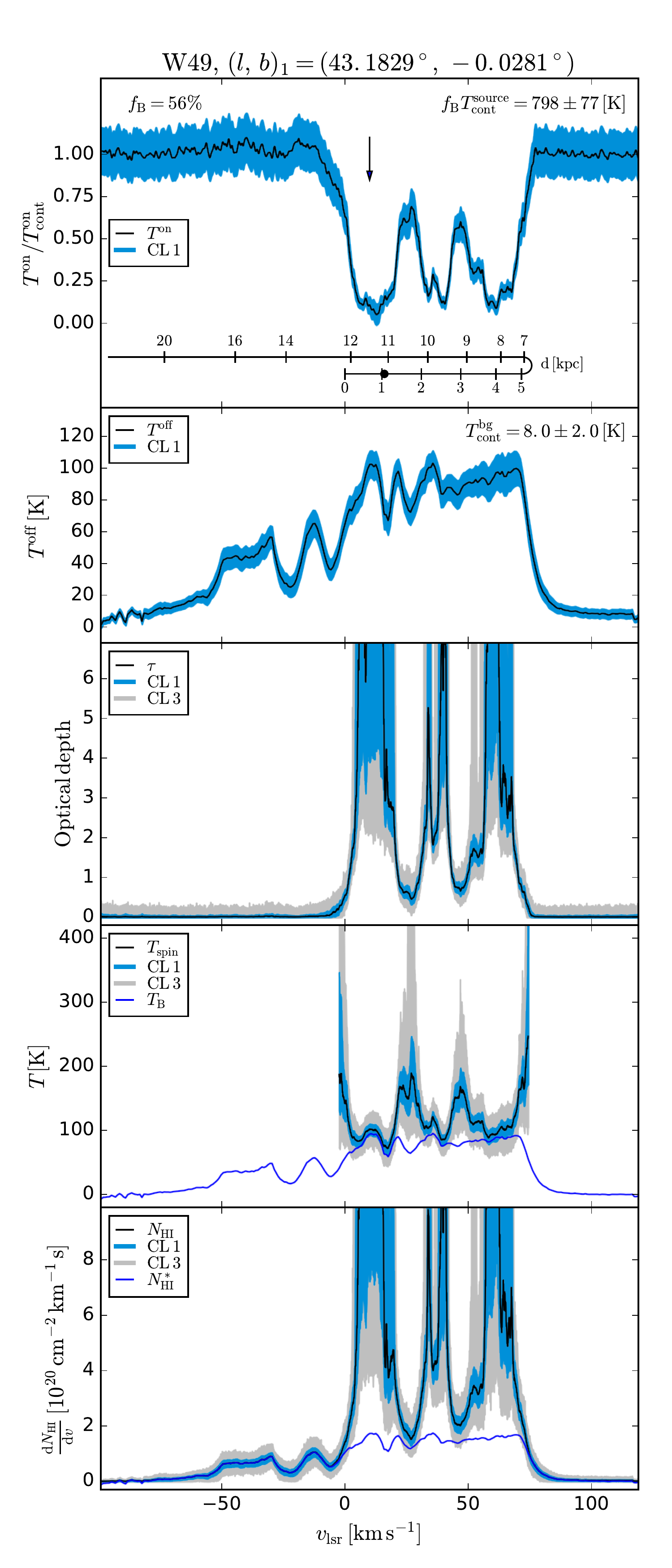}\hfill
\includegraphics[width=0.48\textwidth,viewport=10 10 415 980,clip=]{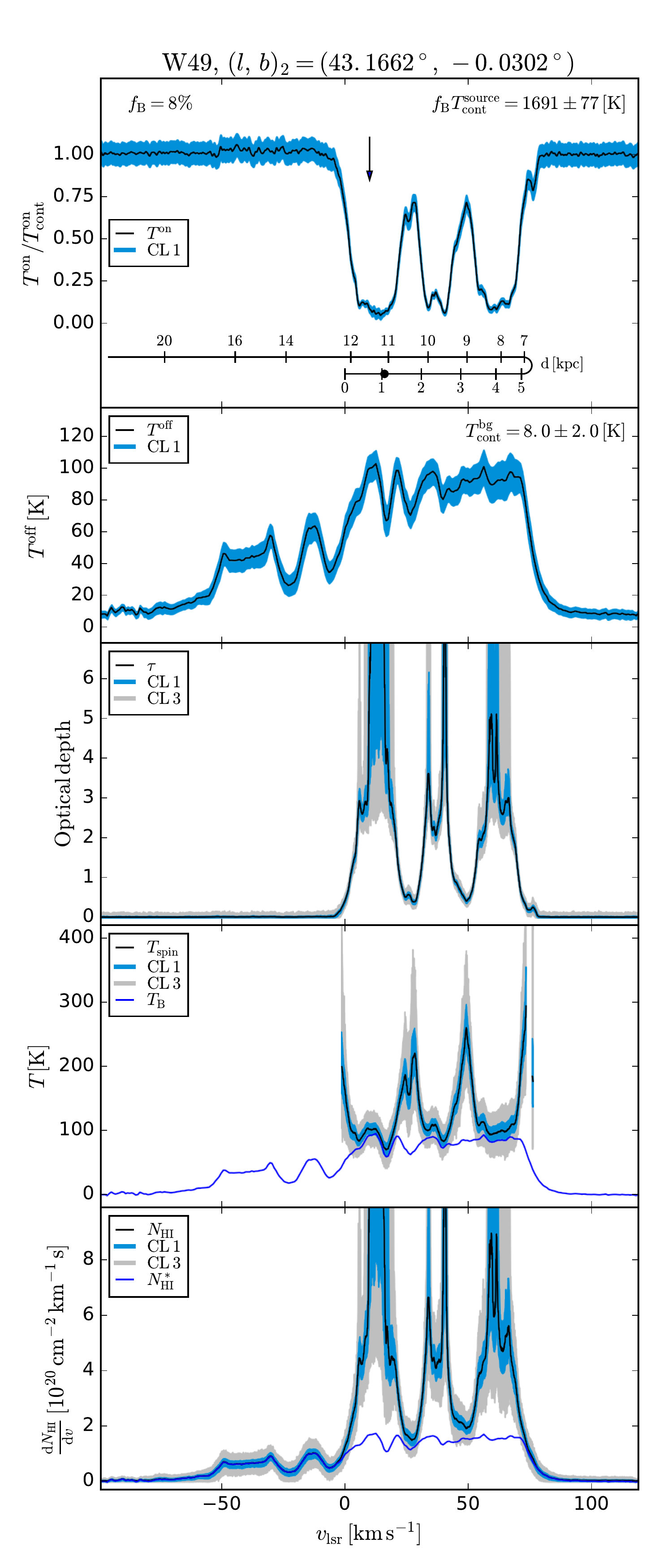}
\caption{As Fig.~\ref{fig:results_sgr_1n2} for source \object{W\,49}, sight lines (1) and (2).}%
\label{fig:results_w49_1n2}%
\end{figure*}

\begin{figure*}[!t]
\centering%
\includegraphics[width=0.48\textwidth,viewport=10 10 415 980,clip=]{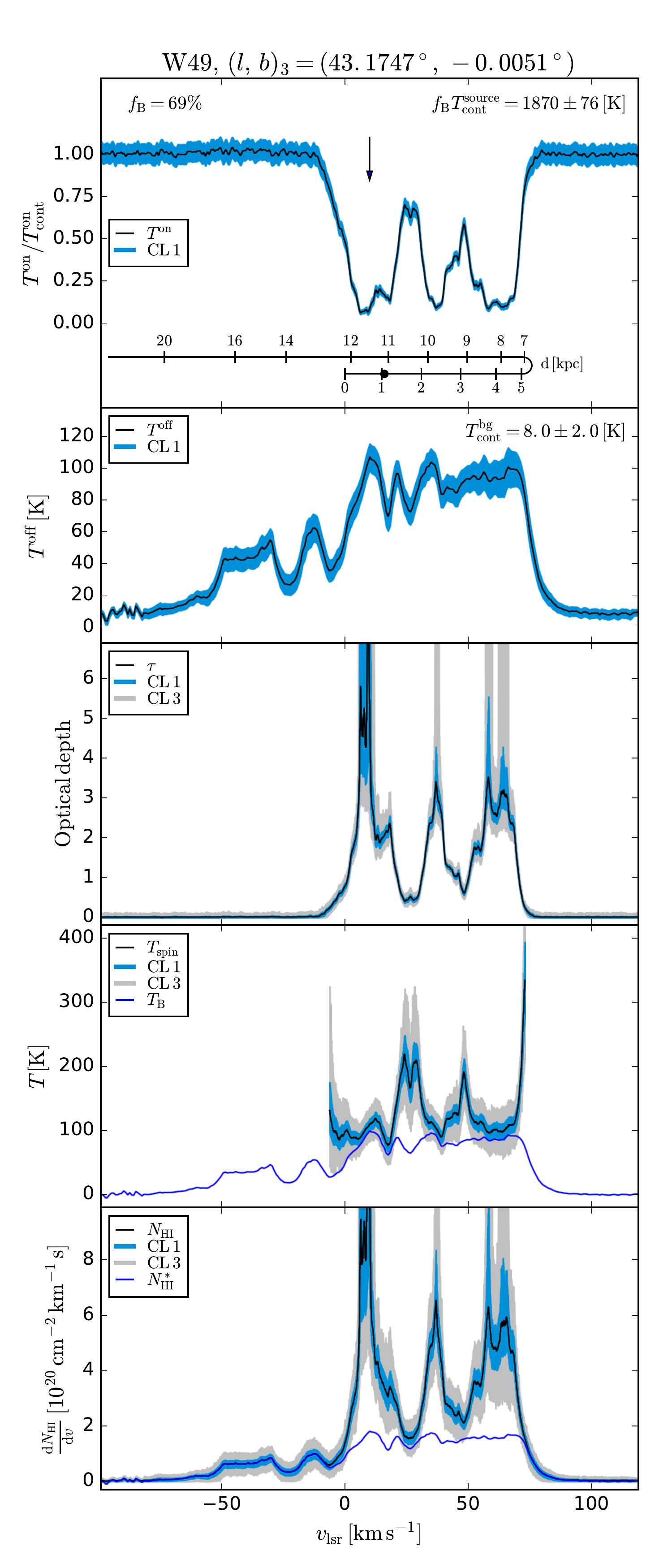}\hfill
\includegraphics[width=0.48\textwidth,viewport=10 10 415 980,clip=]{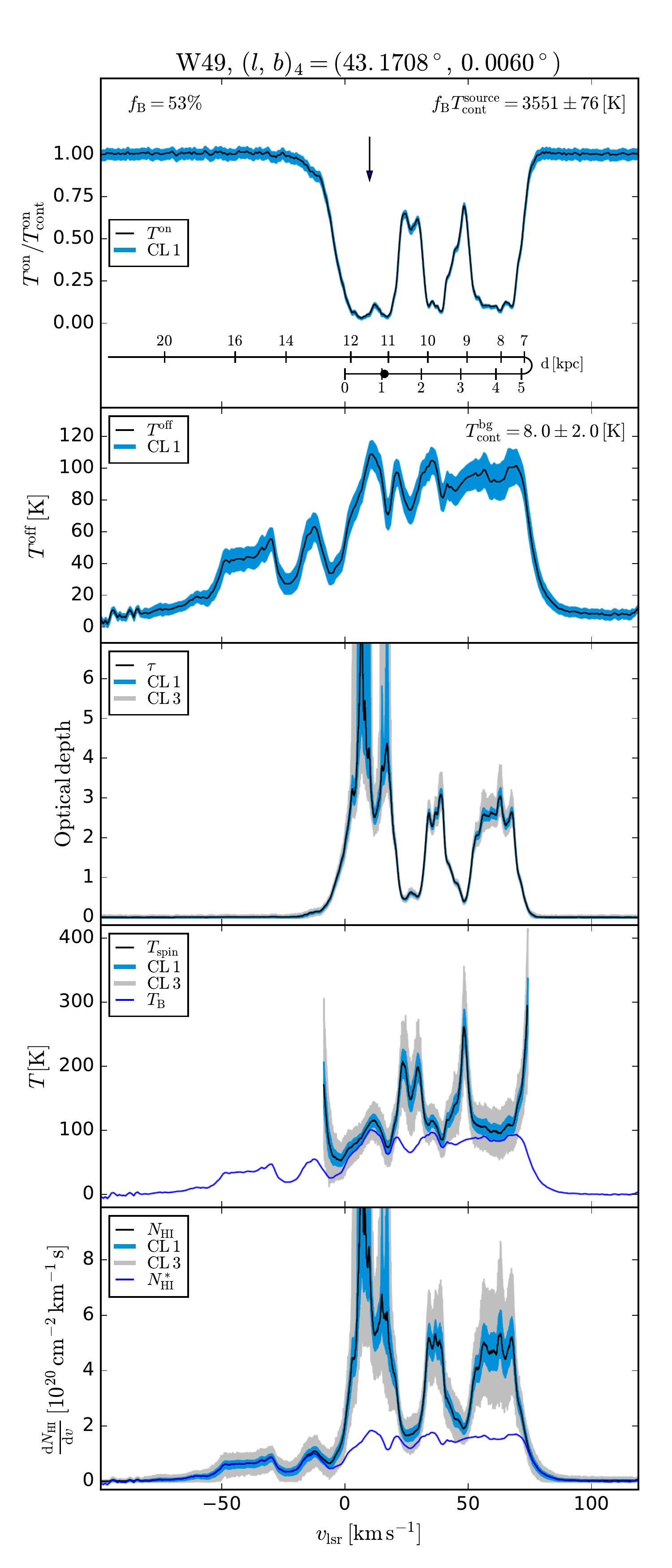}
\caption{As Fig.~\ref{fig:results_sgr_1n2} for source \object{W\,49}, sight lines (3) and (4).}%
\label{fig:results_w49_3n4}%
\end{figure*}

\begin{figure*}[!t]
\centering%
\includegraphics[width=0.48\textwidth,viewport=10 10 415 980,clip=]{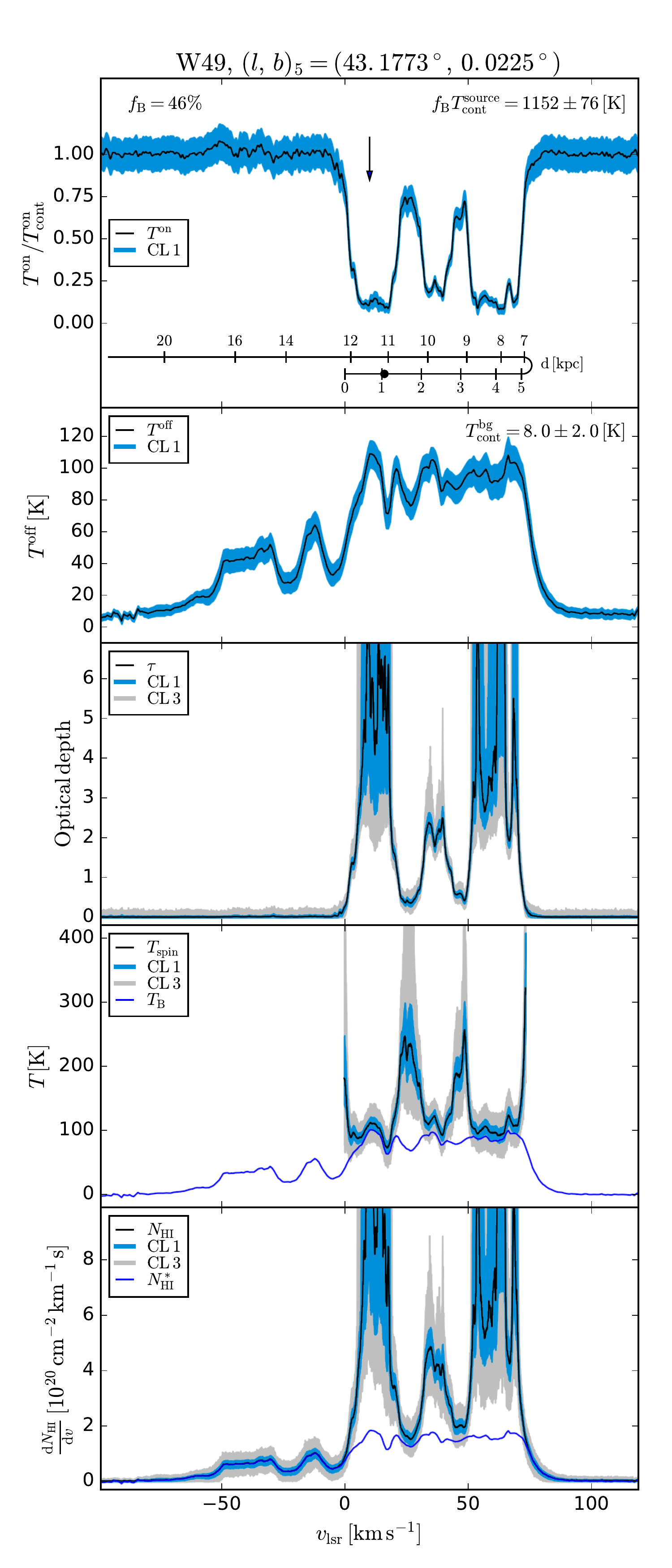}\hfill
\includegraphics[width=0.48\textwidth,viewport=10 10 415 980,clip=]{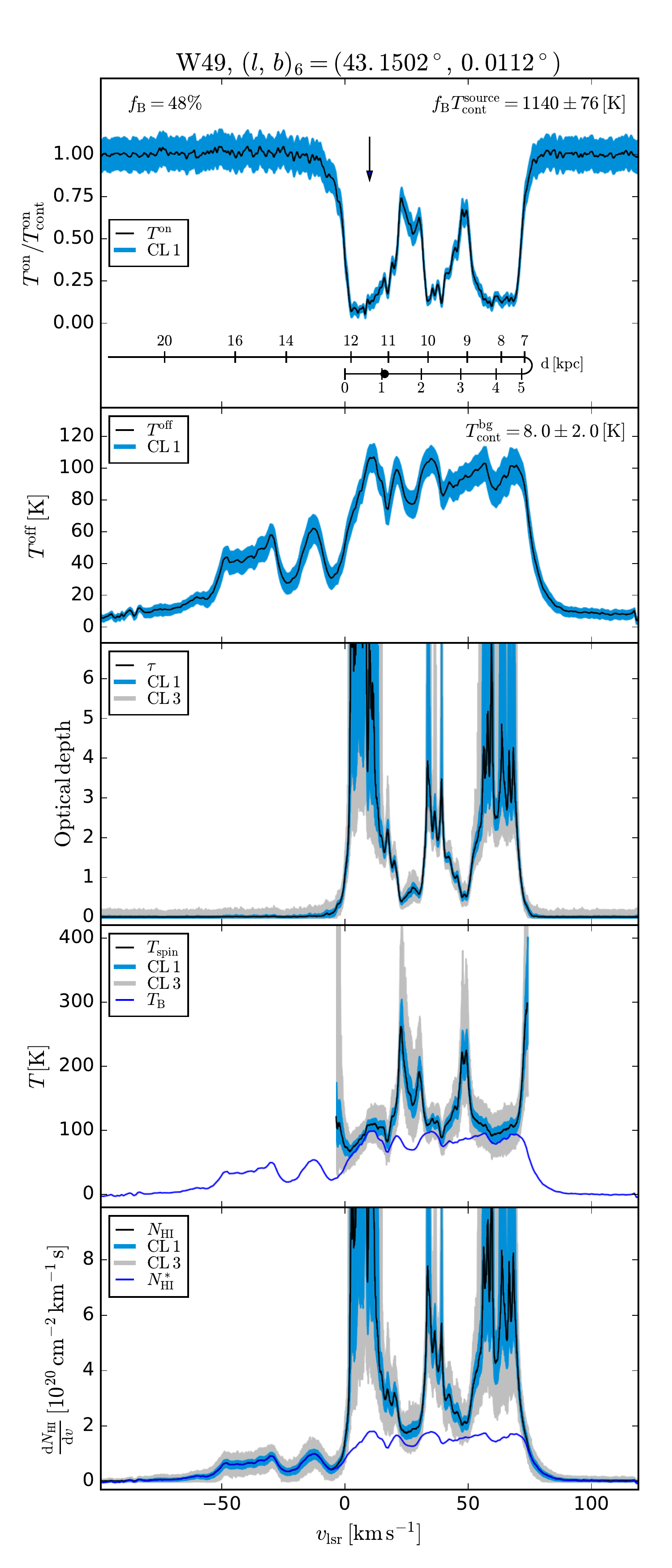}
\caption{As Fig.~\ref{fig:results_sgr_1n2} for source \object{W\,49}, sight lines (5) and (6).}%
\label{fig:results_w49_5n6}%
\end{figure*}

\begin{figure*}[!t]
\centering%
\includegraphics[width=0.48\textwidth,viewport=10 10 415 980,clip=]{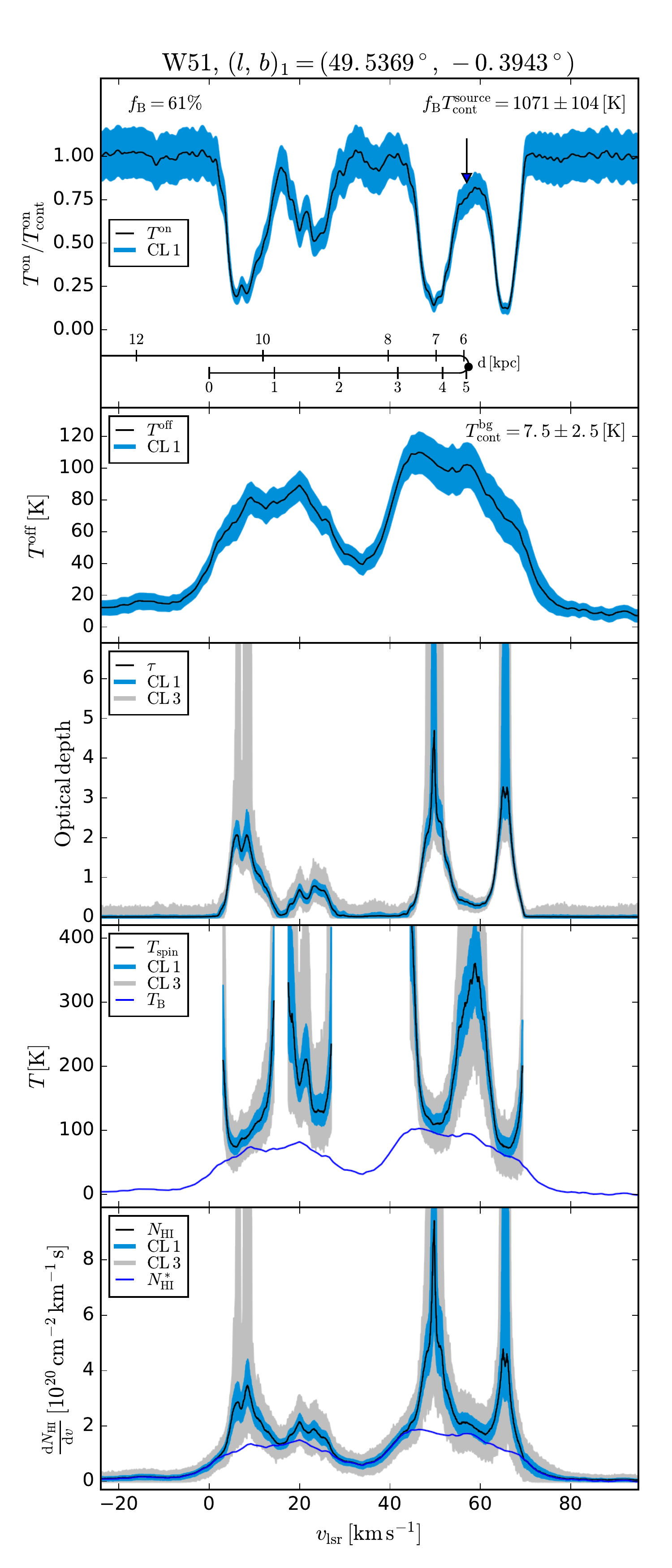}\hfill
\includegraphics[width=0.48\textwidth,viewport=10 10 415 980,clip=]{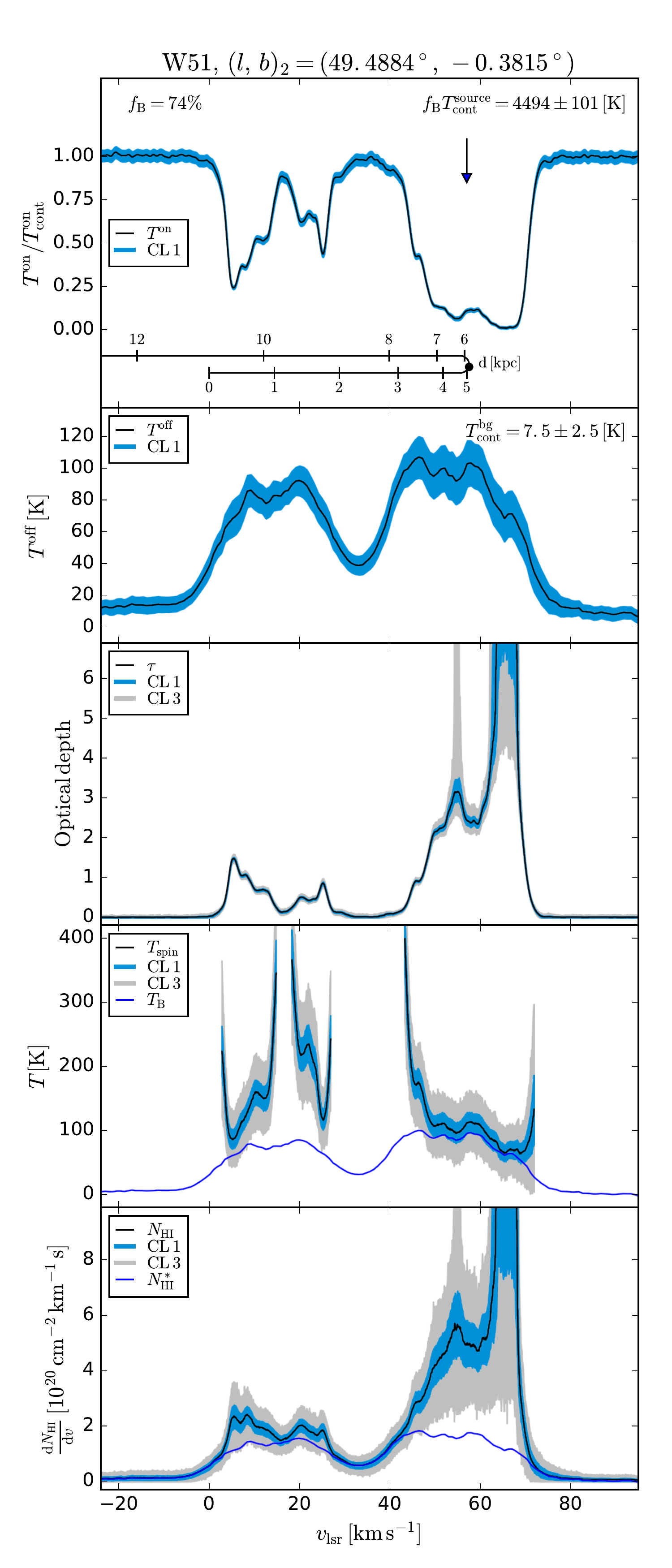}
\caption{As Fig.~\ref{fig:results_sgr_1n2} for source \object{W\,51}, sight lines (1) and (2).}%
\label{fig:results_w51_1n2}%
\end{figure*}

\begin{figure*}[!t]
\centering%
\includegraphics[width=0.48\textwidth,viewport=10 10 415 980,clip=]{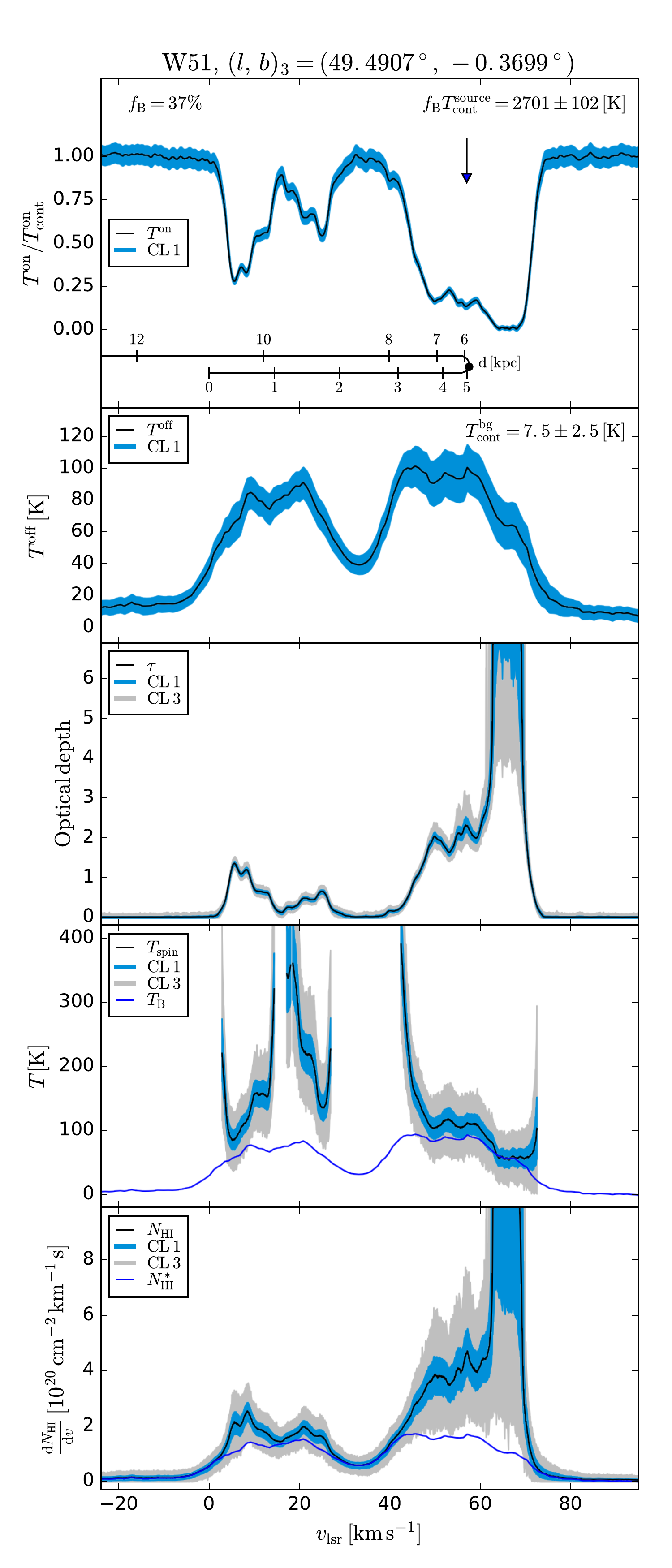}\hfill
\includegraphics[width=0.48\textwidth,viewport=10 10 415 980,clip=]{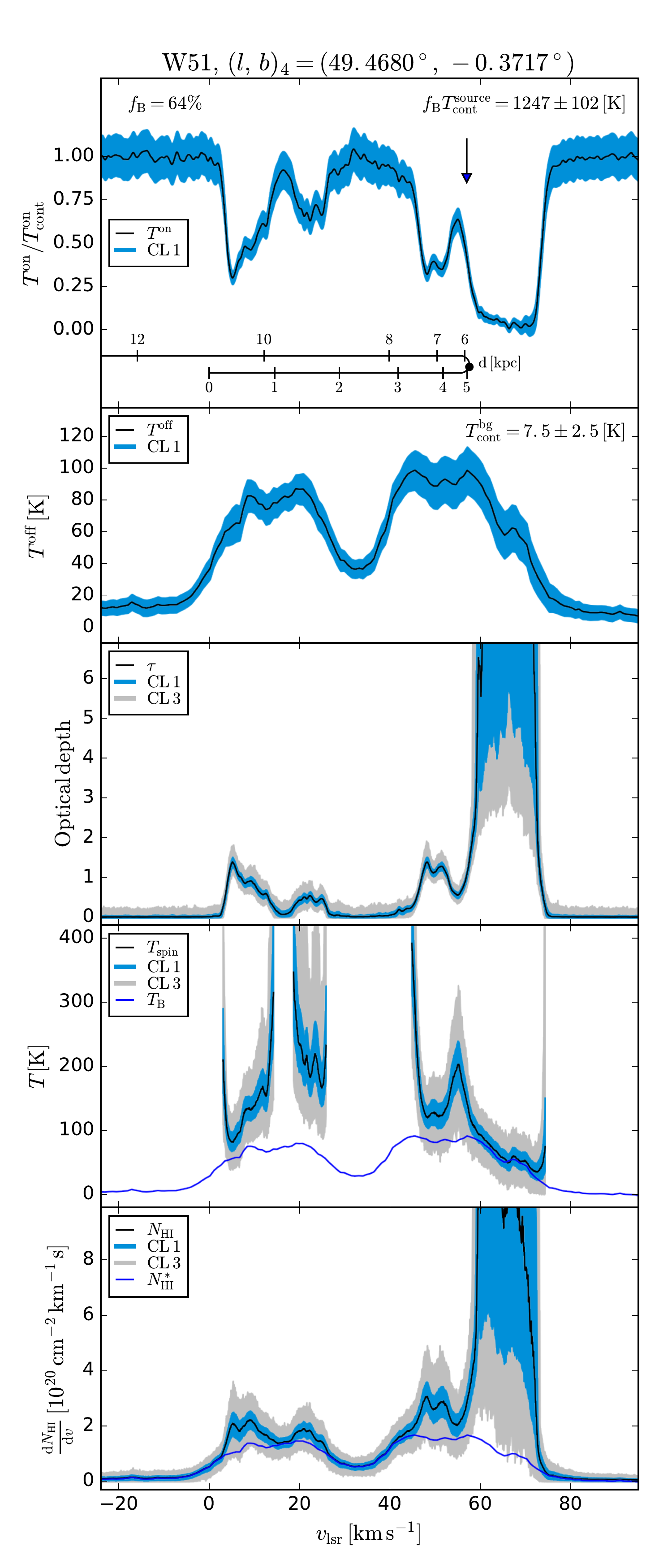}
\caption{As Fig.~\ref{fig:results_sgr_1n2} for source \object{W\,51}, sight lines (3) and (4).}%
\label{fig:results_w51_3n4}%
\end{figure*}

\begin{figure*}[!t]
\centering%
\includegraphics[width=0.48\textwidth,viewport=10 10 415 980,clip=]{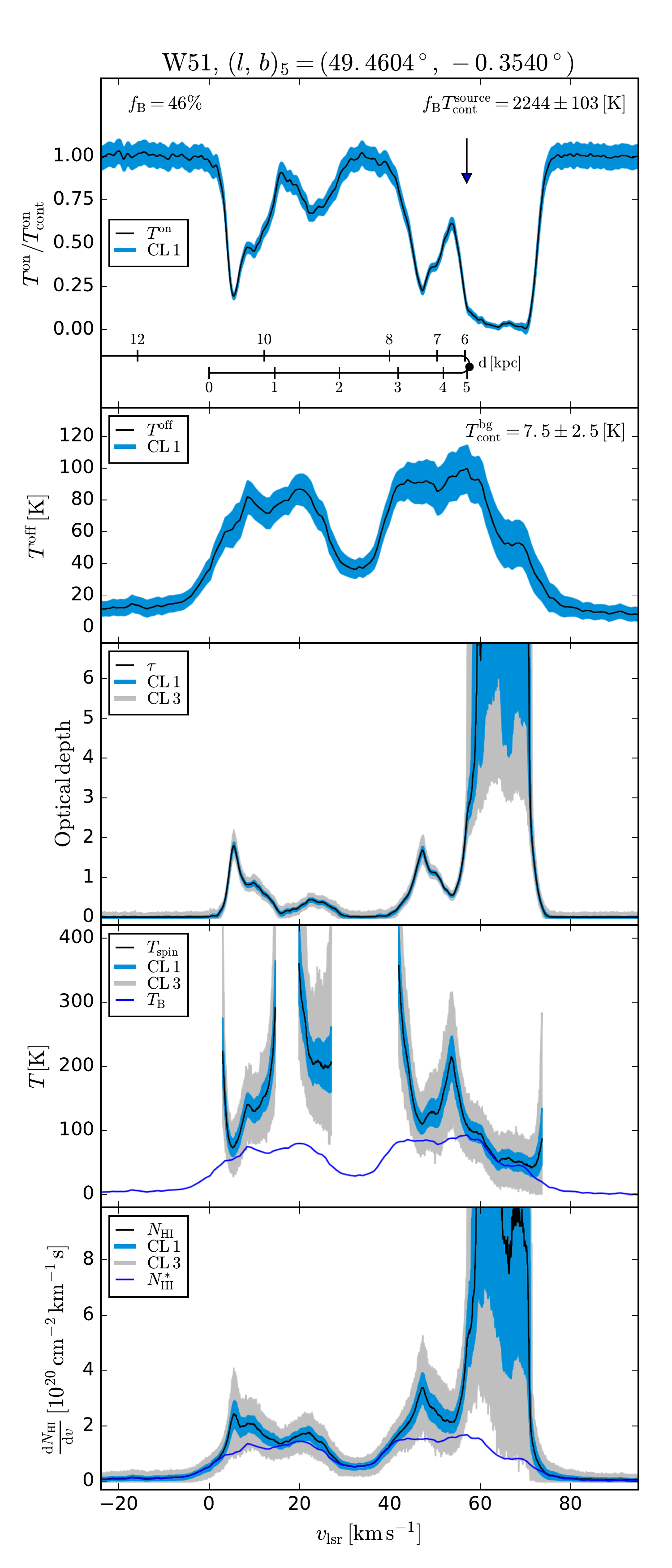}\hfill
\includegraphics[width=0.48\textwidth,viewport=10 10 415 980,clip=]{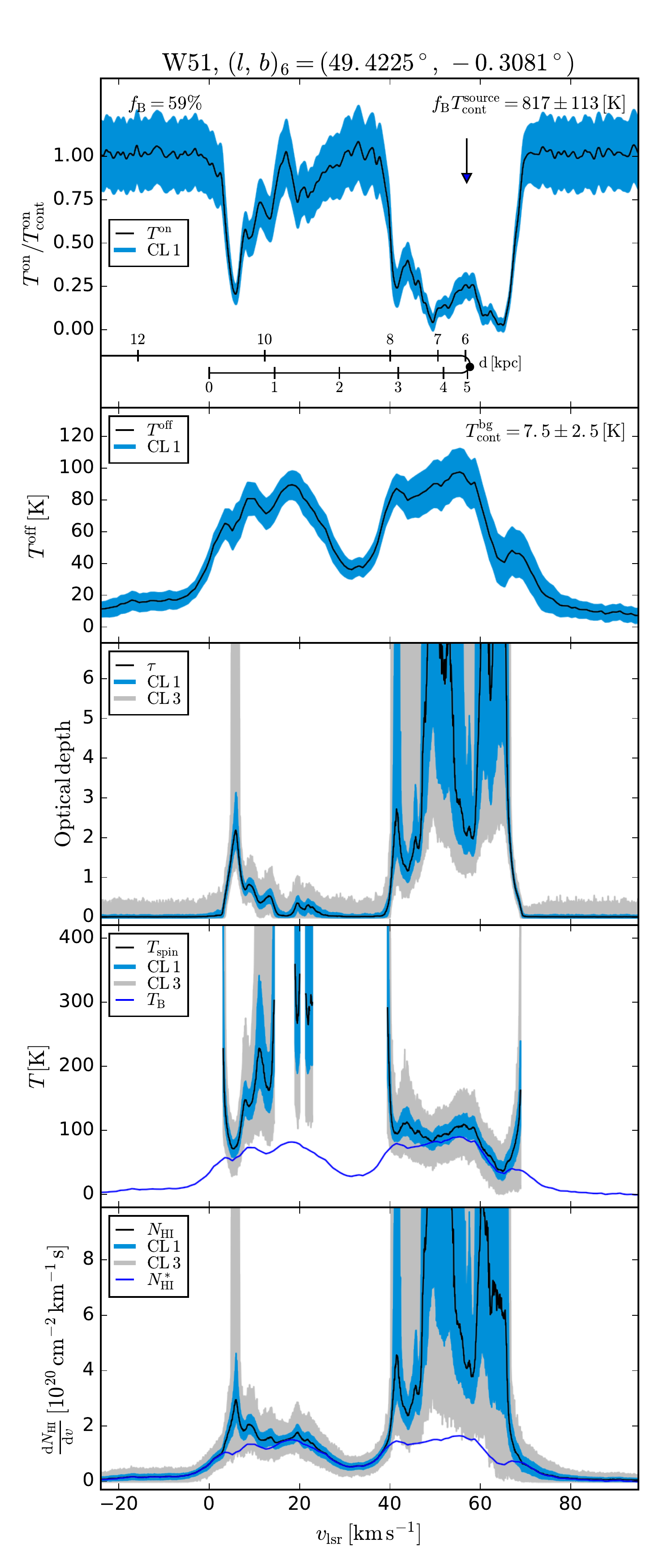}
\caption{As Fig.~\ref{fig:results_sgr_1n2} for source \object{W\,51}, sight lines (5) and (6).}%
\label{fig:results_w51_5n6}%
\end{figure*}

\begin{figure*}[!t]
\centering%
\includegraphics[width=0.48\textwidth,viewport=10 10 415 980,clip=]{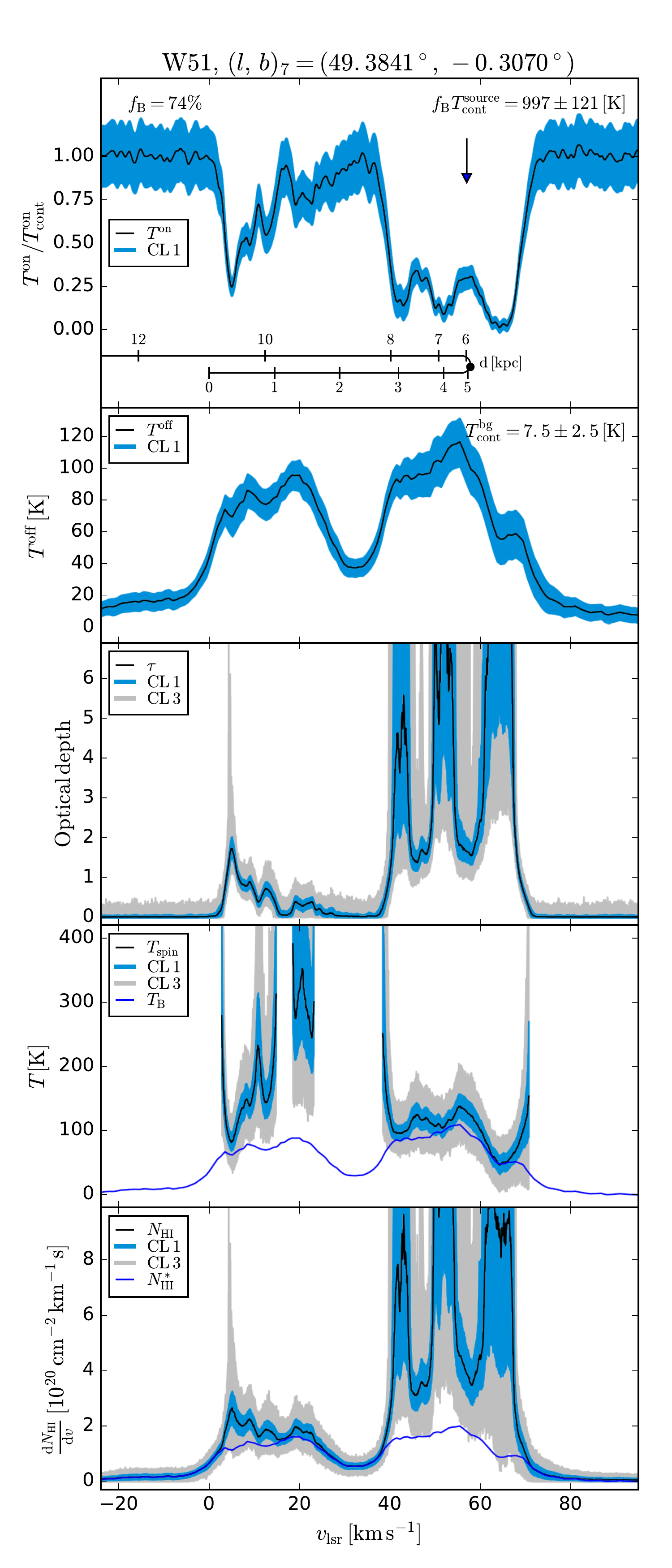}\hfill
\includegraphics[width=0.48\textwidth,viewport=10 10 415 980,clip=]{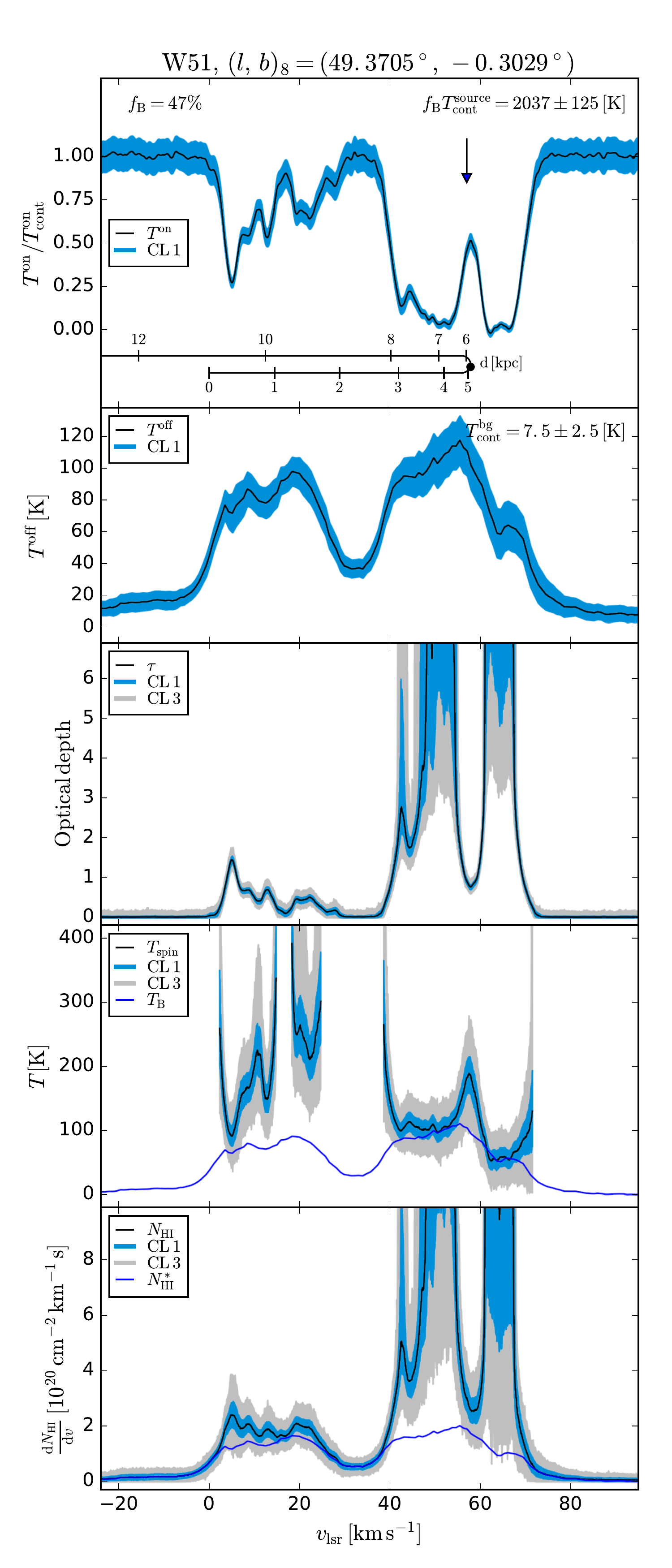}
\caption{As Fig.~\ref{fig:results_sgr_1n2} for source \object{W\,51}, sight lines (7) and (8).}%
\label{fig:results_w51_7n8}%
\end{figure*}

\begin{figure}[!t]
\centering%
\includegraphics[width=0.48\textwidth,viewport=10 10 415 980,clip=]{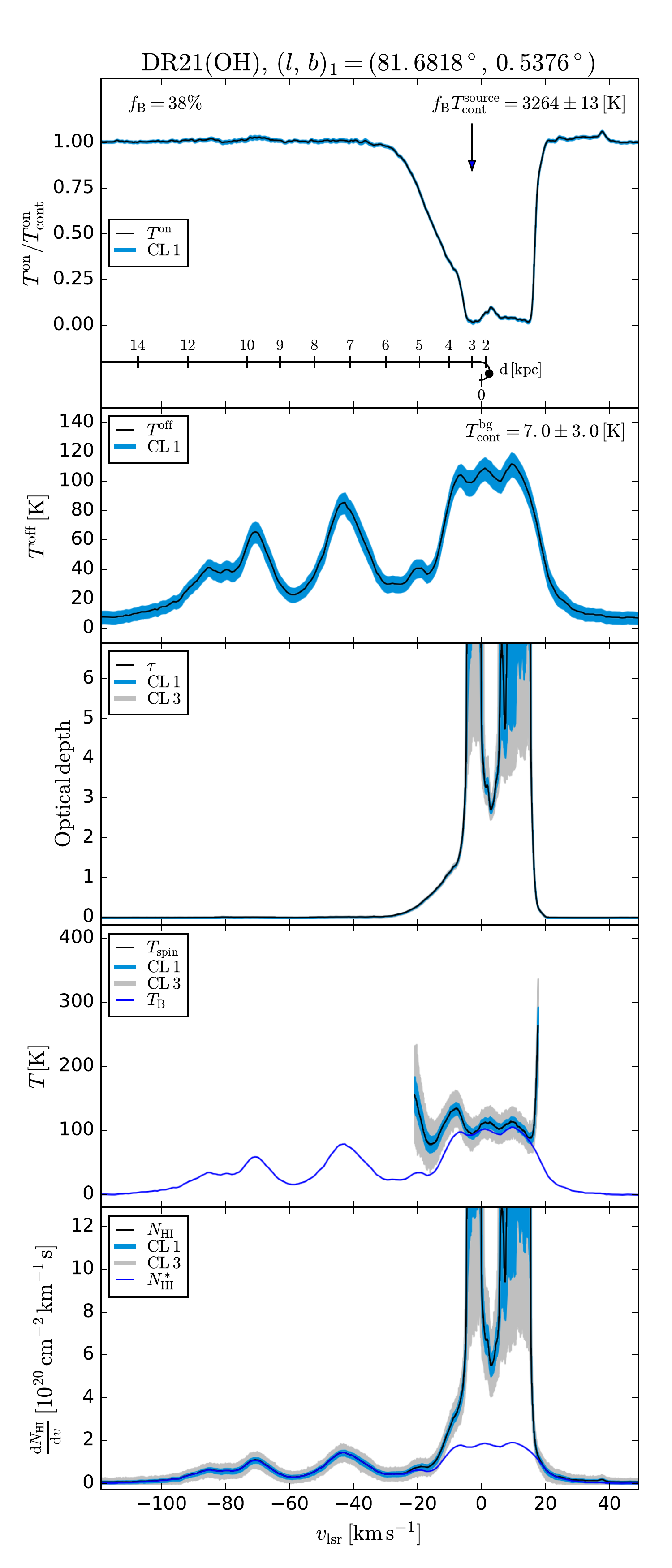}
\caption{As Fig.~\ref{fig:results_sgr_1n2} for source \object{DR\,21}, sight line (1).}%
\label{fig:results_dr21_1}%
\end{figure}

\end{document}